\documentclass[11pt]{article}
\pdfoutput=1
\usepackage[utf8]{inputenc}
\usepackage{cite}
\usepackage{epigraph}
\usepackage[dvipsnames]{xcolor}
\usepackage{amsmath,amssymb,amsbsy,amstext,amsthm,simplewick,amsfonts}
\usepackage{graphicx}
\usepackage{wrapfig}
\usepackage{upgreek}
\usepackage{bm} 
\usepackage{framed}
\usepackage{bbm}
\usepackage{textcomp}
\usepackage{tikz}
\usepackage{pifont}
\usetikzlibrary{matrix,shapes,fit,tikzmark,calc}
\usepackage[labelfont=bf]{caption}
\usepackage{adjustbox}
\usepackage{makecell}
\usepackage{bbold}
\usepackage{tcolorbox}
\usepackage{physics}
\usepackage{empheq}
\usepackage[normalem]{ulem}
\usepackage{enumitem}
\usepackage{braket} 
\usepackage{array}
\usepackage{dsfont}
\usepackage{ulem}
\usepackage{slashed}
\usepackage{titlesec}
\titleformat{\section}{\normalfont\fontsize{12}{16}\bfseries}{\thesection}{1em}{}
\numberwithin{equation}{section}

\def\be{\begin{equation}}
\def\ee{\end{equation}}

\def\cL{{\cal L}}
\def \cV{{\cal V}}
\def\cI {{\cal I}}

\def\ba{\begin{eqnarray}}
\def\ea{\end{eqnarray}}

\def\m{\mu}

\def\bfx{\textbf{x}}
\def\x{\xi}

\def\bfk{\textbf{k}}

\def\bfx{\textbf{x}}

\usepackage{caption}
\usepackage{subcaption}
\usepackage[framemethod=default]{mdframed}
\newmdenv[skipabove=7pt,
skipbelow=7pt,
rightline=false,
leftline=false,
topline=false,
bottomline=false,
backgroundcolor=gray!10,
linecolor=gray,
innerleftmargin=5pt,
innerrightmargin=5pt,
innertopmargin=5pt,
innerbottommargin=5pt,
leftmargin=0cm,
rightmargin=0cm,
linewidth=4pt]{eBox}
\newmdenv[skipabove=7pt,
skipbelow=7pt,
rightline=false,
leftline=false,
topline=false,
bottomline=false,
backgroundcolor=gray!10,
linecolor=gray,
innerleftmargin=5pt,
innerrightmargin=5pt,
innertopmargin=-5pt,
innerbottommargin=5pt,
leftmargin=0cm,
rightmargin=0cm,
linewidth=4pt]{eBox2}

\definecolor{blue3}{RGB}{31,119,180}
\definecolor{red3}{RGB}{214,39,40}
\definecolor{orange3}{RGB}{255,127,14}
\definecolor{green3}{RGB}{44,160,44}

\usepackage{colortbl}
\definecolor{lightgreen}{cmyk}{0.2, 0, 0.2, 0.2}
\definecolor{lightgray}{cmyk}{0.1,0.2,0,0.1}
\definecolor{lightgray2}{cmyk}{0.1,0.1,0,0.1}
\definecolor{pygreen}{RGB}{30, 100, 64}

\usepackage{hyperref}
\hypersetup{colorlinks=true,linkcolor=teal,citecolor=pygreen,urlcolor=green3,pdfencoding=auto}

\makeatletter
\newlength{\apb@width}
\newcommand{\autoparbox}[2][c]{\settowidth{\apb@width}{#2}\parbox[#1]{\apb@width}{#2}}

\makeatother


\def\x{{\bm x}}
\def\y{{\bm y}}

\def\nn{\nonumber}

\def\beq{\begin{equation}}
\def\eeq{\end{equation}}

\allowdisplaybreaks[1]
\begin{document}


\begin{titlepage}
\setcounter{page}{1} \baselineskip=15.5pt

\thispagestyle{empty}

\renewcommand*{\thefootnote}{\fnsymbol{footnote}}

\begin{center}
{\fontsize
{15}{15} \bf  The Massive Flat Space Limit of Cosmological Correlators}
\end{center}

\vskip 18pt
\begin{center}
\noindent
{\fontsize{12}{18}\selectfont Sebastián Céspedes\footnote{\tt s.cespedes-castillo@imperial.ac.uk} and Sadra Jazayeri\footnote{\tt s.jazayeri@imperial.ac.uk}}
\end{center}

\begin{center}
\vskip 8pt
\textit{Department of Physics, Imperial College, London, SW7 2AZ, UK} 
\end{center}


\vspace{1.4cm}

\noindent Identifying useful \textit{flat-space limits} for cosmological correlators, where they can be expressed in terms of observables in Minkowski space is nontrivial due to their scale-invariant nature. In recent years, it has been shown that \textit{momentum-space} correlators encode flat-space amplitudes at specific singularities that emerge in the complex plane of their kinematics after analytical continuation. This flat-space limit is \textit{massless} in the sense that the amplitude corresponds to the ultraviolet regime of the associated flat-space process, where the masses of the internal propagators are effectively zero. In this paper, we introduce a novel \textit{massive flat-space (MFS) limit}, in which the internal masses in the corresponding flat-space Feynman graph remain finite. Our proposal applies to arbitrary graphs with light external legs and heavy internal lines, using a \textit{double-scaling} limit. In this limit, the external energies, treated as independent variables, approach zero in inverse proportion to the propagator masses, which are sent to infinity. We present a general \textit{reduction formula} that expresses diagrams in this limit in terms of amputated Feynman graphs in flat space. Our findings underscore the deep connections between the rich structure of \textit{massive Feynman integrals} and the properties of cosmological correlators involving the exchange of heavy fields. Using this reduction formula, we compute sample one-loop contributions from heavy particles to inflationary correlators in the small sound-speed regime, revealing novel bispectrum shapes. The non-Gaussian signals we uncover, which are especially pronounced around the equilateral configuration, cannot be reproduced by adding local terms to the effective field theory of single-field inflation. Instead, they are captured by incorporating prescribed \textit{spatially non-local} operators into the EFT.
\\


\end{titlepage}


\setcounter{tocdepth}{2}
{
\hypersetup{linkcolor=black}
\tableofcontents
}

\renewcommand*{\thefootnote}{\arabic{footnote}}
\setcounter{footnote}{0} 

\newpage 
\section{Introduction}
The fundamental observables in cosmology are the equal-time correlation functions of quantum fields at the end of inflation, also known as \textit{Cosmological Correlators}. At the practical level, these late time correlators are special because we do not have direct access to the time evolution during inflation. Instead, what we can measure are the statistical properties of matter and radiation distributions at late times, which, following the classical evolution of the universe forward in time, can be derived from the universe's wavefunction on the future boundary of the quasi-de Sitter spacetime during inflation. The correlators derived from the Born rule using this wavefunction provides us with an integrated history of the universe during inflation within a single snapshot. 
At a more conceptual level, boundary correlators are significant because, in the presence of dynamical gravity, observables are sharply defined only on the spacetime asymptotic boundaries, such as the S-matrix in flat spacetime and CFT correlators in AdS. Cosmology is no exception to this principle, as good observables are likewise expected to be defined only on the asymptotic future boundary of the expanding universe \cite{Maldacena:2002vr}. In addition to these theoretical aspects, correlators are also of immense phenomenological importance, as they encode a wealth of information about the microscopic details of inflation, including its particle content, mass spectrum, interactions and potential clues about its UV completion (see, e.g., \cite{Achucarro:2022qrl} and references therein). 

In light of their phenomenological and theoretical significance, cosmological correlators have become a central topic of active study in recent years. Drawing inspiration from the remarkable successes of the scattering amplitude program, significant effort has been dedicated to discovering a set of optimal consistency conditions that enable the \textit{bootstrapping} of these correlators. These conditions often follow from fundamental principles, including unitarity, locality, analyticity, and symmetries.\cite{Maldacena:2011nz,Bzowski:2011ab,Creminelli:2011mw, Mata:2012bx,Bzowski:2012ih, Bzowski:2013sza, Arkani-Hamed:2015bza,Arkani-Hamed:2017fdk, Arkani-Hamed:2018kmz, Arkani-Hamed:2018bjr, Baumann:2019oyu, Benincasa:2019vqr,COT, Cespedes:2020xqq, Baumann:2020dch, Benincasa:2020aoj, Pajer:2020wxk,   Jazayeri:2021fvk, Baumann:2021fxj, Goodhew:2021oqg,  Sleight:2021plv,  Bonifacio:2021azc, Benincasa:2022gtd, Penedones:2023uqc, AguiSalcedo:2023nds, Albayrak:2023hie, Loparco:2023rug, Lee:2024sks, SalehiVaziri:2024joi, Goodhew:2024eup,Stefanyszyn:2024msm,Stefanyszyn:2023qov,Melville:2024ove,Melville:2023kgd}. This perspective, adopted in the \textit{Cosmological Bootstrap} program \cite{Baumann:2022jpr}, has led to an efficient formulation of correlators, circumventing some of the complexities encountered in the traditional in-in formalism, such as the challenging structure of nested time integrals. Using a diverse array of tools—including boundary differential equations \cite{Gomez:2021qfd, Gomez:2021ujt,  Arkani-Hamed:2023bsv, Arkani-Hamed:2023kig, Chen:2023iix, De:2023xue, Hang:2024xas, Baumann:2024mvm,  Benincasa:2024ptf, Grimm:2024mbw, De:2024zic, Chen:2024glu}, cutting rules \cite{Goodhew:2021oqg, Melville:2021lst, Tong:2021wai, Qin:2023bjk, Ghosh:2024aqd}, Mellin transformations \cite{Sleight:2019hfp, Sleight:2019mgd, Sleight:2020obc, Chopping:2024oiu}, recursion relations \cite{Jazayeri:2021fvk}, dispersion relations \cite{Meltzer:2021bmb, Meltzer:2021zin, Salcedo:2022aal, Liu:2024xyi}, and spectral representations \cite{Hogervorst:2021uvp, DiPietro:2021sjt, Xianyu:2022jwk, DiPietro:2023inn, Werth:2024mjg}—the bootstrap approach has greatly advanced our analytical understanding of the structure of cosmological correlators. This approach has also led to useful analytical expressions for correlators that are otherwise exceedingly difficult to obtain within the conventional bulk framework (see, e.g., \cite{Cabass:2021fnw, Jazayeri:2022kjy, Pimentel:2022fsc, Qin:2022fbv, Cabass:2022jda, Bonifacio:2022vwa, Wang:2022eop, Armstrong:2023phb, Qin:2023ejc, Jazayeri:2023xcj, Xianyu:2023ytd, DuasoPueyo:2023kyh,Chakraborty:2023qbp,Chakraborty:2023eoq, Chowdhury:2023khl, Aoki:2024uyi, Qin:2024gtr, Liu:2024str, Chowdhury:2023arc, Chowdhury:2024snc, Anninos:2024fty}).

A powerful consistency condition in the bootstrap program is the so-called \textit{amplitude limit} of correlators. In perturbation theory, it has been demonstrated that as the \textit{total energy} of a perturbative graph approaches zero, the correlator becomes proportional to the on-shell scattering amplitude associated with the same diagram in flat space \cite{Raju:2012zr, Pimentel:2012tw, Arkani-Hamed:2017fdk, Pajer:2020wxk, COT}.
\begin{figure}
    \centering
    \includegraphics[scale=0.55]{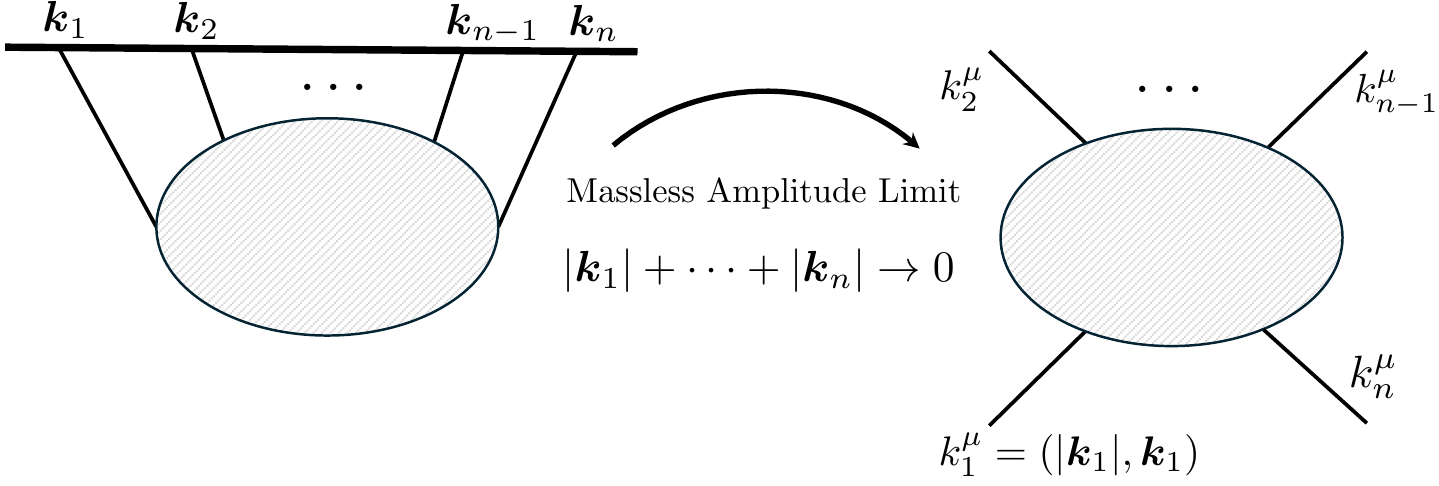}
    \caption{The amplitude limit of correlators, Eq.\,\eqref{amplitudelimit}.}
    \label{masslessFSL}
\end{figure}

In more detail, consider a generic in-in diagram contributing to the $n$-point function of a massless field in momentum space---Figure \ref{masslessFSL}. The blob in the figure represents an arbitrary structure of vertices and internal lines. This $n$-point function can be expressed as
\begin{align} 
\langle \phi(\bm{k}_1)\phi(\bm{k}_2)\dots \phi(\bm{k}_n)\rangle = F_n(\bm{k}_i)(2\pi)^3\delta^3\bigg(\sum_{i=1}^n \bm{k}_i\bigg),
\end{align} 
where $\bm{k}_i$ are the external momenta of the fields. Assuming the Bunch-Davies initial condition, $F_n(\bm{k}_i)$ remains finite for all real-valued external kinematics, i.e., $\bm{k}_i \in \mathbb{R}^3$. However, upon analytic continuation as a function of $k_i = |\bm{k}_i|$ (also referred to as the external energies\footnote{Although energy is not conserved in an expanding background, the terminology remains useful when considering the flat-space limit, where time-translation symmetry is effectively restored, allowing a comoving energy to be assigned to each mode.}), $F_n(\bm{k}_i)$ develops a special type of singularity. This singularity arises when the total energy of the graph, defined as
\begin{align}
    k_T=\sum_{i=1}^n k_i\,,
\end{align}
approaches zero. Near this limit, $F_n$ exhibits the following behavior \cite{COT}:
\begin{align}
\label{amplitudelimit}
    \lim_{k_T\to 0}F_n(\bm{k}_i)\propto \text{Re}\bigg\{\dfrac{i^{n+1}}{(i\,k_T)^{\Delta_T}}{\cal A}(k_1^\mu,\dots,k_n^\mu)\bigg\}\,,
\end{align}
where $k_i^\mu$ are the following set of fictitious null four-momenta:
\begin{align}
    k_i^\mu=(k_i,\bm{k}_i)\,,
\end{align}
which characterize the incoming particles in the flat-space scattering process associated with the same graph. ${\cal A}(k_i^\mu)$ represents the corresponding $n$-point scattering amplitude, and $\Delta_T$ quantifies the degree of divergence of the total energy singularity. Dimensional analysis implies that \cite{Pajer:2020wxk}
\begin{align}
    \Delta_T=1+\sum_{a=1}^V(\Delta_a-4)\,,
\end{align}
where $\Delta_a$ is the mass dimension of the operator acting at the $a-$th vertex\footnote{For $\Delta_T = 0$, $F_n$ exhibits a logarithmic divergence as $k_T \to 0$. In contrast, for $\Delta_T < 0$, while $F_n$ remains finite near $k_T = 0$, its $(-\Delta_T)$-th derivative with respect to $k_T$ eventually becomes singular.}. 

It is important to note that the amplitude appearing as the residue of the singularity in Eq.\,\eqref{amplitudelimit} is the UV limit of the corresponding scattering process in flat space. This implies that the propagators in the internal lines of the associated Feynman graph are massless, allowing the limit to also be termed the \textit{massless flat-space limit} of correlators. 

The aim of this work is to define an alternative flat-space limit of correlators in which the propagator masses survive (see \cite{Marotta:2024sce}, for a recent work in this direction). Retrieving information about the masses in the flat-space limit offers several advantages. Conceptually, it allows us to see how how the rich structure of massive Feynman integrals (see, e.g., \cite{Weinzierl:2022eaz,Boos:1990rg,vanOldenborgh:1990yc,tHooft:1978jhc}) is encoded in cosmological correlators. As discussed previously, in the standard flat-space high-energy limit, all massive exchange processes become degenerate with massless exchanges, which possess a much simpler analytic structure. Thus, identifying an alternative limit that avoids this degeneracy is highly desirable. In addition, from a bootstrap viewpoint, the analytical expressions for correlators involving the exchange of massive fields depend not only on the external momenta, but also on the masses of the internal propagators. Therefore, a flat-space limit that varies with respect to these additional mass parameters would provide a much more powerful constraint compared to the amplitude limit.

Finally, the \textit{massive flat-space limit} developed in this work can be used to extract observational signals imprinted by massive fields on the correlation functions of scalar fluctuations with reduced sound speed, $c_s\ll 1$. In particular, we focus on massive one-loop contributions to inflationary observables, such as the bispectrum. Phenomenologically, calculating one-loop contributions is crucial because they capture the leading-order effects of exchanging various species during inflation, such as fermions and charged bosons, which do not couple to curvature perturbations at tree level\footnote{In this work, we do not discuss one-loop effects induced by light fields and their infrared divergences. The interested reader may refer to, e.g., \cite{Weinberg:2005vy,Senatore:2009cf,Baumgart:2019clc,Gorbenko:2019rza,Green:2020txs,Wang:2021qez,Lee:2023jby,Cespedes:2023aal,Beneke:2023wmt,Ballesteros:2024qqx}.}. Such contributions might also be enhanced by additional color factors. 

Recently, there has been a wide range of studies in the \textit{cosmological collider signal} left by such massive fields on inflationary correlators. This signal captures the quantum oscillations of heavy fields on super-Hubble scales, manifesting as distinctive oscillations in specific soft limits of the bispectrum and trispectrum (see, e.g., \cite{Arkani-Hamed:2015bza, Chen:2015lza,Chen:2016cbe,Lee:2016vti, Chen:2016uwp,Chen:2016hrz, Kehagias:2017cym, Chen:2018sce, Kumar:2019ebj, Liu:2019fag, Wang:2020ioa, Sou:2021juh, Li:2020xwr, Lu:2019tjj, Lu:2021wxu, Pinol:2021aun, Cui:2021iie, Tong:2022cdz, Reece:2022soh,Chen:2022vzh, Qin:2022lva, Craig:2024qgy,Quintin:2024boj,Bodas:2024hih,Gasparotto:2024bku,Pajer:2024ckd}). Despite significant progress in this field, exact results for massive loop diagrams, especially beyond the squeezed limit of the bispectrum or the collapsed limit of the trispectrum, remain limited due to the mathematical complexity of massive exchange processes. For recent works on massive bubble diagrams, see \cite{Xianyu:2022jwk,Qin:2024gtr}.

\begin{figure}
    \centering
    \includegraphics[scale=0.5]{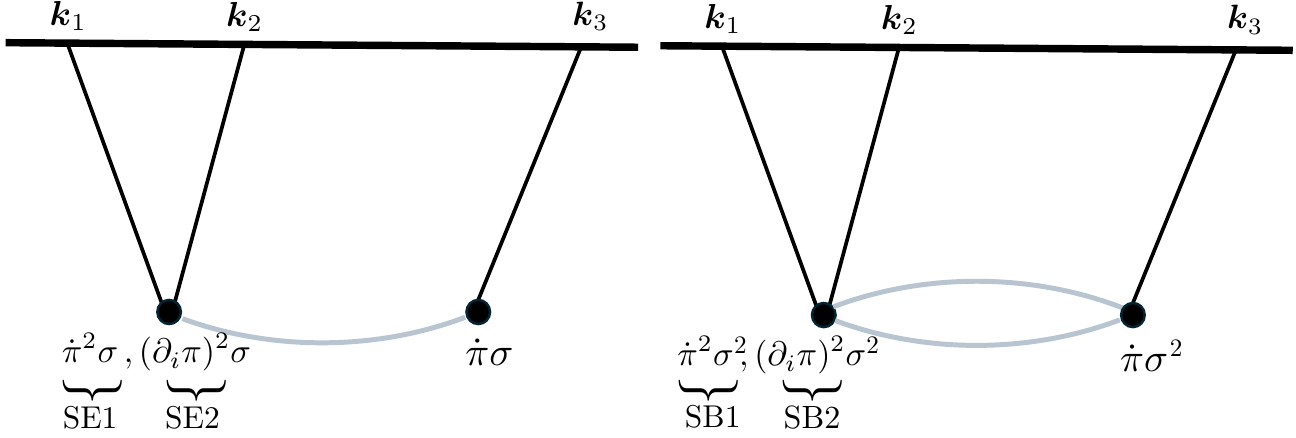}
    \caption{The tree-level and one-loop bubble contributions to the bispectrum, see Eq.\,\eqref{diagramsofinterest}.}
    \label{figdiaginterest}
\end{figure}

In this work, using a reduction formula in our newly proposed flat-space limit, we compute novel one-loop contributions from massive scalars to inflationary correlators (Fig.\,\ref{figdiaginterest}), particularly the bispectrum. Our computation works in the regime where the scalar fluctuations propagate at a small speed of sound, i.e., $c_s\ll 1$, and the mass of the exchanged scalar lies within the window
$H\ll m\lesssim {\cal O}(1)H/c_s$; see \cite{Jazayeri:2022kjy,Jazayeri:2023xcj,Jazayeri:2023kji}, for recent studies of this limit at tree-level. 

Since we focus on the heavy limit of the exchanged field, our computation does not capture the squeezed-limit cosmological collider oscillations, as these are exponentially suppressed by the Boltzmann factor $\exp(-\pi m/H)$. However, we uncover intriguing features in the one-loop-induced bispectrum around the equilateral configuration that cannot be replicated by simply adding local operators to the EFT action. Instead, we demonstrate that these features can be reproduced by adding certain \textit{non-local} operators into the EFT of single-field inflation.

The rest of the paper is organized as follows. In the next subsection, we summarise our \textit{Massive Flat-Space (MFS) Limit} and the elements that goes into its proof for the reader's convenience. These will be discussed in full detail in Section \ref{MFSlimitSection}. In this section, after reviewing the in-in formalism, we present a reduction formula that expresses correlators in the MFS limit in terms of specific massive Feynman diagrams in flat space. In Section \ref{sec:MFSinin}, we prove our reduction formula by directly taking the MFS limit of the in-in expressions for two specific examples, namely the tree-level exchange and the one-loop, bubble graph (in Appendix \ref{app:spectral}, for the one-loop example, we provide another proof of the MFS behaviour based on the Källén–Lehmann representation of composite operators). In Section \ref{MFSfromeffectiveaction}, exploiting the Wilsonian effective action formalism in dS, we provide a general proof for the reduction formula for an arbitrary graph with heavy internal lines. Section \ref{backtoonshellSection} proposes an analogous reduction formula for correlators of scalar fluctuations in the EFT of inflation (relations to a non-local EFT of single-field inflation is discussed in Appendix \ref{EFTintout}). This reduction formula will be exploited in Section \ref{PhononColliderSection} to compute the bispectrum induced by the tree-level and one-loop exchanges of heavy fields, in the $c_s\to 0$ limit.

\subsection{Summary of the results}

\begin{figure}
    \centering
    \includegraphics[scale=0.6]{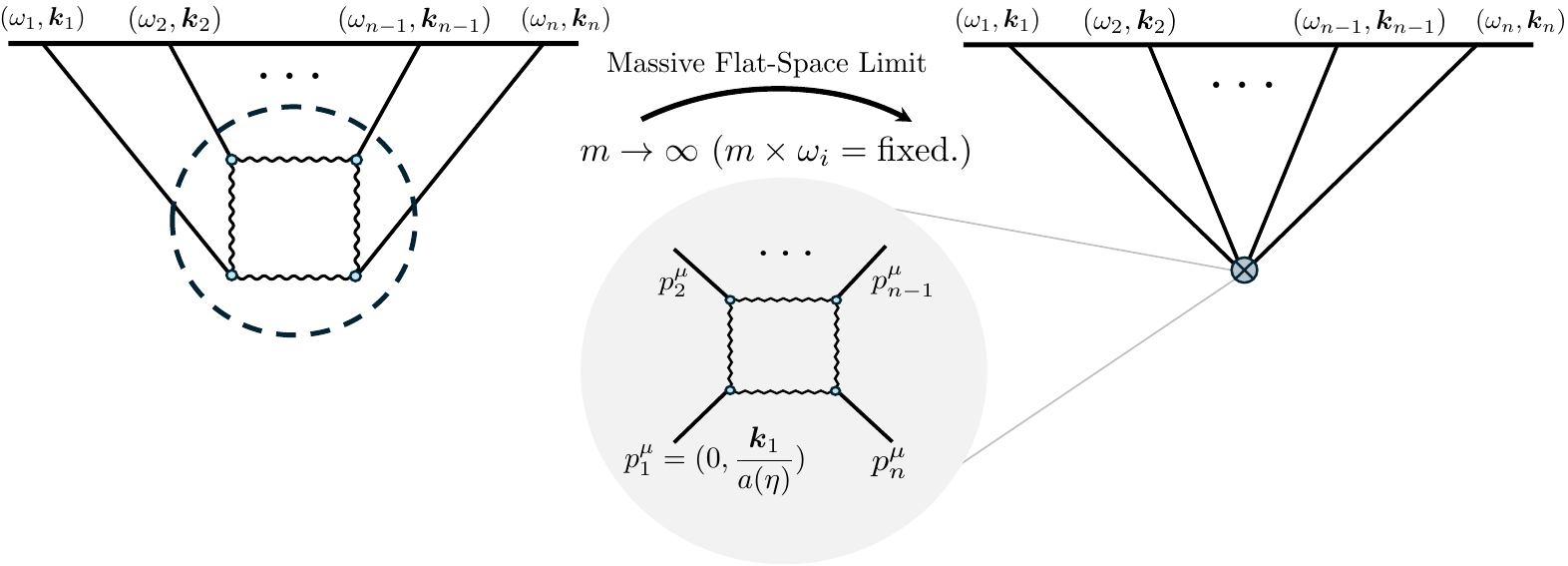}
    \caption{The massive flat-space limit of off-shell diagrams.}
    \label{fig:heavygraphintro}
\end{figure}
Before discussing our massive flat-space limit, let us briefly review why masses vanish in the amplitude limit by studying a tree-level example (see Fig.\,\ref{SESBcc}). Consider the four-point function of a conformally coupled field $\phi$ induced by the single exchange of a heavy scalar $\sigma$, assuming that they interact via the cubic operator $\phi^2\,\sigma$. The $s-$channel component of this diagram is given by
\begin{align}
\label{FSEintro}
&F_4(k_1,k_2,k_3,k_4,s)= \sum_{\pm\pm}(\pm i)(\pm i)\int_{-\infty(1\mp i\epsilon)}^{0} \frac{d\eta}{\eta^2}\frac{d\eta'}{\eta'^2} e^{\pm i(k_1+k_2)\eta} \,e^{\pm i(k_3+k_4)\eta'} G_{\pm\pm}(s, \eta, \eta'),
\end{align}
where $\eta,\eta'$ are the conformal times attached to the vertices, and $G_{\pm\pm}(s,\eta,\eta')$ are the \textit{bulk-to-bulk} propagators associated with the heavy field, given by Eq.\,\eqref{props}, with $s=|\bm{k}_1+\bm{k}_2|$ denoting the exchanged momentum. 

It is useful to expand the integrand around the conformal time corresponding to one of the vertices, e.g., $\eta$, by substituting $\eta' = \eta + \Delta\eta$. As we approach $k_T=0$, the $++$ and $--$ contributions to $F_4$ are the only components that become singular. The singular behavior of these components is dictated by the early-time limit of the integrand, that is, when $\eta\propto k_T^{-1}\to -\infty$, while $\Delta\eta$ is maintained finite. In this early-time limit, 
\begin{align}
    &\lim_{k_T\to 0}F_4 \propto \text{Re}\left[\int^0_{-\infty(1-i\epsilon)} \dfrac{d\eta}{\eta^4}\,e^{ik_T\eta} \left(\int_{-\infty}^{+\infty} d\Delta\eta\,e^{i(k_3+k_4)\Delta\eta}\,G_{++}(s,\eta,\eta+\Delta\eta)\right)\right]\,.
\label{Fnktzerolimit}
\end{align}
Within the bulk-to-bulk propagators $G_{++}$ above, the conformal time $\eta$ is taken to $-\infty$. Therefore, in this limit, the mass of the propagator becomes negligible compared to its blue-shifted physical kinetic energy, defined by $E = H|s\,\eta|$, where $H$ is the Hubble rate. This implies that the Hankel functions appearing in the propagators (Eq.\,\eqref{sigmaplusminus}) can be well approximated by massless plane-waves in flat space. Consequently, as long as we are concerned with the leading order singular behaviour around $k_T=0$, we can effectively replace the time-ordered propagator with
\begin{align}
    G_{++}(s,\eta,\eta+\Delta\eta)\to H^2\eta^2\,G_F(s,0,\Delta\eta)\,,
\end{align}
where $G_F$ is the \textit{massless Feynman propagator} in flat space, 
\begin{align}
    G_F(s,\eta_1,\eta_2)=\dfrac{1}{2s}\left(e^{-is(\eta_1-\eta_2)}\,\theta(\eta_1-\eta_2)+\eta_1\leftrightarrow\eta_2\right)\,.
\end{align}
Substituting the asymptotic limit of the propagator inside the parenthesis ($\dots$) in Eq.\,\eqref{Fnktzerolimit}, along with the incoming plane wave $\exp(i(k_3+k_4)\Delta\eta)$, reproduces the on-shell two-to-two amplitude with vanishing internal and external masses, namely
\begin{align}
    {\cal A}_4=\dfrac{1}{-(k_1+k_2)^2+(\bm{k}_1+\bm{k}_2)^2+i\epsilon}=\dfrac{1}{(k_1^\mu+k_2^\mu)^2+i\epsilon}\,.
\end{align}
Meanwhile, the rest of the integrand, i.e., $\int\frac{d\eta}{\eta^4}\times \eta^2\,e^{ik_T\eta}$, generates the expected singular behaviour, corresponding to $\Delta_T=-1$. \\

\noindent\textbf{Heavy off-shell graphs.} Our proposal for a massive flat-space limit applies to \textit{heavy graphs}, which are defined by in-in diagrams with light external legs and heavy internal lines (see Fig.\,\ref{fig:heavygraphintro}). For simplicity, we assume that all internal lines share the same mass, $m \geq 3H/2$, while the external legs are light, often corresponding to massless or conformally coupled fields.
Such \textit{on-shell} correlators 
are characterized by the momenta associated with its external fields, i.e., $\bm{k}_i$.  To define our flat-space limit, we first define an \textit{off-shell} extension of such graphs by adding a set of fictitious \text{energy variables} $\omega_i$ to the list of the external kinematics (see also \cite{Salcedo:2022aal}). This is naturally achieved by replacing the Schwinger-Keldysh bulk-to-boundary propagators $K^\pm(k_i,\eta)$ with $K^\pm(\omega_i,\eta)$, where $\omega_i$ is an independent variable, i.e., $\omega_i^2\neq \bm{k}_i^2$. To each on-shell correlator $F_n(\lbrace\bm{k}_i\rbrace)$, this replacement assigns a unique function of $\omega_i$, denoted by 
\begin{align}
    F_n(\omega_i,\bm{k}_i;m)\,\quad (i=1,\dots n)\,.
\end{align}
These functions define correlators that are off-shell in the sense that their external propagators do not satisfy the equations of motion, i.e., $\Box (K(\omega_i,\eta)\exp(i\bm{k}_i.\bm{x}))\neq 0$, in contrast with one-shell diagrams with external legs that satisfy $\Box (K(k_i,\eta)\exp(i\bm{k}_i.\bm{x}))=0$. \\

\noindent \textbf{The Massive Flat-Space Limit}. We consider the following \textit{double-scaling limit} of a heavy graph:
\begin{tcolorbox}[colframe=white,arc=0pt] 
\begin{equation}
    m/H\to \infty\,,\quad \dfrac{\omega_i}{|\bm{k}_i|}\to 0\,,\quad \text{while}\qquad \dfrac{\omega_i}{|\bm{k}_i|}\times \dfrac{m}{H}=\text{finite}\,.
\end{equation}
\end{tcolorbox}
\noindent In this limit, we show that $F_n$ asymptotes to 
\begin{align}
    F_n\left(\omega_i,\bm{k}_i; \frac{m}{H}\right)\xrightarrow{\text{MFS}}  F_n^{\text{MFS}}\left(\omega_i,\bm{k}_i; m\right)\left(1+{\cal O}(H/m)\right)\,,
\end{align}
where $F_n^{\text{MFS}}$ is a \text{contact diagram} in de Sitter, given by
\begin{align}
F_n^{\text{MFS}}\left(\omega_i,\bm{k}_i;m \right)=2\,\text{Re}\int_{-\infty(1-i\epsilon)}^0\,\dfrac{d\eta}{H^4\eta^4}\,\textcolor{purple}{G_n\left[p_1^\mu(\eta),\dots p_n^\mu(\eta);m\right]}\,\prod_{i=1}^n\, O_n(\partial_\eta,\eta,\bm{k}_i)\,K^+(\omega_i,\eta)\,.
\end{align}
In this \textit{reduction formula}, $O_n(\dots)$ are derivative operators acting on the external legs, which are inherited from the original in-in diagram. $p_i^\mu$ represent a set of (time-dependent) off-shell four-momenta, given by
\begin{align}
    p_i^\mu(\eta)=\left(0,-H\,\bm{k}_i \eta\right)\,\qquad  (i=1,\dots n)\,,
\end{align}
which characterize $n$ external legs of a dual Feynman graph in flat space (see Fig.\,\ref{fig:heavygraphintro}). This Feynman graph is obtained by amputating the external lines of the original in-in diagram, and is given by $\textcolor{purple}{G_n(p_i^\mu;m)}$. The rules for computing $G_n$ are the familiar Feynman rules in flat space: four-momentum is conserved at each vertex; each undetermined momentum brings a factor of $\int d^4q_l$ (where $q_l$ is the loop four-momentum), and each internal line is assigned a massive Feynman propagator $\frac{-i}{q^2+m^2}$. 

For the tree-level exchange example above (Eq.\,\eqref{FSEintro}), the MFS reduction formula gives
\begin{align}
    \lim_{\text{MFS}}F_4(\omega_i,k_i,s)=2\,\text{Im}\int_{-\infty(1-i\epsilon)}^0 d\eta\,\dfrac{1}{H^2\,s^2\eta^2+m^2}\exp(i\omega_T\,\eta)\,.
\end{align}
As another example, consider the one-loop, four-point function induced by the $\phi^2\sigma^2$ interaction (see Fig.\,\ref{SESBcc} and Eq.\,\eqref{FSB}). The MFS limit of this loop diagram is given by
\begin{align}
    \lim_{\text{MFS}}F_4(\omega_i,k_i,s)&=\dfrac{-1}{16\pi^2}\,\text{Re}\int_{-\infty}^0 d\eta\,\left(\int_0^1 dx\,\log\left[\dfrac{m^2+x(1-x)H^2s^2\eta^2}{\mu^2}\right]\right)\,\exp(i\omega_T\,\eta)\\ \nn
    &+\dfrac{1}{8\pi^2}\left(-\dfrac{1}{(d-4)}+\dfrac{1}{2}\log(4\pi\,e^{-\gamma_E})\right)\dfrac{1}{\omega_T}\,,
\end{align}
where the UV divergence of the associated loop graph $G_4$ is regulated in dim reg, and $\mu$ is the corresponding renormalization scale. In the $d\to 4$ limit, the second line above is a divergent local contribution that is canceled by adding a $g_4\phi^4$ counter-term to the action, using the $\bar{\text{MS}}$ scheme.

In summary, the reduction formula states that in the MFS limit, any heavy graph, regardless of its internal structure, simplifies to a contact diagram. The time-dependent vertex of this diagram is determined by the corresponding amputated flat-space graph with red-shifted external momenta.
\\

\noindent\textbf{Elements of the proof.} Let us sketch the proof of the reduction formula, which is based on two observations. First of all, in the heavy limit, the mode functions of the massive field can be replaced by its WKB approximation, namely
\begin{align}
\sigma^{\text{WKB}}_+(s,\eta)\propto\dfrac{1}{a(\eta)}\,\left(s^2+\dfrac{m^2}{H^2\eta^2}\right)^{-1/4}\exp\left(- i\int_{-\infty}^{\eta}\,d\eta'\sqrt{s^2+\dfrac{m^2}{H^2\eta'^2}}\right)\,,
\end{align}
For large masses, this formula can be used to simplify the in-in integrals because the WKB approximation is valid during the entire evolution, from $\eta=-\infty$ all the way to $\eta=0$. The fast oscillations of the WKB mode function as $m\to \infty$ implies that the multi-dimensional in-in integral peaks at a region where the vertices ($\eta_a$) are clustered around a central conformal time $\eta$, i.e., 
\begin{align}
    \left\vert\dfrac{1}{H\,\eta}(\eta_a-\eta)\right\vert\lesssim {\cal O}(1)\dfrac{1}{m}\,.
\end{align}
This is the underlying reason why taking the MFS limit collapses the internal structure of the diagram into a single vertex.

The second observation is that, in the MFS limit, sending $\omega_T=\sum \omega_i \to 0$ plays a similar role as sending $k_T \to 0$ in the amplitude limit: it effectively pushes all the vertices $\eta_a$ to $-\omega_T^{-1}\to -\infty$, with the crucial different that, in this limit, the mass of the propagators cannot be ignored. This is because, in the early time limit, the physical kinetic energy of each internal line scales inversely with $1/\omega_T$, hence proportionally to the mass. Specifically,
\begin{align}
    E=-H s\,\eta\to \dfrac{H s}{\omega_T}\propto m\,, 
\end{align}
where $s$ is the norm of the internal line comoving momentum. In Section \ref{MFSlimitSection} and Appendix \ref{app:spectral} we use these two ingredients to prove the reduction formula in three different ways, namely ($i$) by directing taking the MFS limit of the in-in integrals, ($ii$) using the Wilsonian effective action formalism and finally, for the bubble diagram, ($iii$) using a spectral decomposition approach.  
\\

\noindent \textbf{Conventions.} We chart the de Sitter Poincaré patch using the following coordinates
\begin{align}
    ds=a^2(\eta)(-d\eta^2+d\bm{x}^2)\,,\qquad a(\eta)=-\dfrac{1}{\eta\,H}\,,
\end{align}
where $H$ is the Hubble constant, and $\eta\in(-\infty,0)$ is the conformal time. We sometimes also use the FLRW format of the metric, expressed as 
\begin{align}
    ds^2=-dt^2+a^2(t)d\bm{x}^2\,,
\end{align}
where in de Sitter we have $a=\exp(H t)$. Prime on fields, e.g. $\pi'$, denotes derivative with respect to $\eta$, while dot will stand for derivative with respect to $t$. We use the following convention for the Riemann and Ricci tensors: 
$R^\mu_{\,\,\nu\alpha\beta}=\partial_\alpha \Gamma^\mu_{\beta\nu}+\dots $, and $R_{\mu\nu}=R^{\alpha}_{\,\,\mu\alpha\nu}$. We use bold letters to refer to spatial vectors, e.g., $\bm{x}$ for spatial coordinates and $\bm{k}$ for spatial momentum. We also use the notation
\begin{align}
    k_{ij}=k_i+k_j\,,
\end{align}
where $k_i=|\bm{k}_i|$. 
A massive scalar with mass $m$ in $\text{dS}_{d}$ is characterized by the following conformal weights
\begin{align}
    \Delta_{\pm}=\dfrac{d-1}{2}\pm i \mu\,,\qquad 
    \mu^2=\frac{m^2}{H^2}-\frac{(d-1)^2}{4}\,.
\end{align}
The principal series (heavy fields) corresponds to $\mu>0$, and complementary series (light fields) corresponds to $\mu=i\nu$ with $\nu>0$. We collectively refer to the former fields with $\sigma$ and the latter with $\phi$, which includes both the conformally coupled field ($\Delta_+=\frac{d-2}{2}$) and the massless field ($\Delta_+=0$). We also use the same symbol $\mu$ to refer to the renormalization scale, but the context will make the difference clear. A prime on a correlator indicate that the overall momentum-conserving delta function has been stripped off, i.e., 
\begin{align}
    \langle \phi(\bm{k}_1)\dots\phi(\bm{k}_n) \rangle=(2\pi)^3\delta^{(3)}\left(\sum_{i=1}^n \bm{k}_i\right)\,\langle \phi(\bm{k}_1)\dots\phi(\bm{k}_n) \rangle' \,.
\end{align}
\section{The Massive Flat-Space Limit}
\label{MFSlimitSection}
\subsection{Recap: the Schwinger-Keldysh formalism}
Let us begin by discussing how to obtain correlation functions from the path integral (see also \cite{Chen:2017ryl}). The most practical approach involves defining a partition function by introducing external sources (or currents) to the path integral and taking functional derivatives with respect to these sources. In cosmology, where observations are made at finite times, the most relevant framework for this purpose is the \textit{in-in} or Schwinger-Keldysh formalism.

In this formalism, the theory is assumed to start in the standard Bunch-Davies (BD) vacuum\footnote{Other vacua are possible, but we focus exclusively on the Bunch-Davies choice.}, with observations made at a specific finite time $\eta_0$, usually taken to be the end of inflation. Fields evolve forward in time from the initial vacuum state to $\eta_0$ along the upper branch of the Schwinger-Keldysh contour and then reverse-evolve back to the vacuum along the lower branch. The partition function is constructed by placing currents on each branch to act as sources for the fields. The evolution is governed by two path integrals: one over the bulk fields, describing their forward and backward evolution, and another over the field profiles at $\eta_0$.

For a single scalar field with the action $S[\phi]$, the field evolves forward on the upper branch from the BD vacuum at  $\eta \to -\infty$ all the way to the end of inflation, when it takes the profile  $\phi(\eta_0) = \phi_0$. On the lower branch, the field evolves backward from the same profile  $\phi_0$ , returning to the BD vacuum. To distinguish between the solutions on the two branches, we label them with  $+$  (upper branch) and  $-$  (lower branch) superscripts. The partition function is given by:
\begin{align}
Z[J_+, J_-] = \int D\phi_0 \int_{\mathrm{BD}}^{\phi_0} D\phi_+ \int_{\mathrm{BD}}^{\phi_0} D\phi_- \, e^{iS[\phi_+] - iS[\phi_-] + i\int d^4x \, (\phi_+ J_+ - \phi_- J_-)}\,,
\end{align}
where, $J_+$ and $J_-$ are external sources coupled to the fields on the respective branches.
We assume that the  bulk path integral is dominated by the classical solution $\phi_{\mathrm{cl}}(\eta)$, which must satisfy the appropriate boundary conditions imposed by the path integral.

In order to build the propagators we will need the mode functions obtained from the action $S_0[\phi]$. For example, the positive-frequency mode functions for massless and conformally coupled scalar fields on the upper branch are:
\begin{align}
\phi_+(k,\eta) = \frac{iH}{\sqrt{2 k^3}}(1 + ik\eta) e^{-ik\eta} \quad (\text{massless}), \quad
\phi_+(k,\eta) = -\frac{H}{\sqrt{2 k}}\eta\, e^{-ik\eta} \quad (\text{conformally coupled}).
\end{align}
On the lower branch, the solutions are the complex conjugates of the upper branch mode functions. 

The propagators can be derived from the free partition function by solving for the classical solutions sourced by the external currents on each branch. After substituting these solutions into the path integral and integrating over the field profile, we obtain: 
\begin{align}
    Z_0&=\exp\left[-\frac{1}{2}\int_{\bm{k}}\int d\eta a^4(\eta)\int d\eta'a^4(\eta') \left(J_+(\eta)G_{++}(k,\eta,\eta')J_+(\eta')-J_+(\eta)G_{+-}(k,\eta,\eta')J_-(\eta')\right.\right.\nonumber\\
  &\qquad\qquad\qquad\qquad\qquad\qquad\qquad  \left.\left.-J_-(\eta)G_{-+}(k,\eta,\eta')J_+(\eta')+J_-(\eta)G_{--}(k,\eta,\eta')J_-(\eta')\right)\right]
  \label{free_partition}
\end{align}
where the propagators expressed in terms of the  mode functions are,
\begin{align}
\nn
G_{++}(k, \eta, \eta') &= \phi_{-}(k, \eta')\phi_{+}(k, \eta)\theta(\eta - \eta') + \phi_{-}(k, \eta)\phi_{+}(k, \eta')\theta(\eta' - \eta), \\ 
\nn
G_{+-}(k, \eta, \eta') &= \phi_{+}(k, \eta')\phi_{-}(k, \eta), \\
\nn
G_{--}(k, \eta, \eta') &= \phi_{+}(k, \eta')\phi_{-}(k, \eta)\theta(\eta - \eta') + \phi_{+}(k, \eta)\phi_{-}(k, \eta')\theta(\eta' - \eta), \\
G_{-+}(k, \eta, \eta') &= \phi_{-}(k, \eta')\phi_{+}(k, \eta).
\label{props}
\end{align}
Here, $G_{++}$ and $G_{--}$ are the time-ordered and the anti-time-ordered propagators on their respective branches, while $G_{+-}$ and $G_{-+}$ are Wightman propagators that mix the two branches. To include interactions, we split the action into a free part $S_0$ and an interacting part $S_{\mathrm{int}}$. The full partition function can then be written as:
\begin{align}
Z[J_+, J_-] = e^{iS_{\mathrm{int}}\left[\frac{\delta}{\delta J_+}\right] - iS_{\mathrm{int}}\left[\frac{\delta}{\delta J_-}\right]} Z_0[J_+, J_-],
\end{align}
where the free partition function is given in \eqref{free_partition}.

This formalism can be easily extended to include fields with different masses and spins. In addition to a massless or a conformally coupled field, we will also consider a massive scalar field $\sigma$ with mass $m$. In four dimensions the corresponding mode functions are given by:
\begin{align}
\nn
\sigma_+(k\eta) &= \frac{\sqrt{\pi} H}{2}e^{-\pi\mu/2 + i\pi/4}(-\eta)^{3/2}H_{i\mu}^{(1)}(-k\eta), \\ \label{sigmaplusminus}
\sigma_-(k\eta) &= \frac{\sqrt{\pi} H}{2}e^{\pi\mu/2 - i\pi/4}(-\eta)^{3/2}H_{i\mu}^{(2)}(-k\eta),
\end{align}
where \(\mu = \sqrt{\frac{m^2}{H^2} - \frac{9}{4}}\). Fields in the principal series representation have \(m^2 \geq \frac{9H^2}{4}\), such that \(\mu\) is real, whereas for the complementary series \(m^2 \leq \frac{9H^2}{4}\), \(\mu\) is imaginary. 

In general, a systematic set of Feynman rules can be set up for computing cosmological correlators. External lines are represented by the bulk-to-boundary propagators as given in  $K^+$ and  $K^-$ ,
\begin{align}
K^+(k, \eta) = \phi_+(k, \eta_0)\phi_-(k, \eta), \quad K^-(k, \eta) = \phi_-(k, \eta_0)\phi_+(k, \eta),
\label{BtBdary}
\end{align} while internal lines correspond to the bulk-to-bulk propagators as defined in \eqref{props}. For each vertex, a volume factor of  $a^4(\eta)$  is included, along with a time integral that runs from $\eta \to -\infty$ up to the observation time $\eta_0$. These rules can be readily generalized to interactions involving derivatives by applying the appropriate differential operators directly to the corresponding propagators.

We focus on in-in diagrams with external lines representing light fields (e.g., massless or conformally coupled fields) and internal lines corresponding to heavy fields (with $\mu > 0$). Such diagrams will be referred to as \textit{heavy graphs}. To calculate these diagrams, we consider a partition function that depends on four external sources, namely $J^\phi_\pm,J^\sigma_\pm$. A generic $n$-point correlator of the light field, which we denote by $\phi$, can then be expressed as follows:
\begin{align}
\langle \phi(\bm{k}_1)\dots \phi(\bm{k}_n) \rangle =\left.(-i)^n\frac{\delta^n}{\delta J_+^\phi(\bm{k}_1)\dots\delta J_+^\phi(\bm{k}_n)}Z[J^\phi_+,J^\sigma_+,J^\phi_-,J^\sigma_-]\right\vert_{J_{\pm}^{\phi,\sigma}=0}
\end{align}

The derivatives with respect to the external sources will bring down bulk-to-boundary propagators of $\phi$ whereas the interaction appearing in the partition function will bring down bulk-to-boundary propagators of $\sigma$. The contribution from a heavy diagram with $\cV$ vertices and $\cI$ internal lines then becomes 
\begin{tcolorbox}[colframe=white,arc=0pt]
\begin{align}
    \langle \phi(\bm{k}_1)\dots\phi(\bm{k}_n) \rangle'&\supset \sum_{\pm_1,\pm_2,\dots \pm_\cV}\left(\prod_{a=1}^\cV (\pm_a i\lambda_a)\int d\eta_a a^4(\eta_a)\right) \prod_{i=1}^n O_i(\bm{k}_i,\dfrac{\partial}{\partial \eta_{d_i}},\eta_{d_i}) K^{\pm_{d_i}}_\phi(k_i,\eta_{d_i})\nonumber\\
    &\times \left(\prod_{l=1}^\cL\int \dfrac{d^3\bm{q}_l}{(2\pi)^3} \right) \prod_{\text{internal lines}(b,c)} O_{bc}(\bm{q}_{bc},\eta_b,\eta_c,\partial_{\eta_b},\partial_{\eta_c})G^{\sigma}_{\pm_b,\pm_c}(q_{bc},\eta_b,\eta_c),
    \label{heavy_graph}
\end{align}
\end{tcolorbox}
\noindent where we have indexed the vertices by $a=1,\dots, \cV$, $\lambda_a$ are the associated coupling constants, and $1\leq d_i\leq \cV\,(i=1,\dots,n)$ are a set of integers that specifies the vertex to which the $i-$th external leg is connected. Vertices can be of $+$ or $-$ types, depending on which an appropriate bulk-to-bulk or bulk-to-boundary propagators is included. $\bm{q}_l\,(l=1,\dots \cL)$ represent the loop momenta, with $\cL={\cal I}-{\cal V}+1$ denoting the number of loops. The three-momenta associated with the internal line connecting vertex  $b$ to vertex $c$ are denoted by $\bm{q}_{bc}$, which are determined by momentum conservation in terms of the external and loop momenta. The operators $O_n$ describe derivatives at the vertices acting on the external legs, while $O_{ab}$ captures any derivatives acting on the internal lines.

Notice that the ordering of the time integrals is not arbitrary but is dictated by the causal structure of the bulk-to-bulk propagators. For simplicity, we drop the $\sigma$ label from the bulk-to-bulk propagators, as internal lines always involve $\sigma$.
\begin{figure}
    \centering
    \includegraphics[scale=0.8]{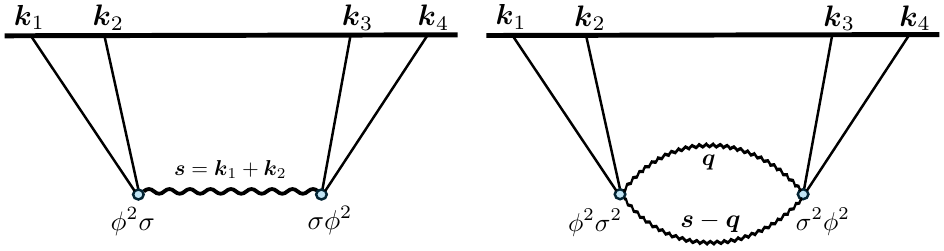}
    \caption{The scalar-exchange (SE) and scalar one-loop, bubble diagrm (SB) diagrams contributing to the four-point function of the conformally coupled field. }
    \label{SESBcc}
\end{figure}\\

\noindent \textbf{Four-point functions of the conformally coupled field.} Using the Feynman rules introduced above, we write the in-in expression for two specific diagrams of interest: the single-exchange diagram and the one-loop bubble diagram, both contributing to the four-point function of a conformally coupled (cc) field (see Figure \ref{SESBcc}). To construct these diagrams, consider the following cubic and quartic interactions between a cc field $\phi$ and a heavy scalar $\sigma$:
\begin{align}
    {\cal L}_{\text{int}}=a^4(\eta)\left(-g\,\phi^2\,\sigma-g'\phi^2\sigma^2\right)\,.
\end{align}
Using these vertices we can form a single-exchange and a one-loop, bubble graph. The corresponding four-point functions in the $s-$channel can be expressed as 
\begin{align}
\nn
    \langle \phi(\bm{k}_1)\phi(\bm{k}_2)\phi(\bm{k}_3)\phi(\bm{k}_4) \rangle'_{\text{SE}} &=\frac{\eta_0^4}{4k_1 k_2 k_3 k_4}g^2F_{\text{SE}}(k_{12}, k_{34}, s) + t, u\text{-channels}\qquad (\text{Scalar-Exchange)}\,,\\ \langle\phi(\bm{k}_1)\phi(\bm{k}_2)\phi(\bm{k}_3)\phi(\bm{k}_4) \rangle'_{\text{SB}} &=\frac{\eta_0^4}{2k_1 k_2 k_3 k_4}g'^2 F_{\text{SB}}(k_{12}, k_{34}, s) + t, u\text{-channels}\qquad (\text{Scalar-Bubble). }
\end{align}
\noindent where $F_{\text{SE},\text{SB}}$, which are functions of the three variables $k_{12}=k_1+k_2$, $k_{34}=k_3+k_4$ and $s=|\bm{k}_1+\bm{k}_2|$, are given by
\begin{tcolorbox}[colframe=white,arc=0pt]
\begin{align}
    F_{\text{SE},\text{SB}}(k_{12},k_{34},s) &\equiv F_{\text{SE},\text{SB}}^{++} + F_{\text{SE},\text{SB}}^{+-} + F_{\text{SE},\text{SB}}^{-+} + F_{\text{SE},\text{SB}}^{--},\nn
\end{align}
with 
\begin{align}
\label{FSE}
&F_{\text{SE}}^{\pm\pm} = (\pm i)(\pm i)\int_{-\infty(1\mp i\epsilon)}^{0} \frac{d\eta}{\eta^2}\frac{d\eta'}{\eta'^2} e^{\pm ik_{12}\eta} \,e^{\pm ik_{34}\eta'} G_{\pm\pm}(s, \eta, \eta'),\\ 
\label{FSB}
    &F_{\text{SB}}^{\pm\pm} = (\pm i)(\pm i)\int_{-\infty(1\mp i\epsilon)}^{0} \frac{d\eta}{\eta^2}\frac{d\eta'}{\eta'^2} e^{\pm ik_{12}\eta}\,e^{\pm ik_{34}\eta'} \int \dfrac{d^3\bm{q}}{(2\pi)^3}G_{\pm\pm}(|\bm{q}|,\eta,\eta')G_{\pm\pm}(|-\bm{q}+\textbf{s}|,\eta,\eta')\,.
\end{align}
\end{tcolorbox}
\noindent In Section \ref{bispectrumSection}, we will use the above four-points to evaluate the bispectrum of curvature perturbations by applying appropriate weight-shifting operators, transforming the external cc fields into massless ones. 
\subsubsection*{The Effective Action}

When considering heavy graphs introduced above, a natural expectation is that as the internal masses become very large, the complicated structure of in-in time integrals should simplify. Drawing from the more familiar in-out computations, integrating out a massive field $\sigma$ is expected to yield an effective action. At low energies, this effective action typically contains a series of local operators with an increasing number of derivatives, suppressed by the mass scale of the integrated-out field.

Extending the same EFT concept to the in-in formalism introduces subtleties, particularly in how boundary conditions are imposed on the path integral \cite{Salcedo:2022aal,Salcedo:2024smn,Burgess:2024eng,Burgess:2024heo,Green:2024cmx}. The key difference between integrating out massive fields in the in-in and in-out formalisms arises from the structure of the propagators. While the effective action in the in-out formalism involves only time-ordered propagators, the in-in framework also includes non-time-ordered propagators. These additional components account for non-unitary effects arising from interactions between the two branches of the in-in contours and must be incorporated into the effective action.

 To illustrate this, we consider the interaction $g'\phi^2\sigma^2$. To construct the correct effective action for $\phi$, one must solve the path integral for $\sigma$ while accounting for both branches of the in-in contour. The effective action is then formally expressed as:

\begin{align}
e^{iS_{\mathrm{eff}}[\phi_+,\phi_-]}=\int D\sigma\int^{\sigma}_{BD} D\sigma_+\int^{\sigma}_{BD}D\sigma_-\,e^{iS[\phi_+,\sigma_+]-S[\phi,_-\sigma_-]}.
\end{align}

Notice that the path integrals over $\sigma_+$ and $\sigma_-$ are identified at their endpoints, leading to an effective action that generally mixes terms containing $\phi_+$ and $\phi_-$. This type of action, often referred to as “non-unitary,” cannot be derived from a conventional Hamiltonian. For the example considered here, the resulting effective action contains terms like:
\begin{align}
S_{\mathrm{eff}}[\phi_+,\phi_-]\supset \int a^4(\eta)d^3\bm{x}\int a^4(\eta')d^3\bm{y}\,\sum_{\pm,\pm}
\phi_\pm(\bm{x},\eta)^2 G_{\pm,\pm}(\vert\x-\y
\vert,\eta,\eta')^2\phi_\pm(\bm{y},\eta')^2\,,
\label{effective_actioninin}
\end{align}
which captures the complete propagator structure, maintaining consistency with the in-in formalism and accurately reproducing the correlators in Eq. \eqref{FSB}. 

The effective action \eqref{effective_actioninin} in its current form is non-unitary and fully non-local, making it not very useful for simplifying the correlator, i.e., in this case the scalar bubble graph. We will demonstrate that certain simplifications arise in the massive flat space limit. First, by taking the mass of the exchanged field to infinity, we argue that the contributions from the non-time-ordered propagators $G_{\pm\mp}$ become exponentially small due to the rapid oscillations of the massive propagators in the WKB approximation, so we can use the unitary part of this effective action. Furthermore, in the MFS limit, we show that the corresponding bubble graph receives its dominant contribution from near the diagonal $\eta=\eta'$ of the in-in time integral, allowing the effective action to also be expanded around $\eta=\eta'$, making it local in time, although still non-local in space. We clarify these points in Section \ref{MFSfromeffectiveaction}.

\subsection{A reduction formula in the massive flat-space limit}
In this section, we define a new flat space limit for diagrams involving heavy internal lines. We then provide a universal reduction formula that determines the behavior of correlators in this limit. Additionally, we study a tree-level example and highlight the distinction between our limit and the conventional amplitude limit.

\subsubsection*{Off-shell correlators and the MFS limit}
To establish our limit, it is convenient to work with a set of \textit{off-shell} correlators. These are defined within perturbation theory using the same diagrammatic rules outlined in the previous section, with one crucial modification: a fictitious energy variable \(\omega_i\) is introduced into the bulk-to-boundary propagators, treated as independent of \(\bm{k}_i\), i.e., \(\omega_i^2 \neq \bm{k}_i^2\). For related constructions of off-shell wavefunction coefficients, see \cite{Salcedo:2022aal}\footnote{We thank Scott Melville for insightful discussions on the off-shell extension of cosmological correlators.}.
In practice, this amounts to the following substitution inside the in-in integrand
\begin{align}
    K^+(k_i,\eta)\to K^+(\omega_i,\eta)\,,
\end{align}
while keeping the bulk-to-bulk propagators unchanged.
Following this substitution, to each $n-$point diagram one can assign a unique function of the external energies $\omega_i$, given by 
\begin{align}
    F_n(\bm{k}_i)\,\rightarrow \,F_n(\omega_i,\bm{k}_i)\,,
\end{align}
which we will refer to by the off-shell correlator associated with that graph. The terminology is appropriate because the plane-wave defined by $K^+(\omega,\eta)\exp(i\bm{k}.\bm{x})$ does not satisfy the equation of motion for the free theory, unless we take the on-shell limit by setting $\omega_i=k_i$. 

\begin{figure}
    \centering    
    \includegraphics[scale=0.8]{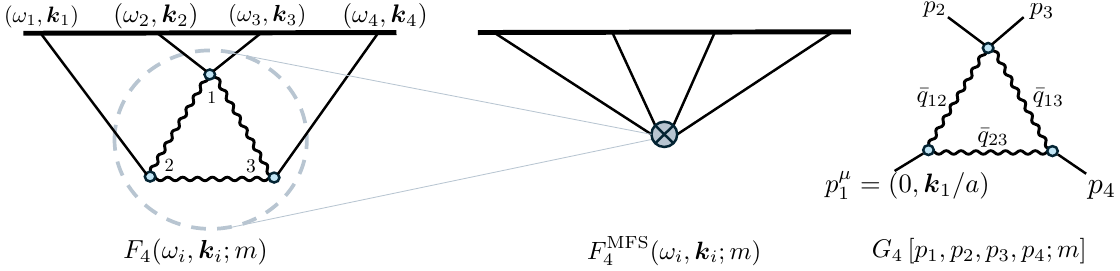}
    \caption{An illustration of the elements involved in the MFS limit reduction formula \eqref{flatspace} for a one-loop graph of the four-point function with three vertices (\textit{left}). $p_i^\mu = (0, \bm{k}_i / a)$ ($i = 1, \dots, 4$) represent a set of fictitious four-momenta that characterize the amputated graph in flat-space (\textit{right}). $\bar{q}_{bc}$ ($b, c = 1, 2, 3$) denote the internal line four-momenta of this graph, which can be expressed in terms of the loop momentum $\bar{q}$ and the $p_i$'s, i.e., $\bar{q}_{12} = q$, $\bar{q}_{13} = p_2 + p_3 - q$, and $\bar{q}_{23} = q + p_1$. In the MFS limit, the correlator reduces to a contact diagram (\textit{middle}), with its vertex corresponding to the amputated Feynman diagram.}
    \label{fig:trianglegraphMFS}
\end{figure}

We now focus on \textit{heavy graphs}, defined as graphs with light external legs and vertices connected solely by heavy internal lines, where the masses of the internal lines lie within the de Sitter principal series. For simplicity, we consider scalar fields on both the internal and external legs and assume that all internal lines have the same mass $m$. However, we expect our results to extend to graphs with internal lines of varying spins and masses, as well as to external legs with light spinning fields such as gluons and gravitons.

For these heavy graphs, we introduce the \textit{Massive Flat-Space (MFS)} limit, defined as:
\begin{tcolorbox}[colframe=white,arc=0pt] 
\begin{align}
    \dfrac{\omega_i}{|\bm{k}_i|}\to 0^+\,,\quad \dfrac{m}{H}\to +\infty\,,\quad \text{while}\quad \dfrac{m}{H}\times \dfrac{\omega_i}{|\bm{k}_i|}=\lambda_i<\infty\,. 
\end{align}
\end{tcolorbox}
\noindent \noindent 
In the MFS limit, we propose that an off-shell heavy graph $F_n$
reduces to a contact diagram with a vertex that is determined by a Feynman diagram in flat space, which is obtained by amputating the external legs (see Fig.\,\ref{fig:trianglegraphMFS}).

In more detail, in the MFS limit $F_n$ asymptotes to
\begin{align}
    F_n\left(\omega_i,\bm{k}_i; \frac{m}{H}\right)\xrightarrow{\text{MFS}}  F_n^{\text{MFS}}\left(\omega_i,\bm{k}_i; m\right)\left(1+{\cal O}(H/m)\right)\,,
\end{align}
where $F_n^{\text{MFS}}$ is given by the following \textit{reduction formula}:
\begin{tcolorbox}[colframe=white,arc=0pt]
\begin{align}
\label{flatspace}
F_n^{\text{MFS}}\left(\omega_i,\bm{k}_i;m \right)=2\,\text{Re}\int_{-\infty(1-i\epsilon)}^0\,d\eta\,a^4(\eta)\,\textcolor{purple}{G_n\left[p_1^\mu(\eta),\dots p_n^\mu(\eta);m\right]}\,\prod_{i=1}^n\, O_n(\partial_\eta,\eta,\bm{k}_i)\,K^+(\omega_i,\eta)\,.
\end{align}
\end{tcolorbox}
\noindent Here, the $O_n$'s are the derivative operators acting on the external legs of the original graph, as carried over from Eq.\,\eqref{heavy_graph}. The time-dependent momenta entering the vertex factor $G_n$ are given by 
\begin{align}
    p_i^\mu(\eta)=\left(0,\dfrac{\bm{k}_i}{a(\eta)}\right)\,\qquad  (i=1,\dots n)\,.
\end{align}
$G_n$ is an amputated Feynman graph in flat-space, computed using the following Feynman rules:
\begin{enumerate}
    \item For each vertex, insert a factor of $+i\lambda_a (a=1,\dots \cV)$. 
    \item Conserve the four-momentum at each vertex. This fixes the four-momentum $\bar{q}^\mu_{bc}$, associated with the internal line connecting vertices $b$ and $c$, in terms of the external four-momenta $p_i^\mu$ and loop momenta $\bar{q}^\mu_j (j=1,\dots \cL)$ (with $\cL$ denoting the number of loops). 
    Notice that the total four-momentum of the graph is guaranteed to vanish because $\sum_{i=1}^n p_i^\mu=(0,\sum_{i=1}^n \bm{k}_i)=0$. 
    \item Assign a Feynman propagator $\dfrac{-i}{\eta_{\mu\nu}(\bar{q}_{bc})^\mu\,(\bar{q}_{bc})^\nu+m^2+i\epsilon}$ to each internal line.
    \item For each loop, insert a factor of $\displaystyle\int \dfrac{d^4\bar{q}}{(2\pi)^4}$. 
    \item For each spacetime derivative $a^{-1}(\eta)\partial_\mu \sigma$ acting on the heavy field, insert a factor of $i (\bar{q}_{bc})_\mu$, where $\bar{q}_\mu=\eta_{\mu\nu}\bar{q}^\nu$. 
\end{enumerate}
Because $G_n$ is an amputated diagram, derivatives acting on the external legs have no impact on its structure. Depending on the type of vertices, Lorentz indices appearing in $G_n$ might be contracted with those appearing in $O_n$. We also emphasize that the external energies $p^0_i$ appearing in the flat-space graph $G_n$ are identically set to zero, while $\omega_i$'s have to be kept finite inside the bulk-to-boundary propagators $K^+(\omega_i,\eta)$ appearing in the integrand of the MFS reduction formula. 

Before delving into the derivation of the MFS limit reduction formula, let us  highlight a few remarks. At first glance, one might expect that in the $m/H \to \infty$ limit, a heavy graph would reduce to a series of contact terms in an effective field theory. This EFT would result from integrating out the heavy field and would be expressed as an expansion in powers of $\nabla/m$. However, this expectation does not hold due to the double-scaling nature of the MFS limit: As we increase $m$, we simultaneously send $\omega_i/k_i \to 0$, inversely proportional to $m$. This interplay implies that at the characteristic time $|\eta_c| \sim \mathcal{O}(1/\omega)$ (where $\omega$ represents the typical size of the external energies) the kinetic energy of the massive field becomes of the same order as its mass, specifically $|\bm{k}_i|/a(\eta_c) \sim \mathcal{O}(m)$. Therefore, around $\eta\sim\eta_c$, corresponding to the moment when the \textit{physical} energies of the external legs ($=\omega_i/a(\eta)$) become comparable to the Hubble scale $H$\footnote{For linear dispersion relations, i.e., $\omega = c_s |\bm{k}|$, $\eta_c$ corresponds to sound-horizon crossing.}, it is not justified to expand in spatial derivatives. This shows that we are in a limit where the effective action is spatially \textit{non-local}, a point we come back to in Section \ref{MFSfromeffectiveaction}. 

%
In the MFS limit, an important simplification arises in the structure of the propagators. As previously discussed, the in-in graph $F_n$ includes contributions from all possible $\pm$ vertex combinations in its original definition. However, in the heavy mass limit, only terms associated with time-ordered or anti-time-ordered propagators survive, while non-time-ordered contributions become exponentially suppressed in mass, hence negligible. At the level of the reduction formula \eqref{flatspace}, the all-plus contributions are captured by the $G_n$ vertex factor, while all-minus terms appear in $G_n^*$. 

It is important to emphasize that the MFS limit also applies to loop diagrams, which can exhibit UV divergences requiring consistent regularization. In this work, we employ dimensional regularization to regulate the loop integrals in both the correlators ($F_n$) and their associated Feynman graphs ($G_n$). When UV divergences are present, it is assumed that both sides of the the MFS limit reduction formula are expressed in $d$-dimensions (with $d\neq 4)$. We will elaborate on this point further when discussing bubble graphs below. 

Finally, note that a similar reduction formula could be derived for diagrams containing  \textbf{heavy subgraphs}, defined as parts of the graph that include vertices connected exclusively by massive internal lines. The remaining internal lines in such diagrams are assumed to be light. To define a useful massive flat space limit for these diagrams, one must, in addition to the external energy variables, assign a set of energies to the internal lines entering the heavy subgraph. These energies should be treated as independent from the comoving momenta carried by the corresponding internal lines. In the MFS limit, these additional energy variables are also scaled to zero, inversely proportional to the mass. As a result, one would obtain a reduction formula analogous to Equation \eqref{flatspace}, except that the right-hand side will feature multiple time integrals: one layer of integration will correspond to the amputated heavy subgraph considered in flat space, while the remaining layers capture other vertices in the original graph. Moreover, within the integrand, the bulk-to-boundary and bulk-to-bulk propagators of the light fields will appear with the same structure as in the original diagram.
\\

\noindent \textbf{Example: the single-exchange diagram.} 
Let us inspect the MFS limit of a simple diagram, namely the tree-level exchange four-point function $F_{\text{SE}}$ (Fig.\,\ref{SESBcc}), which is described by Eq.\,\eqref{FSE}. The off-shell version of this diagram is simply obtained by setting $k_{12}\to \omega_{12}$ and $k_{34}\to \omega_{34}$, where $\omega_{ij}=\omega_i+\omega_j$. 

The reduction formula implies that 
\begin{align}
   F_{\text{SE}}(\omega_i,\bm{k}_i)\xrightarrow{\text{MFS}}  -2\,\text{Im}\int_{-\infty(1-i\epsilon)}^0 d\eta\,\dfrac{1}{s^2\eta^2+\frac{m^2}{H^2}}\exp(i\omega_T\eta)\,.
   \label{linear_MFS_corr}
\end{align}
The amputated diagram in this case is a single-exchange four-point function in flat-space, given by
\begin{align}
\label{feyn}
    G_4(p_i^\mu)=\dfrac{i}{\eta_{\mu\nu}(p_1+p_2)^\mu(p_1+p_2)^\nu+m^2+i\epsilon}=\dfrac{i}{H^2 s^2\eta^2+m^2+i\epsilon}\,,
\end{align}
in which 
\begin{align}
    p_1+p_2=\bigg(0,\dfrac{\bm{k}_1+\bm{k}_2}{a(\eta)}\bigg)\,,
\end{align}
is the four-momentum exchanged in the $s-$channel of the flat-space graph. 

For this specific graph, the time integral can be directly performed, leading to
\begin{align}
    F^{\text{MFS}}_{\text{SE}}=\dfrac{1}{s}\frac{H}{2m}\text{Re}\Big\lbrace\exp(-\frac{\omega_T}{s}\frac{m}{H})\text{Ei}\left(\frac{\omega_T}{s}\frac{m}{H}+i\epsilon\right)-\exp(\frac{\omega_T}{s}\frac{m}{H})\text{Ei}\left(-\frac{\omega_T}{s}\frac{m}{H}+i\epsilon\right)\Big\rbrace\,,
    \label{correl:singleExchangeMFS}
\end{align}
where $\text{Ei}(z)$ is the Exponential Integral function. 

This formula, clearly demonstrates that the limit cannot be replicated by the local quartic operator \( \frac{g^2}{m^2} \phi^4 \), which would result from integrating out the massive field at tree level. Such EFT operator generates a contact diagram that is given by $F_{\text{SE}}\propto \frac{g^2}{m^2}\frac{1}{\omega_T}$, which is distinct from the function above, which is a 
non-rational function of \( \frac{\omega_T}{s} \frac{m}{H} \) (\( \sim {\cal O}(1) \)). This mismatch exemplifies the non-local nature of the MFS limit.

It is instructive to assess the accuracy of the MFS formula by comparing it with the exact analytical result available for this particular tree-level, off-shell diagram. This is given by Equation (4.34) of the reference \cite{Jazayeri:2022kjy}, which was obtained by solving the bootstrap equations for the four-point function in unphysical configurations. 

\begin{figure}
    \centering 
    \includegraphics[scale=1.0]{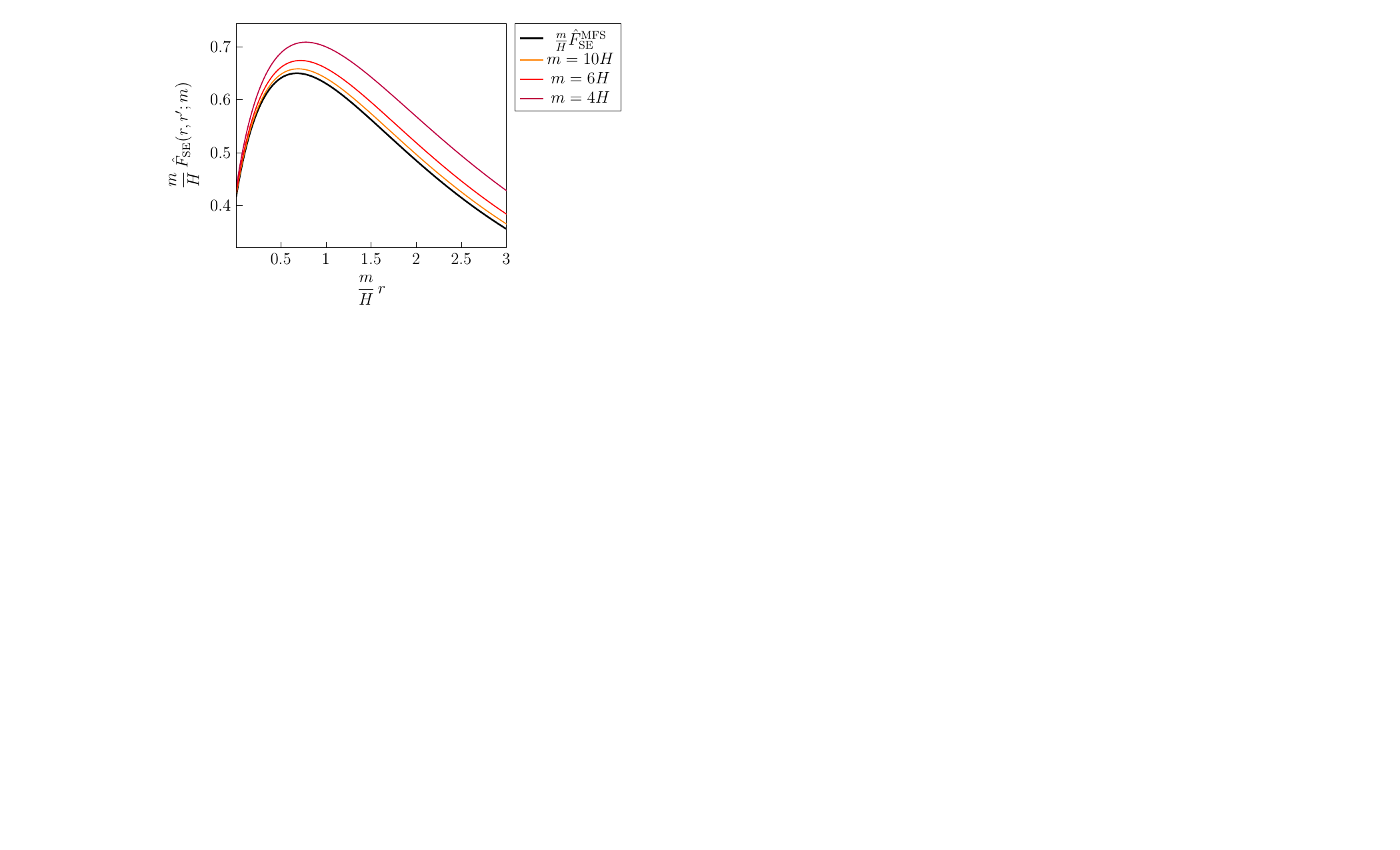}
    \caption{Comparison between the exact single-exchange diagram $\hat{F}_{\text{SE}}$ with varying masses (the colored curves) and their massive flat-space limits $\hat{F}^{\text{MFS}}_{\text{SE}}$ (the black curve), all plotted as functions of $r\times (m/H)$ with a fixed value of $r'\times (m/H) = 0.2$.}
    \label{comparisonFSE}
\end{figure}

To simplify the comparison, we plot both the exact four-point function and the asymptotic MFS limit, using the following conventions. The exact four-point can be written as  
\begin{align}
   F_{\text{SE}}=\dfrac{1}{s}\hat{F}_{\text{SE}}(r,r';m)\,,
\end{align}
where it is parametrized in terms of
\begin{align}
    r=\dfrac{\omega_{12}}{s}\,,\qquad r'=\dfrac{\omega_{34}}{s}\,.
\end{align}
In the MFS limit, $r\times (m/H)$ and $r'\times (m/H)$ are of order one. Therefore, it is reasonable to plot $\hat{F}_{\text{SE}}(r, r'; m)$ against $\hat{F}_{\text{SE}}^{\text{MFS}} \equiv s\,F_{\text{SE}}^{\text{MFS}}$ as a function of $\frac{m}{H}r$, while keeping $\frac{m}{H}r'$ fixed, as shown in Fig.~\ref{comparisonFSE}. 

The plot demonstrates excellent agreement between $\hat{F}_{\text{SE}}$ and $\hat{F}_{\text{SE}}^{\text{MFS}}$ for large masses, confirming the validity of the MFS-limit approximation. Similar comparisons were made in \cite{Jazayeri:2022kjy}, albeit with slightly different notations. 
\\

\noindent \textbf{Comparison to the amplitude limit at $k_T=0$.}  Finally, we compare our MFS limit to the conventional amplitude limit of correlators near their respective total energy singularity, $k_T=0$ \cite{Raju:2012zr,Pimentel:2012tw,Arkani-Hamed:2017fdk, Pajer:2020wxk,COT}. In this limit, the total energy singularity of a graph is proportional to the high-energy limit of the scattering process described by the same graph in flat space. This means that propagators' masses in the final amplitude go to zero. 

Conversely, the propagators of the amputated flat-space Feynman graph appearing in \eqref{flatspace} are massive.  The difference arises because, in the ordinary flat-space limit, as $k_T\to0$, the ratio $m/H$ remains fixed. Sending $k_T\to 0$ has the effect of pushing all the vertices of the diagram to $-\infty$, where the kinetic energies of the internal lines go to infinity, rendering the masses comparatively negligible. In contrast, in the MFS limit, $m$ is simultaneously sent to infinity, as $\omega_i's$ approach zero, in an inversely proportional manner. As such, while by sending $\omega_i\to 0$, the vertices are sent to early times, the mass and the kinetic energies associated with the bulk-to-bulk propagators remain proportional. 

Another key difference is that the amplitude limit only fixes the leading-order behavior of the correlator near \( k_T = 0 \), whereas the MFS limit provides an approximation where the correlators are given as smooth functions of the finite variable \( \omega_i \times m/H \). In this way, the MFS limit contains infinitely more information than the amplitude limit, which isolates only the singular part of the correlator.

\subsection{From in-in integrals to the MFS limit}
\label{sec:MFSinin}
We now demonstrate how to formally derive the formula \eqref{flatspace} by directly taking the MFS limit of the time integral representation of $F_n$, as given in Eq.\,\eqref{heavy_graph}. To make the discussion concrete, we focus on two specific examples: a single-exchange tree-level diagram and a one-loop bubble diagram. For simplicity, we assume the external legs correspond to massless or conformally coupled fields, but the derivation can be easily generalised to other fields in the complementary series. 
\subsubsection*{The MFS limit of tree-level exchange diagrams}
To analyze the asymptotic behavior of the heavy field’s propagators in the large mass regime, we start with the free field equation of motion for $\sigma$ in momentum space:
\begin{align}
    \eta^2\partial_\eta^2\sigma-2\eta\partial_\eta\sigma+\left(s^2\eta^2+\frac{m^2}{H^2}\right)\sigma=0\,.
\end{align}
For sufficiently large masses, this equation is amenable to a WKB approximation, which is valid from early times, $\eta\to -\infty$, all the way to the end of inflation, $\eta \to 0$. The solution can be expressed as:
\begin{align}
\sigma^{\text{WKB}}_+(s,\eta)=\sum_{\pm}\,c_\pm(\eta_*)\,\dfrac{1}{a(\eta)}\,\left(s^2+\dfrac{m^2}{H^2\eta^2}\right)^{-1/4}\exp(\mp i\int_{\eta_*}^{\eta}\,d\eta'\sqrt{s^2+\dfrac{m^2}{H^2\eta'^2}})\,,
\label{eq:modefunctionWKB}
\end{align}
where $\eta_*$ is an arbitrary integration constant that we choose to satisfy $|s\eta_*|\gg \frac{m}{H}$, without loss of generality. 
The coefficients $c_\pm$ can be determined by matching the WKB solution to the Bunch-Davis vacuum at past infinity, where $\sigma_+$ behaves as
\begin{align}
    \lim_{\eta\to -\infty}\sigma_+(s,\eta)=-\dfrac{H \eta}{\sqrt{2s}}\exp(-is\eta)\,. 
\end{align}
This implies 
\begin{align}
    c_+\approx \dfrac{1}{\sqrt{2}}\exp(-is\eta_*)\,,\qquad c_-=0\,.
\end{align}
It is important to note that the WKB solution in Eq.\,\eqref{eq:modefunctionWKB} does not account for particle production effects at late times. Such effects lead to a mixing of positive and negative frequency solutions, resulting in a small nonzero value for $c_-\propto e^{-\pi m/H}$. However, this contribution is exponentially suppressed in the heavy mass limit and can be safely neglected. \\

\begin{figure}
    \centering
    \includegraphics[scale=0.8]{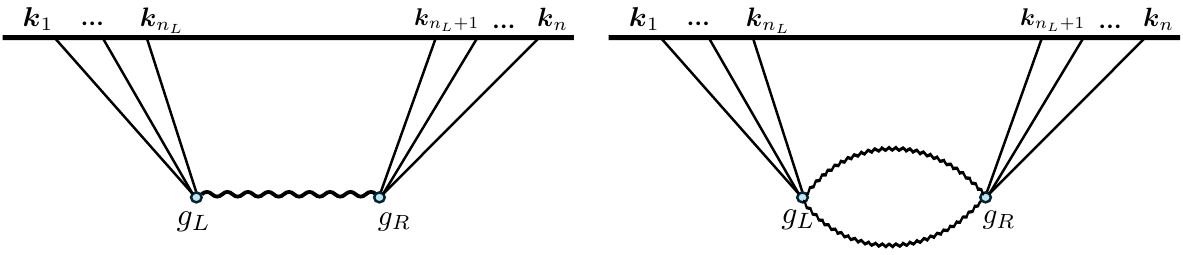}
    \caption{Generic single-exchage and one-loop, bubble graphs contributing to the $n-$point function of $\phi$.}
    \label{generalGSESB}
\end{figure}
\noindent \textbf{A single-exchange diagram}. Now, consider a single-exchange diagram involving an arbitrary number of external conformally coupled fields (see Figure \ref{generalGSESB}). In this case, the in-in expression takes the following form: 
\begin{align}
\nn
    F_{\text{SE}}(\lbrace\omega_i\rbrace, \lbrace \bm{k}_i\rbrace; m)&=\sum_{\pm\pm}(\pm i g_L)(\pm i g_R)\int \dfrac{d\eta_1}{\eta_1^4\,H^4}\,\dfrac{d\eta_2}{\eta_2^4\,H^4}\,G_{\pm\pm}(s,\eta_1,\eta_2)\\\nn
    &\left(\prod_{i=1}^{n_L}O_i\left(\eta_1,\partial_{\eta_1},\bm{k}_i\right)K^{\pm}(\omega_i,\eta_1)\right)\,\left(\prod_{i=n_L+1}^{n}\,O_i\left(\eta_2,\partial_{\eta_2},\bm{k}_i\right)K^{\pm}(\omega_i,\eta_2)\right)
    \\ \label{SEdiag}
    &=\sum_{\pm\pm}(\pm i g_L)(\pm i g_R)\int d\eta_1\,d\eta_2\, f_{\pm\pm}(\eta_1,\eta_2)\, e^{\pm i\omega_{L}\eta_1}\,e^{\pm i\omega_{R}\eta_2}\,G_{\pm\pm}(s,\eta_1,\eta_2)\,,
\end{align}
where $s=|\sum_{i=1}^{n_L} \bm{k}_i|=|\sum_{i=n_{L+1}}^{n} \bm{k}_i|$, $g_{L}$ and $g_R$ are the coupling constants at the left and right vertex, respectively. In the second line above, $f_{\pm\pm}$ is a polynomial in $\eta_{1,2}$ whose  specific form is not important for the following argument, and we have defined $\omega_{L}=\sum_{i=1}^{n_{L}}\omega_i$ and $\omega_{R}=\sum_{i=1}^{n_{R}}\omega_{i+n_L}$, denoting the total external energy of the lines entering the left and right vertices, respectively, with $n_{L,R}$ denoting the number of external legs attached to each vertex. 

It is useful to exchange $\eta_{1,2}$ for the following dimensionless variables
\begin{align}
    x_{1,2}=\frac{H}{m} s\,\eta_{1,2}\,,
\end{align}
and rewrite $F_{\text{SE}}$ as
\begin{align}
\nn
    & F_{\text{SE}}(\lbrace\omega_i\rbrace, \lbrace \bm{k}_i\rbrace; m)=\left(\dfrac{m}{H}\right)^2 \dfrac{1}{s^2}\sum_{\pm\pm}(\pm i g_L)(\pm i g_R)\times \\ \label{SEexpress}
    &\int_{-\infty(1\mp i\epsilon)}^0 dx_1\,dx_2\, \tilde{f}_{\pm\pm}(x_1,x_2)\, \exp(\pm i\frac{m}{H}\,\dfrac{\omega_{L}}{s}\,x_1)\,\exp(\pm i\frac{m}{H}\,\dfrac{\omega_{R}}{s}\,x_2)\,G_{\pm\pm}\left(s,\frac{m}{H}\,x_1,\frac{m}{H}\,x_2\right)\,,
\end{align}
where we have defined
\begin{align}
    \tilde{f}_{\pm\pm}=f_{\pm\pm}\left(\dfrac{m}{H}\dfrac{1}{s}\,x_1,\dfrac{m}{H}\dfrac{1}{s}x_2;s,\omega_i\right)\,.
\end{align}
The above change of variable makes the highly oscillating parts of the integral manifest. Specifically, in the MFS limit, the first two exponential factors in Eq.\,\eqref{SEexpress} are slowly varying because 
\begin{align}
   \frac{m}{H}\frac{\omega_{R}}{s}\sim  \frac{m}{H}\frac{\omega_{L}}{s}\sim {\cal O}(1)\qquad (\text{MFS limit})\,.
\end{align}
Using the solutions from  \eqref{eq:modefunctionWKB} the bulk-to-bulk propagators are given by 
\begin{align}
\nn
    G_{\pm\pm}&=\dfrac{1}{2s^3}\left(\dfrac{m}{H}\right)^2\,\dfrac{x_1}{(1+x_1^{-2})^{1/4}}\,\dfrac{x_2}{(1+x_2^{-2})^{1/4}}\,\exp(\mp\,i\frac{m}{H}\int_{x_1}^{x_2}\,dx'\,(1+\frac{1}{x'^2})^{1/2})\theta(x_2-x_1)\\ \nn
    &+x_1\leftrightarrow x_2
    \,,\\ 
    G_{\pm\mp}&=\dfrac{1}{2s^3}\left(\dfrac{m}{H}\right)^2\,\dfrac{x_1}{(1+x_1^{-2})^{1/4}}\,\dfrac{x_2}{(1+x_2^{-2})^{1/4}}\,\exp(\pm\,i\frac{m}{H}\int_{x_1}^{x_2}\,dx'\,(1+\frac{1}{x'^2})^{1/2})\,.
\end{align}
Conversely to the first two exponential factors in \eqref{SEexpress}, these propagators exhibit rapidly oscillating phases.

We first study the $+-$ and $-+$ contributions to $F_{\text{SE}}$: The corresponding integrals over $\int dx_1\,dx_2\,$ factorise, and each integral becomes exponentially small in the mass due to the fast oscillations of $G_{\pm\mp}$. 
This can be seen
by noting that each integral takes the following schematic form 
\begin{align}
\label{YuhangIntegral}
    {\cal I}_\pm=\int_{-\infty(1\mp i \epsilon)}^0 dx\,\text{Poly}(x)\,\exp\left(\pm i\frac{m}{H}\frac{\omega}{s}x\right)\exp(\mp i \frac{m}{H}\int_{x_0}^x\dfrac{dx'}{x'}(1+x'^2)^{1/2})\,, 
\end{align}
where $x_0$ is an arbitrary constant that drops out of $F^{\pm\mp}_{\text{SE}}$(=${\cal I}_+\times {\cal I}_-$), and $\omega=\omega_L$ or $\omega_R$. The second exponential factor above has the following convergent expansion around $x=0$:
\begin{align}
\nn
    &\exp\left(\pm i \frac{m}{H}\int_{x_*}^x\dfrac{dx'}{x'}(1+x'^2)^{1/2}\right)=\\ \label{expansionWKB}
    &(-x)^{\pm i m/H}e^{\pm iC(x_0)}\left[1\pm\dfrac{i}{4}\frac{m}{H}x^2-\dfrac{1}{32}(\pm i\frac{m}{H}+\frac{m^2}{H^2})x^4+\dots\right]\,,
\end{align}
where $C(x_0)=(1-x_0-\log(2))\,m/H$, and $e^{\pm iC(x_0)}$ is a pure $x$-independent phase. Inserting this expansion in ${\cal I}_\pm$, we find
\begin{align}
   {\cal I}_\pm=\sum_n c^\pm_n(m) \int_{-\infty(1\mp i\epsilon)}^0 dx\,\text{Poly}(x)\,x^n\,\exp\left(\pm i\frac{m}{H}\frac{\omega}{s}x\right)\,(-x)^{\pm i\, m/H}\,e^{\pm iC(x_0)}\,,
\end{align}
where $c^\pm_n(m)$ are the mass-dependent coefficients appearing within the bracket, in \eqref{expansionWKB}, which are polynomials in $m/H$. Sending mass to infinity causes all the terms above to vanish exponentially because they include a factor of the form
\begin{align}
    \int_{-\infty(1\mp i\epsilon)}^0 (-x)^{k\pm i m/H}\,\exp\left(\mp i\frac{m}{H}\frac{\omega}{s}x\right)\,dx=\Gamma(1+k\mp im/H)\,\left(\pm i \dfrac{m}{H}\frac{\omega}{s}\right)^{-1-k\pm im/H}\,\quad (k\in \mathbb{Z})\,,
\end{align}
which is indeed exponentially small in the mass\footnote{A more direct method involves using the saddle point approximation to evaluate the integral \ref{YuhangIntegral}. Due to the branch cut of the square root, it is somewhat nontrivial to show that the integral is indeed exponentially suppressed, which becomes evident after performing an appropriate contour deformation. We thank Yuhang Zhou for their correspondence on this point and refer the reader to \cite{upcomingYuhang} for details.}.

In contrast, the $++$ and $--$ components receive contributions from the vicinity of the diagonal $x_1=x_2$, which are not exponentially suppressed. Within these blocks, at leading order in $H/m$, the propagators $G_{++}$ and $G_{--}$ effectively resemble Dirac delta functions of the form $h(x_1,x_2)\delta(x_1-x_2)$, where $h$ is an appropriate multiplicative function. This can be seen by expanding the $\pm\pm$ propagators around the diagonal $x_1=x_2$: 
\begin{align}
\nn
    G_{\pm\pm}&\sim \dfrac{1}{2s^3}\left(\dfrac{m}{H}\right)^2\,\dfrac{x_1^3}{\sqrt{1+x_1^2}}\\ 
    & \times\exp(\mp\,i\frac{m}{H}(1+\frac{1}{x_1^2})^{1/2}\Delta x)\left(1+{\cal O}(\Delta x)\right)\theta(\Delta x)+(\Delta x\to -\Delta x)\,,
\end{align}
where $\Delta x=x_2-x_1$. The $\pm\pm$ contributions take the following schematic form
\begin{align}
    {\cal F}_{\pm\pm}=\int_{-\infty(1\mp i\epsilon)}^0 dx\,\int_0^{\infty(1\pm i\epsilon)}d(\Delta x)\left(1+{\cal O}(\Delta x)\right)J_{\pm}(x,\Delta x)\exp(\mp\,i\frac{m}{H}(1+\frac{1}{x^2})^{1/2}\Delta x)\,,
\end{align}
where $J_{\pm}$ are slowly varying functions of $x$ and $\Delta x$. In the $m/H\to \infty$ limit, this integral simplifies to: 
\begin{align}
\label{deltaapprox}
    \lim_{m\to \infty}{\cal F}_{\pm\pm}=\int_{-\infty(1\mp i\epsilon)}^0 dx\,\left(\pm\,i\frac{m}{H}(1+\frac{1}{x^2})^{1/2}\right)^{-1}\,J_{\pm}(x,\Delta x=0)\left(1+{\cal O}(H/m)\right)\,,
\end{align}
where the ${\cal O}(H/m)$ terms come from higher order corrections in $\Delta x$. The approximation above can be replicated by the following replacements:  
\begin{align}
    G_{\pm\pm}\to \pm i\dfrac{1}{s^3}\dfrac{m}{H}\,\dfrac{x_1^3}{1+x_1^2}\delta(x_1-x_2)=\pm i\dfrac{\eta^4}{\frac{m^2}{H^2}+s^2\eta^2}\,\delta(\eta-\eta')\,.
    \label{eq:MFSFeynamnnProp}
\end{align}
Note that both the $i\epsilon$ prescription and the step functions $\theta(\pm\Delta x)$ in the propagators were crucial to reaching this conclusion. 

Using the asymptotic behavior of the $\pm\pm$ propagators derived above and ignoring the exponentially suppressed $\pm\mp$ branches, we ultimately arrive at the MFS limit of the single-exchange diagram. This corresponds to Eq.\,\eqref{flatspace} with the following flat-space amputated $n-$point diagram:
\begin{align}
\nn
G_n=g_L\,g_R\,\dfrac{i}{s^2/a^2(\eta)+m^2}=g_L\,g_R\,\dfrac{i}{(p_1(\eta)+\dots+p_{n_L}(\eta))^2+m^2}\,,
\end{align}
where $p^\mu_i(\eta)=(0,\bfk_i/a(\eta))$ are the fictitious, time-dependent four-momenta associated with the external legs. 
\subsubsection*{The MFS limit of one-loop, bubble diagrams}
Consider the one-loop bubble diagram in Figure \ref{generalGSESB}, which is described by the following expression:
\begin{align}
\nn
    F_{\text{Bubble}}(\lbrace\omega_i\rbrace, \lbrace \bm{k}_i\rbrace; m)&=\sum_{\pm\pm}(\pm i g_L)(\pm i g_R)\int \dfrac{d\eta_1}{(\eta_1\,H)^{d}}\,\dfrac{d\eta_2}{(\eta_2\,H)^{d}}\\\nn
    & \times \mu^{4-d}\,\int \dfrac{d^{d-1}\bm{q}}{(2\pi)^{d-1}}G_{\pm\pm}(|\bm{q}|,\eta_1,\eta_2)\,G_{\pm\pm}(|-\bm{q}+\textbf{s}|,\eta_1,\eta_2)\\ \label{bubblegraph}
&\,\,\times\left(\prod_{i=1}^{n_L}O_i\left(\eta_1,\partial_{\eta_1},\bm{k}_i\right)K^{\pm}(\omega_i,\eta_1)\right)\,\left(\prod_{i=n_L+1}^{n}\,O_i\left(\eta_2,\partial_{\eta_2},\bm{k}_i\right)K^{\pm}(\omega_i,\eta_2)\right)\,.
\end{align}
The diagram is regulated using dimensional regularization \cite{Senatore:2009cf,Ballesteros:2024qqx,Lee:2023jby,Bhowmick:2024kld}. For consistency with dim. reg., the mode functions associated with both the external and internal lines have to be continued to $d$ dimensions, though in this case continuing the bulk-to-boundary propagators to general dimensions is inconsequential. Additionally, the renormalization scale $\mu$ is included to adjust the dimension of the loop integral. 

In the heavy mass limit $m/H \to \infty$, similar to the tree-level case, the $+-$ and $-+$ contributions in the second line of Eq.\,\eqref{bubblegraph} yield factorized contributions (in $\eta_1$ and $\eta_2$), each of which is exponentially suppressed due to the rapid oscillations of $\sigma_{\pm}(q, \eta_{1,2}) \sigma_{\pm}(|\mathbf{s} - \mathbf{q}|, \eta_{1,2})$. In contrast, using the approximate solutions from Eq.\,\eqref{eq:modefunctionWKB}, the squares of the $++$ and $--$ propagators take the following form:
\begin{align}
\nn
 &G_{\pm\pm}(|\bm{q}|,\eta_1,\eta_2)\,G_{\pm\pm}(|\textbf{s}-\bm{q}|,\eta_1,\eta_2)=\frac{1}{2}H(\eta_1,\eta_2)\theta(\eta_2-\eta_1)\\ \label{fastoscil}
 &\times\exp\left(\mp\,i\int_{\eta_1}^{\eta_2}d\eta\,(\bm{q}^2+\frac{m^2}{H^2\eta^2})^{1/2}\right) \exp\left(\mp\,i\int_{\eta_1}^{\eta_2}d\eta\,(|\textbf{s}-\bm{q}|^2+\frac{m^2}{H^2\eta^2})^{1/2}\right)+\eta_1\leftrightarrow \eta_2\,,
\end{align}
where 
\begin{align}
\nn
    H(\eta_1,\eta_2)&=\dfrac{1}{2\,a^{d-2}(\eta_1)\,a^{d-2}(\eta_2)}\left(q^2+\frac{m^2}{H^2\eta_1^2}\right)^{-1/4}\,\left(q^2+\frac{m^2}{H^2\eta_2^2}\right)^{-1/4}\\ 
    &\times \left((\textbf{s}-\bm{q})^2+\frac{m^2}{H^2\eta_1^2}\right)^{-1/4}\,\left((\textbf{s}-\bm{q})^2+\frac{m^2}{H^2\eta_2^2}\right)^{-1/4}\,.
\end{align}
In analogy to the tree-level case, for a fixed $\mathbf{q}$, the rapid oscillations of the phases in the large-mass regime suppress the time integrals, except within a small strip around the diagonal $\eta_1 = \eta_2$. Consequently, the squares of the propagators effectively behave like distributions that are proportional to $\delta(\eta_1 - \eta_2)$. 

The functions appearing in front of these Dirac delta distributions can be derived similarly to Eq.\,\eqref{loopdelta}, resulting in:
\begin{align}
\label{loopdelta}
    &G_{\pm\pm}(|\bm{q}|,\eta_1,\eta_2)\,G_{\pm\pm}(|\textbf{s}-\bm{q}|,\eta_1,\eta_2)\sim \\ \nn
    &\pm\,i\,\left((q^2+\frac{m^2}{H^2\,\eta_1^2})^{1/2}+((\textbf{s}-\bm{q})^2+\frac{m^2}{H^2\,\eta_1^2})^{1/2}\right)^{-1}\,H(\eta_1,\eta_1)\,\delta(\eta_1-\eta_2)\,,\\ \nn
    &=\pm i\,\dfrac{1}{2}(-H\eta_1)^{2d-1}\,\dfrac{1}{E_q(\eta_1)\,E_{|\textbf{s}-\bm{q}|}(\eta_1)}\dfrac{1}{E_q(\eta_1)+E_{|\textbf{s}-\bm{q}|}(\eta_1)}\,\delta(\eta_1-\eta_2)\,,
\end{align}
where in the last line we have introduced the time-dependent energy variables 
\begin{align}
    E_{k}(\eta)=\left(\frac{k^2}{a^2(\eta)}+m^2\right)^{1/2}\,.
\end{align}
We emphasize that the effectiveness of replacing $G_{\pm\pm}^2$ with a Dirac delta function remains unchanged as we vary the loop momentum. In fact, for ultraviolet modes, where $|\bm{q}|\to \infty$, this approximation improves further due to the phase factor in Eq.\,\eqref{fastoscil}, which grow with $|\bm{q}|$.

We proceed with simplifying the expression in \eqref{loopdelta}: The last line can be recast as an an integral over a fictitious energy variable $\bar{q}^0$, namely we can write: 
\begin{align}
\nn
    &\dfrac{1}{2}\dfrac{1}{E_q(\eta_1)\,E_{|\textbf{s}-\bm{q}|}(\eta_1)}\dfrac{1}{E_q(\eta_1)+E_{|\textbf{s}-\bm{q}|}(\eta_1)}=\\ 
    &-\int_{-\infty}^{+\infty}\,\dfrac{d\bar{q}^0}{2\pi i}\,\dfrac{1}{-(\bar{q}^0)^2+E_q^2(\eta_1)+i\epsilon}\,\dfrac{1}{-(\bar{q}^0)^2+E_{|\textbf{s}-\bm{q} |}^2(\eta_1)+i\epsilon}\,.
\end{align}
By combining $\bar{q}^0$ and $\bm{q}/a$, it is natural to form a full four-momentum in $d$ dimension, namely 
\begin{align}
    \bar{q}^\mu=(\bar{q}^0,\bm{q}/a(\eta))\,,
\end{align}
in terms of which the measure of the momentum integrals can be re-expressed as
\begin{align}
    d\bar{q}^0\,d^{d-1}\bm{q}=d^{d}\bar{q}\,(-H\eta)^{1-d}\,. 
\end{align}
Using the new four-momentum $\bar{q}$ we can rewrite the middle term in Eq.\,\eqref{bubblegraph} as
\begin{align}
\nn
    &\mu^{4-d}\,\int \dfrac{d^{d-1}\bm{q}}{(2\pi)^{d-1}}G_{\pm\pm}(|\bm{q}|,\eta_1,\eta_2)\,G_{\pm\pm}(|-\bm{q}+\textbf{s}|,\eta_1,\eta_2)=\\ \label{similifiedGG}
    &\mp\mu^{4-d}(-\eta_1 H)^{d}\int \dfrac{d^d \bar{q}}{(2\pi)^d}\dfrac{1}{\bar{q}^2+m^2+i\epsilon}\dfrac{1}{(\bar{q}-\bar{s})^2+m^2+i\epsilon}\delta(\eta_1-\eta_2)\,,
\end{align}
where we have introduced the artificial four-momentum
\begin{align}
    \bar{s}^\mu=(0,\textbf{s}/a(\eta))\,.
\end{align}

Substituting the simplified expression \eqref{similifiedGG} in \eqref{bubblegraph}, we recover the MFS reduction formula for the one-loop diagram, with 
\begin{align}
\label{simplifiedloop}
    G_n(p_1(\eta),\dots,p_n(\eta))=(ig_L)\,(ig_R)\,\mu^{4-d}\int \dfrac{d^{d}\bar{q}}{(2\pi)^{d}}\dfrac{-i}{\bar{q}^2+m^2+i\epsilon}\dfrac{-i}{(\bar{q}-\sum_{i=1}^{n_L}p_i(\eta))^2+m^2+i\epsilon}\,.
\end{align}
As explained in Section \ref{seedcorrelatorssection}, the UV divergent  part of $G_n$, as 
$d\to 4$, must be canceled by an appropriate local counterterm in the action or, equivalently, by adding a contact diagram. We discuss this point further in Section \ref{seedcorrelatorssection}.

 In Appendix \ref{app:spectral}, we provide an alternative proof of the MFS limit of the bubble diagram based on its spectral decomposition. Starting with the Källén–Lehmann representation of the composite operator $\sigma^2$, we express the bubble diagram as an integral over an infinite continuum of tree-level exchange diagrams, each weighted by a spectral density\cite{Xianyu:2022jwk}. In the heavy mass limit, $m \to \infty$, we show that the spectral density in this representation reduces to its Minkowski-space counterpart, with corrections suppressed by powers of $H/m$. The asymptotic behavior of the spectral density, combined with the reduction formula for the tree-level diagram \eqref{linear_MFS_corr}, reproduces the expected MFS limit of the bubble graph.

\subsection{The MFS limit from the effective action in curved spacetime}
\label{MFSfromeffectiveaction}
In this section, we derive the MFS limit reduction formula directly from the path integral by examining how the effective action arises after integrating out internal massive lines. The key insight is that when the Compton wavelength of a massive field in a curved background is much smaller than the characteristic scale set by the background curvature \(H\), the field can be integrated out using an \textit{in-out} path integral. This viewpoint is standard in much of the QFT literature on curved spacetime, where background curvature is treated perturbatively within the in-out effective action relative to other mass scales in the problem (see, e.g., \cite{Parker:2009uva}). Not surprisingly, this simplification overlooks non-perturbative effects due to particle production induced by curvature, which cannot be captured at any order in the in-out effective action.

As argued earlier, in the large-mass limit, the contributions from non-time-ordered propagators, which correspond to mixed terms connecting the upper and lower branches of the Schwinger-Keldysh contour, become exponentially small. This suppression arises from the rapid oscillatory behavior of the Wightman functions in the heavy-mass regime. Consequently, the dominant contributions to the effective action and resulting correlators are well-approximated by terms constructed solely from time-ordered propagators and their complex conjugates. Under these assumptions, the in-out path integral is adequate for computing the effective action, with interactions governed by time-ordered Feynman propagators, thereby simplifying the analysis.

The in-out effective action is expressed as:
\begin{align} 
\label{intout} 
e^{iS_{\text{eff}}(\phi)} = \int \mathcal{D}\sigma\, e^{iS(\phi, \sigma)}\,,
\end{align}
where $\sigma$ is the heavy field, and $\phi$ denotes the light degrees of freedom.
\begin{figure}
    \centering
    \includegraphics[scale=1]{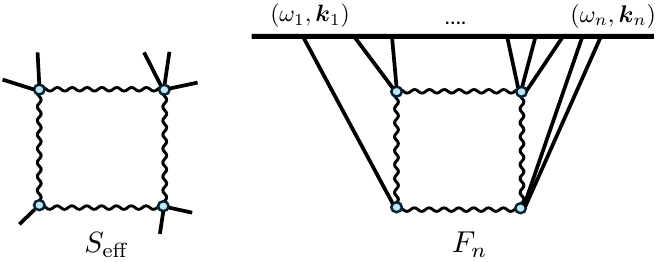}
    \caption{A sample amputated diagram contributing to the effective action of $\phi$ obtained by integrating out the heavy field, perturbatively. After dressing its external legs with bulk-to-boundary operators, the same diagram contributes to the $n-$point correlator of $\phi$ (in the notation of Eq.\,\eqref{eftcurved}, here we have $n=8,\,\cV=r=4, \cL=1,\,t_1=1,\,t_2=t_4=2$ and $t_3=3$).}
    \label{seffdiag}
\end{figure}
Typically, to compute correlation functions of $\phi(x)$ at sufficiently large separations, $S_{\text{eff}}$ is expressed as a sum over a series of \textit{local} operators. However, as we will show below, in our case, the spatial derivatives are not small compared to the mass of $\sigma$. As a result, $S_{\text{eff}}$ must be regarded as a non-local effective action (see also \cite{Baumann:2011nk,Achucarro:2012sm,Achucarro:2012yr,Jazayeri:2023kji}).

We will divide the proof into two. First, we show that for very heavy fields, the Feynman propagator can be expressed in terms of the flat-space propagator up to curvature corrections suppressed by powers of $H/m$. Second, we demonstrate that in the MFS limit, vertices can be explicitly made local in time while maintaining non-locality in space. 

Let us start by computing the Wilsonian in-out effective action. Since the fields $\phi$ and $\sigma$ are weakly coupled, the path integral in Eq.\,\eqref{intout} can be evaluated perturbatively. For simplicity, we focus on the case where all vertices are polynomial and of the form  $\phi^n \sigma^m$, without derivatives. Extending the proof to include vertices with derivatives is straightforward and does not alter the core arguments.

A connected graph with $n$ external legs and $\cV$ vertices, connected exclusively by heavy internal lines, contributes to the effective action a term of the schematic form:
\begin{align}
\label{eftcurved}
iS_{\text{eff}}\supset \left(\prod_{a=1}^\cV\,\int i\lambda_a\sqrt{-g(x_a)}d^dx_a\right)\,\left(\prod_{\text{internal lines (b,c)}}\Delta_F(x_b,x_c)\right)\phi^{t_1}(x_1)\phi^{t_2}(x_2)\dots \phi^{t_r}(x_r)\,,
\end{align}
where $r\leq \cV$ is the number of vertices attached to the external fields, $\lambda_a$'s are the coupling constants, $t_a (a=1,\dots r)$ is the number of external legs attached to the $\cV$-th  vertex (the total number of external legs is then $n=\sum_{i=1}^\cV t_a$), and $\Delta_F$ is the Feynman propagator in curved space attributed to the scalar $\sigma$, namely 
\begin{align}
\Delta_F(x,x')=\langle\text{VAC}|T\lbrace\sigma(x)\sigma(x')\rbrace|\text{VAC} \rangle\,.
\end{align}
We have also expressed the action in $d\neq 4$ to accommodate for potential UV divergences in the diagram, which will be regulated using dim-reg.  
Notice that the effective action in \eqref{eftcurved} is unitary as a result of having dropped the non-time-ordered contributions, such as those arising from $\pm\mp$ propagators.\\

\noindent \textbf{Curvature expansion of the propagator.} The Feynman propagator in curved space satisfies the following PDE:  
\begin{align}
    (\Box-m^2)\Delta_F(x,x')=\dfrac{i}{\sqrt{-g(x)}}\delta^4(x-x')\,.
\end{align}
 When $m \gg H$, it is expected that the massive propagators mediating between two external light fields can be locally expanded in powers of the spacetime curvature. This is because, in the large mass limit, the Compton wavelength of the massive field becomes much smaller than the curvature scale set by the Hubble parameter $H$.  To make this notion precise, we introduce Riemann normal coordinates $\xi$ in a neighborhood around the point $x'$. In these coordinates, the metric near $x'=0$ can be expressed as:
\begin{align}
\bar{g}_{\mu\nu} = \eta_{\mu\nu} - \frac{1}{3} R_{\mu\alpha\nu\beta}(x') \xi^\alpha \xi^\beta + \mathcal{O}(\xi^3)\,,    
\end{align}
where $R_{\mu\alpha\nu\beta}(x')$ is the Riemann curvature tensor at $x'$ and  $\eta_{\mu\nu}$ is the flat-space Minkowski metric.
 We can use these coordinates around $x'$ and express the propagator in a local momentum space, following the approach introduced by Bunch and Parker~\cite{Bunch:1979uk}\footnote{The curvature expansion of the propagator in \eqref{schematic:MFS_propagat} can be derived by various other methods, including the heat-kernel formulation (see, e.g., \cite{Vassilevich:2003xt} and references therein).}. In this coordinate system, the Feynman propagator satisfies the following  equation:
\begin{align}
\label{expandedKlein}
    \left[\eta^{\mu\nu}\partial_\mu\partial_\nu-(m^2-R(x')/6)-\dfrac{1}{3}\eta^{\mu\nu} R_{\alpha\mu}(x')\xi^\alpha\partial_\nu+\dots\right]((-\bar{g}(\xi))^{\frac{1}{4}}\Delta_F(\xi;x'))=i\,\delta^d(\xi)\,.
\end{align}
In de Sitter spacetime, the Ricci tensor takes the form $R_{\mu\nu} = (d-1)H^2 g_{\mu\nu}$, and $\bar{g}$, the determinant of the metric in Riemann normal coordinates, is expanded as $ -1 + \frac{1}{3}R_{\mu\nu} \xi^\mu \xi^\nu + \dots$. The terms represented by dots in the equation correspond to contributions involving three or more derivatives of the metric. Within the brackets, indices are raised and lowered using the flat-space metric $\eta_{\mu\nu}$.  In de Sitter spacetime, curvature corrections scales as  $\nabla^m\,R^n\sim {\cal O}(H^{m+2n})$, making  $H |\xi|$ and $H/m$ suitable small expansion parameters. Solving the propagator equation iteratively in curvature yields:

\begin{align}
\nn
  \Delta_F(\xi;x')&=-i (-\bar{g}(\xi))^{-\frac{1}{4}}\int \dfrac{d^d\bar{q}}{(2\pi)^d}\dfrac{1}{\bar{q}^2+m^2+i\epsilon}e^{i\bar{q}_\mu \xi^\mu}\left(1+\frac{R(x')}{6}\dfrac{1}{\bar{q}^2+m^2+i\epsilon}+\dots\right)\,\\ \label{expansionDelta}
  &=\Delta_{\text{flat}}(\xi)\left[1-\dfrac{1}{12}R_{\mu\nu}\xi^\mu\xi^\nu-\dfrac{d-2}{12}\dfrac{R}{m^2}-\dfrac{1}{12}\dfrac{R}{m^2}\dfrac{F'(\sqrt{-\xi^2}m)}{F(\sqrt{-\xi^2}m)}\sqrt{-\xi^2}m+\dots\right]\,,
\end{align}
where $\xi^2=\eta_{\alpha\beta}\xi^\alpha\xi^\beta$, and  $F$ is defined through,
\begin{align}
    \Delta_{\text{flat}}(\xi)=-i\,m^{d-2}F(\sqrt{-\xi^2}m)=\int\dfrac{d^d\bar{q}}{(2\pi)^d}\dfrac{-i}{\bar{q}^2+m^2+i\epsilon}e^{i\bar{q}_\mu \xi^\mu}\,.
\end{align}
More explicitly, the function $F$ takes the form

\begin{align}
F = \left(i\,m\sqrt{-\xi^2}\right)^{(2-d)/4} K_{d/2-1}(i\,m\sqrt{-\xi^2}) \,
\end{align}
where $K_{d/2-1}$ is the modified Bessel function of the second kind. We have that  $F$ is of order one for $\xi\sim {\cal O}(1/m)$, while it behaves as $\exp(-m\sqrt{-\xi^2})$ at long distances, which is either exponentially damping or highly oscillating.  This asymptotic behavior implies that, in the heavy mass limit, the effective action $S_{\text{eff}}$ receives its dominant contribution from regions where the physical distance between the vertices $x_a$ is of order $1/m$ or smaller. Larger separations are exponentially suppressed, confirming that the interactions mediated by the heavy field are highly localized in space. In the large-mass limit, it is therefore valid to substitute the expansion \eqref{expansionDelta} into the effective action. This is because the dominant contributions to $S_{\text{eff}}$ arise from regions where $\xi\sim \mathcal{O}(1/m)$,  much smaller than the spacetime curvature scale $1/H$ in de Sitter. At leading order in $H/m$, we retain only the leading term, which allows us to use the Minkowski space representation of the Feynmann propagator $\Delta_{\text{flat}}$ to further simplify the calculations.

If needed, curvature corrections can be included systematically by perturbatively adding higher-order terms from the expansion within the bracket in Eq.\,\eqref{expansionDelta}. 
In general,  the Feynman propagator $\Delta_F(x, x')$ can be written as the flat-space Feynman propagator $\Delta^{\text{flat}}(\xi)$, plus curvature-dependent corrections:
\begin{align}
    \Delta_F(x,x')=\Delta_{\text{flat}}(\xi^\mu)\left(1+\sum_{n+l-2k\geq 0} \nabla^{n+l-2k} R^k(x')\,m^{-n}\xi^{l}\,\Delta_{n,l,k}(m\,\xi^\mu)\right)\,,
    \label{schematic:MFS_propagat}
\end{align}
where $R$  represents the components  the Riemann tensor $R_{\mu\nu\alpha\beta}$ and its contractions, $\nabla$ is the covariant derivative (with suppressed indices).  
$\Delta_{n,l,k}$ are functions of the dimensionless combination $m\xi^\mu$, generically of order one.

It will be useful to rewrite the leading-order term of \eqref{expansionDelta} by expressing the Feynman propagator in a locally defined Fourier space. To do this, we first relate the Riemann normal coordinates $\xi$ to the original coordinates $(\eta, \bm{x})$. At leading order, the transformation is given by:
\begin{align}
    \eta-\eta'&=\dfrac{1}{a(\eta')}\left[\xi^0-\dfrac{1}{2}H\xi^i\xi^i+{\cal O}(\xi^3)\right]\,,\\ \nn
    x^i-x'^i &=\dfrac{1}{a(\eta')}\left[\xi^i-H\xi^0\xi^i+{\cal O}(\xi^3)\right]\,.
\end{align}
Inverting this transformation and substituting it into \eqref{expansionDelta}, we obtain the following expression for the Feynman propagator:
\begin{align}
\nn
    &\Delta_F(x,x')=\int \dfrac{d^d\bar{q}}{(2\pi)^d}\dfrac{-i}{\bar{q}^2+m^2+i\epsilon}\exp\left(-i\,\bar{q}^0 a(\eta')(\eta-\eta')+ i\,\bar{q}^i a(\eta')(x^i-x'^i)\right)\left[1+{\cal O}(H \xi,H/m)\right]\,\\ \label{DeltaFlocal}
    &=\dfrac{1}{a^{d}(\eta')}\int \dfrac{d^dq}{(2\pi)^d}\dfrac{-i}{q^2/a^2(\eta')+m^2+i\epsilon}\exp(-iq^0(\eta-\eta')+iq^i(x^i-x'^i))\left[1+{\cal O}(H \xi,H/m)\right]\,,
\end{align}
where the non-linear powers of $\eta$ and $\bm{x}$ appearing in the exponent $\exp(iq_\mu\xi^\mu)$ have been absorbed schematically into the ${\cal O}(H \xi,H/m)$ corrections, and $q=\bar{q}\,a(\eta')$ is the comoving four-momentum. \\

\noindent \textbf{Correlators from the effective action.}
We will now derive the MFS limit using the effective action given in Eq.\,\eqref{eftcurved}. As discussed earlier, the in-out effective action can be used to compute correlation functions, assuming that boundary terms are negligible in the MFS limit. This implies that the heavy graph introduced in Eq.\,\eqref{heavy_graph} should match, in the MFS limit, with the diagram constructed using the effective action (see Figure \ref{seffdiag}). The latter can be expressed as:
\begin{align}
\nn
    &F'_n\left(\omega_i,\bm{k}_i;\frac{m}{H}\right)=2\, \text{Re}\left(\prod_{a=1}^\cV \int i\lambda_a\,a^d(\eta_a)\,d\eta_a d^d\bm{x}_a\right)\,\prod_{\text{internal lines (b,c)}}\Delta_{F}(\eta_b,\bm{x}_b;\eta_c,\bm{x}_c)\\ \label{flatFeynman}
    &\times \prod_{i=1}^{t_1} \tilde{K}^+(\omega_{i}-i\epsilon,\bm{k}_{i};\eta_1,\bm{x}_1)\prod_{i=t_1+1}^{t_2+t_1} \tilde{K}^+(\omega_{i}-i\epsilon,\bm{k}_{i};\eta_2,\bm{x}_2)\dots \prod_{i=t_{r-1}+1}^{n} \tilde{K}^+(\omega_{i}-i\epsilon,\bm{k}_{i};\eta_r,\bm{x}_r)\,,
\end{align}
where we have defined $F'_n=F_n (2\pi)^3 \delta^3(\sum_{i=1}^n \bm{k}_i)$, and the objects $\tilde{K}^+$ are defined by the following positive-energy plane wave: 
\begin{align}
    \tilde{K}^+(\omega,\bm{k};\eta,\bm{x})\equiv K^+(\omega,\eta)\exp(-i\bm{k}.\bm{x})\,.
\end{align}
The $i\epsilon$ term inserted in the argument of the bulk-to-boundary propagators ensures the convergence of the integrals in the $\eta_a\to -\infty$ limit. Note that the formula above corresponds to only one permutation out of a total of $t_1! \dots t_r!$.

The derivation of the MFS limit begins by observing that the dominant contributions to the effective action and $F_n$ arise from configurations where the vertices are closely spaced, as characterized by the conditions \(a(\eta_1)(\eta_a - \eta_1) \lesssim \mathcal{O}(1)/m\) and \(a(\eta_1)|\Delta \bm{x}_a| \lesssim \mathcal{O}(1)/m\). This proximity allows one to consider a local expansion of the bulk-to-boundary propagators, $\tilde{K}^+(\omega, \bm{k}; x)$, around a reference vertex $x_1 = (\eta_1, \bm{x}_1)$. For conformally coupled external fields, the propagators simplify to $\tilde{K}^+ \propto \eta \exp(i\omega \eta)$, and their expansion introduces an overall oscillatory factor $\exp(i\omega_T \eta_1)$ in the integrand. Consequently, the integral over $\eta_1$ is dominated by the region \(|\eta_1| \lesssim \omega_T^{-1}\), while contributions from $|\eta_1| \gg \omega_T^{-1}$ are suppressed by the rapid oscillations regulated by the $i\epsilon$-prescription.

Within this regime, it is important to distinguish the behavior of time and spatial derivative expansions. The time separation, $\Delta \eta_a = \eta_a - \eta_1$, can be treated perturbatively since $|\Delta \eta_a| \sim \mathcal{O}(1)/m$. This allows for a controlled expansion in time derivatives, with higher-order corrections suppressed by powers of $H/m$. On the other hand, spatial separations $\Delta \bm{x}_a = \bm{x}_a - \bm{x}_1$ cannot be expanded in $\bm{k} \cdot \Delta \bm{x}_a$ due to the scaling behavior $|\bm{k} \cdot \Delta \bm{x}_a| \sim \mathcal{O}(1)$ in the MFS limit. Specifically, for $a(\eta_1)\Delta \bm{x}_a \sim \mathcal{O}(1)/m$, the combination $\frac{H}{m}\frac{k}{\omega_T} \sim \mathcal{O}(1)$ becomes non-negligible. This precludes a spatial derivative expansion of the bulk-to-boundary propagators in terms of $\bm{k} \cdot \Delta \bm{x}_a$.

To address this, we use the MFS form of the propagators, as expressed in Eq.\,\eqref{DeltaFlocal}. The translation invariance of the propagators simplifies the spatial integrals over $\Delta \bm{x}_a$, ensuring momentum conservation at each vertex. After performing these integrals, the internal 3-momenta, $\bm{q}_{bc}$, are expressed as linear combinations of external and loop momenta. The time integrals over $\Delta \eta_a$ yield $\mathcal{V} - 1$ energy-conserving delta functions of the form $(2\pi)\delta(\sum_b q_{bc}^0)$, where the sum runs over the energies flowing into the vertex c. External legs do not contribute to these sums as their energies are effectively set to zero

Corrections to the approximation $a(\eta_a) \approx a(\eta_1)$ in the propagators are suppressed by $\mathcal{O}(H/m)$. Expanding the scale factor as $a(\eta_a) = a(\eta_1)(1 + \mathcal{O}(\Delta \eta_a / \eta_1))$ and incorporating these terms into the propagator shows that higher-order corrections are negligible. This results in a further simplification of the integrand, ensuring that time-dependent effects beyond leading order do not affect the dominant contribution in the MFS limit.
Finally, integrating over the internal energy components $q^0_{bc}$, except for $\mathcal{L}$  undetermined loop energies $q_j^0$, leads to the final form:
\begin{align}
\nn
        F^{\text{MFS}}_n =2\,\text{Re}\,&\left\lbrace i^\cV \lambda_1\dots \lambda_\cV\int_{-\infty}^0  a^d(\eta_1)\,d\eta_1 \left(\prod_{i=1}^{n} K(\omega_{i}-i\epsilon,\eta_1)\right)\right.\\ \label{FinalMFS}
        &\left.\times\left(\dfrac{1}{(a^d(\eta_1))^L}\prod_{j=1}^L \dfrac{d^{d}q_j}{(2\pi)^{d}}\right)\prod_{\text{internal lines (b,c)}}\dfrac{-i}{-(q_{bc}^0/a(\eta_1))^2+\bm{q}_{bc}^2/a^2(\eta_1)+m^2}\right\rbrace\,,
\end{align}
This result matches the reduction formula for $F_n^{\text{MFS}}$ given in Eq.\,\eqref{flatspace}, with the amputated flat-space diagram given by
\begin{align}
  G_n(p_i^\mu)=(i\lambda_1)\dots (i\lambda_V) \left(\prod_{j=1}^L \dfrac{d^{d}\bar{q}_j}{(2\pi)^{d}}\right)\prod_{\text{internal lines (b,c)}}\dfrac{-i}{\bar{q}^2_{bc}+m^2}\,,
\end{align}
where the 4-momenta associated with the external legs are defined as $p_i^\mu=(0,\bm{k}_i/a(\eta_1))$, the loop physical momenta are defined by $\bar{q}_j=q_j/a(\eta_1)$, and the internal lines physical four-momenta $\bar{q}_{bc}$ ($=q_{bc}/a(\eta_1)$) are determined in terms of the $p_i$'s and $\bar{q}_j$'s through energy momentum conservation at each vertex.

 Finally, we address the inclusion of derivative interactions. Adding derivatives to the external legs does not alter the validity of the proof outlined above. This is because the derivation primarily relied on two key elements: the flat-space limit of the heavy field two-point function and the observation that the dominant contributions to the in-in time integrals arise from the region \(|\eta_a| \lesssim 1/\omega_T\). These properties remain unchanged when derivatives are applied to the external fields. Consequently, Eq.\,\eqref{FinalMFS} continues to hold, with the modification that $\prod_{i=1}^n K$ should be replaced by $\prod_{i=1}^n O_n(\bm{k}_i, \partial_{\eta})K(\omega_i, \eta)$, where $O_n$ are the appropriate derivative operators acting on the external legs.

For the internal lines, the momentum representation of the propagator in Eq.\,\eqref{DeltaFlocal} makes it clear that spatial derivatives introduce factors of $\pm i\bm{q}_i$. Specifically, a derivative $\partial / \partial \bm{x}$ acting on $\Delta_F(x, x')$ introduces a factor of$ +i\bm{q}_i$, while a derivative $\partial / \partial \bm{x}'$ introduces $-i\bm{q}_i$. Similarly, a time derivative with respect to $\eta$ introduces a factor of $iq^0$, and a derivative with respect to $\eta'$ gives:
\begin{align}
    \partial_{\eta'}\Delta_F\approx \int \dfrac{d^d\bar{q}}{(2\pi)^d}\dfrac{-i}{\bar{q}^2+m^2}\exp(-ia(\eta')\bar{q}^0 \Delta\eta+ia(\eta')\bm{\bar{q}}.\Delta\bm{x})\left(i\bar{q}^0 a(\eta')-i\bar{q}^0 a(\eta')\Delta\eta/\eta'\right). 
\end{align}
The second term in the bracket arises from the derivative acting on the scale factor $a(\eta')$. However, it can be neglected in the MFS limit, where $|a(\eta')\Delta\eta| < \mathcal{O}(1)/m$. This ensures that the dominant contributions come from the first term.

In conclusion, each spacetime derivative with respect to $x$ (or $x'$) acting on the propagator $\Delta_F(x, x')$, which always appears in the scale-invariant combination $a^{-1}(\eta)\frac{\partial}{\partial x}$ 
(or $a^{-1}(\eta')\frac{\partial}{\partial x'}$), introduces an additional factor of $i(\bar{q}_{bc})^\mu$ or $-i(\bar{q}_{bc})^\mu$ in the final MFS reduction formula, Eq.\,\eqref{FinalMFS}. 

\subsection{Back to on-shell correlators with a reduced sound speed}
\begin{figure}
    \centering
    \includegraphics[scale=0.8]{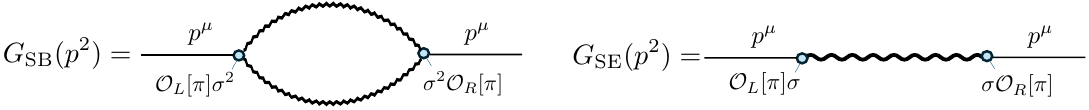}
    \caption{The building blocks of the scalar exchange and scalar bubble diagrams contributing to the correlators of $\pi$.}
    \label{gsesb}
\end{figure}
\label{backtoonshellSection}
To generalize the reduction formula \eqref{FinalMFS} to on-shell diagrams—where the external energies $\omega_i$ and momenta $\bm{k}_i$ are no longer independent but instead satisfy a specific dispersion relation—we consider massless fields propagating at a reduced speed while interacting with heavy fields. A notable example arises in the EFT of inflation, where scalar fluctuations generically acquire a subluminal sound speed.

The free field Lagrangian for such species is given by:
\begin{align}
\nn
    S_2 &=\int a^3(t)\,dt\,d^3\bm{x}\,\left(\dfrac{1}{2}\dot{\pi}_c^2-\dfrac{1}{2}\dfrac{c_s^2}{\,a^2(t)}(\partial_i \pi_c)^2\right)=\int a^2(\eta)\,d\eta\,d^3\bm{x} \left(\dfrac{1}{2}\pi_c'^2-\dfrac{1}{2}c_s^2(\partial_i \pi_c)^2\right)\,,
\end{align}
where $\pi_c$ denotes the canonically normalized phonon $\pi$ appearing in the EFT of inflation. The mode function for $\pi_c$ is given by
\begin{align}
    \pi_{c,\pm}(|\bfk|,\eta)=\dfrac{H}{\sqrt{2c_sk}}(1\pm i\,c_s k)\exp(\mp i\,c_s k \eta)\,,
\end{align}
We also denote the corresponding bulk-to-boundary propagator by $K_\pi$.

Although the action for $\pi_c$ strongly breaks the de Sitter boost symmetries, it preserves dilatation, $\pi_c(\eta, \bm{x}) \to \pi_c(\lambda \eta, \lambda \bm{x})$.
In the next chapter, we review how to rigorously write down the EFT for $\pi$ and couple it to matter fields, using the underlying symmetries. For simplicity, we assume that the free actions of the heavy matter fields remain de Sitter invariant, corresponding to unit sound speeds.

Within the on-shell correlators of $\pi_c$, the limit $c_s \to 0$ effectively mimics the behavior of setting $\omega_i \to 0$ in the off-shell correlators analyzed previously. This correspondence arises because the sound speed $c_s$ directly controls the dispersion relation, tying the external energies and momenta. Consequently, in an alternative MFS limit defined by:
\begin{align}
    c_s\to 0\,,\qquad m/H\to \infty\,,\qquad \text{with}\,\qquad \alpha=\frac{c_s\,m}{H}<\infty\,,
\end{align}
the on-shell correlators of $\pi$ enjoy a similar simplification. In this limit, an arbitrary on-shell $n$-point function of $\pi_c$ with heavy internal lines reduces to:
\begin{align}
\label{csreduction}
F_n^{\text{MFS}}(\lbrace\bm{k}_i\rbrace,\frac{c_s\,m}{H};m)=2\,\text{Re}\int_{-\infty(1-i\epsilon)}^0\,d\eta\,a^4(\eta)\,G_n\left[p_1^\mu(\eta),\dots p_n^\mu(\eta);m\right]\,\prod_{i=1}^n\, O_n\,\,K_\pi^+(|\bm{k}_i|,\eta)\,. 
\end{align}
This expression is similar to Eq.\,\eqref{flatspace}, except that the argument of the bulk-to-boundary propagator is substituted with the original momentum size $|\bm{k}_i|$, rather than $\omega_i$, which does not appear above.
The four-momentum $p_i$ remains unchanged, retaining its definition as $p_i = (0, \bm{k}_i / a(\eta))$. Finally, introducing a sound speed for the external legs does not change the definition of $G_n$, which still describes a flat-space diagram obtained by amputating the external legs. \\

\begin{figure}
    \centering
    \includegraphics[scale=1.1]{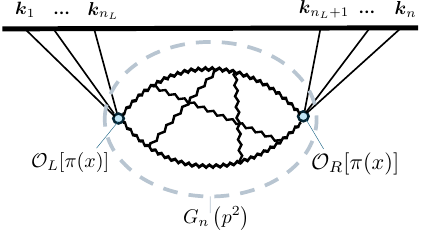}
    \caption{Two vertex, heavy graphs contributing to $n-$point correlators of phonons $\pi$.}
    \label{twovertexOLOR}
\end{figure}

\noindent \textbf{Correlators of $\pi$ from two-vertex, heavy graphs.} We will be interested in simpler diagrams containing only two vertices (see Figure \ref{twovertexOLOR}). In these diagrams, the corresponding amputated contribution depends solely on the momenta at one of the vertices, denoted by $p^\mu = \sum_{i=1}^{n_L}p_i^\mu = -\sum_{i=n_L+1}^n p_i^\mu $, and can be expressed as  $G_n(p^2)$. Within the MFS limit, the effective action for the light field $\pi$ corresponding to such diagrams schematically looks like:
\begin{align}
\label{MFStwovertexaction}
    S_{\text{eff}}=\int a^4(\eta)\,d\eta\,d^3\bfx\,{\cal O}_L[\pi(\eta,\bfx)]\,i\,G_n(\nabla^2/a^2(\eta))\,{\cal O}_R[\pi(\eta,\bfx)]\,.
\end{align}
where ${\cal O}_{L,R}$ are operators at the left and right vertices that couple to the heavy sector. These operators typically take the schematic form $\partial^{\#}\pi^{n_{L,R}}$, with $n_{L,R}$ being the number of external legs attached to each vertex. We emphasize again that $G_n$ is generically non-analytic in $p^2$, and therefore the above action is inherently \textit{non-local} in space, even though it remains local in time.

Among all possible types of vertices, we will be especially interested in the following menu: 
\begin{align}
\nn
   &{\cal O}_{L,R}[\pi(x)]\sigma\, \qquad \,\,\,\,\quad\text{scalar single-exchange (SE)}\,,\\ \nn
   &{\cal O}_{L,R}[\pi(x)]\sigma^2\, \,\,\quad\qquad \text{scalar bubble (SB)}\,, 
\end{align}
using which we can form single-exchange and one-loop, bubble graphs with external $\pi$ legs (Figure \ref{gsesb}). The corresponding amputated diagrams for these cases are given by:

\begin{align}
\label{GSEGSB}
    G_{\text{SE}}(p^2) &=\dfrac{i}{p^2+m^2}\,,\\ \nn
    G_{\text{SB}}(p^2)&=\mu^{4-d}\int \dfrac{d^4\bar{q}}{(2\pi)^4}\dfrac{1}{\bar{q}^2+m^2+i\epsilon}\dfrac{1}{(\bar{q}-p)^2+m^2+i\epsilon}\\ \nn
    &=\dfrac{i\,\mu^{4-d}}{(4\pi)^{d/2}}\Gamma(2-d/2)\int_0^1 dx\,(m^2+p^2x(1-x))^{d/2-2}\\ \label{GSB}
    &\to 
    \dfrac{i}{16\pi^2}\int_0^1 dx\,\log\left[\dfrac{m^2+x(1-x)p^2}{\mu^2}\right]+\dfrac{i}{8\pi^2}\left(-\dfrac{1}{(d-4)}+\dfrac{1}{2}\log(4\pi\,e^{-\gamma_E})\right)\,.
\end{align}
We will use the $\bar{\text{MS}}$ scheme to regularize the bubble graphs, for which we will add counter terms of the form ${\cal O}_L[\pi]{\cal O}_R[\pi]$ to cancel the second term in the last line.  

Finally, we discuss how to extend the effective action beyond the leading-order massive limit. The full effective action is expected to have the following form
\begin{align}
\label{MFScorrected}
    S_{\text{eff}}=\int a^4(\eta)\,d\eta\,d^3\bfx\,{\cal O}_L[\pi(\eta,\bfx)]\,i\,\tilde{G}_n(\Box)\,{\cal O}_R[\pi(\eta,\bfx)]\,,
\end{align}
in which the D'almbertian operator should be understood as 
\begin{align}
    \Box=\dfrac{1}{a^2(\eta)}\nabla^2+\hat{\delta}\,, \quad \text{where}\quad \hat{\delta}=\eta^2\partial_{\eta}^2-2\eta\partial_{\eta}\,,
\end{align}
and the operator $\hat{\delta}$, which includes derivatives with respect to time, should be treated perturbatively. For the single-exchange diagram (where $G_n \propto (p^2 + m^2)^{-1}$), we simply have $\tilde{G}_n = G_n$. In Appendix \ref{EFTintout}, we show that in this case, organizing $G_n$ as a systematic expansion in $\frac{H}{m}\hat{\delta}$ corresponds to adding operators with higher time-derivatives to the lowest-order non-local effective action in \eqref{MFStwovertexaction}, see \cite{Jazayeri:2022kjy,Jazayeri:2023xcj}. At loop level, we generally expect $\tilde{G}_n \neq G_n$ due to the presence of curvature terms in the effective action, in addition to the terms arising from the flat-space limit by replacing ordinary derivatives with covariant ones. Computing these higher-order corrections requires accounting for sub-leading contributions in the propagator expansion, as well as previously neglected higher-order terms in $\Delta x_a$ and $\Delta \eta_a$, which we leave for future work\footnote{Alternatively, the one-loop effective action can be computed using the Schwinger-De Witt asymptotic expansion, which streamlines the inclusion of curvature corrections; see, e.g., \cite{Avramidi:1986mj}.}.
\section{Cosmological Phonon Collider}
\label{PhononColliderSection}
In this section, we apply the MFS reduction formula to compute sample tree-level and one-loop contributions to the bispectrum of curvature perturbations. These diagrams involve the exchange of heavy fields, which can be as massive as $H/c_s\gg H$, yet they leave imprints that cannot be captured by any local single-field theory for inflation.

\subsection{EFT of inflation and energy scales}
Single clock inflation can be thought of as a state of matter in which time translation symmetry is spontaneously broken. From this perspective, the long-wavelength scalar fluctuation during inflation is the Goldstone mode that non-linearly realizes the broken time translation. 
More specifically, under an arbitrary time diffeomorphism $t\to t-\xi$, the Goldstone field $\pi(t,\bfx)$ transforms as $\pi\to \pi(t-\xi,\bfx)+\xi$. In order to construct an effective field for the fluctuations, it is convenient to start from the unitary gauge, defined by $\pi=0$. In this gauge, the EFT should be invariant only under the action of spatial diffs. As a result, at leading order in derivatives, the EFT building blocks comprise $g^{00}$, the extrinsic curvature of constant time hypersurfaces $K_{\m\nu}$ as well as fully covariant quantities constructed out of the Riemann tensor.  
The action then takes the following form:  
\begin{align}
\label{actiondeltag}
S =\int d^4x\,\sqrt{-g}\, &\left(\dfrac{1}{2}M_P^2\,R+M_P^2 \dot{H}g^{00}-M_P^2 (3H^2+\dot{H})\right.\\ \nn
&\left. +\dfrac{1}{2}M_2^4\delta g_{00}^2+\dfrac{1}{2}M_3^4\delta g_{00}^3+\dots-\dfrac{1}{2}\bar{M}_2^4\delta K^{\mu}_\mu{}^2+\dots\right)\,,
\end{align}
where, using the background equation of motion, the Lagrangian has been fixed up to terms that are quadratic or higher in perturbations, namely in $\delta g_{00}=1+g_{00}$ and $\delta K_{\mu\nu}=K_{\mu\nu}-a^2 h_{\mu\nu}$ (with $h_{\mu\nu}$ standing for the induced metric on the constant time hypersurfaces). To leading order in slow-roll parameters, the model-dependent coefficients $M_i$ and $\bar{M}_i$ can be taken as time independent constants. 

The full four-dimensional covariance of the action \eqref{actiondeltag} can be restored via the Stueckelberg trick, namely by sending $t\to t-\pi(t,\bfx)$. We will be interested in the so called decoupling limit of the system, where we take
\begin{align}
M_\mathrm{Pl}\to \infty\,,\qquad |\dot{H}|\to 0\, \qquad (M_\mathrm{Pl}^2|\dot{H}|=\text{fixed})\,.
\end{align}
In this limit, the $\pi$ sector decouples from metric perturbations, and the action for scalar perturbations simplifies to
\begin{align}
\label{EFTpi}
    S_{\pi}=\int dt\,d^3 x\,a^3\,\dfrac{M_\mathrm{Pl}^2|\dot{H}|}{c_s^2}\Big [\dot{\pi}^2-c_s^2(\tilde{\partial}_i \pi)^2+(1-c_s^2)\left(\dot{\pi}(\tilde{\partial}_i \pi)^2+\dfrac{A}{c_s}\dot{\pi}^3\right)+ \dots\Big ]\,,
\end{align}
where $\tilde{\partial}_i=a^{-1}\partial_i$, and dots stand for terms with higher derivatives or powers of $\pi$. Most notable in this Lagrangian is the universal cubic operator $\dot{\pi}(\partial_i \pi)^2$ with its coefficient uniquely fixed by the speed of sound $c_s$, as opposed to the competing cubic interaction $\dot{\pi}^3$ with an independent coefficient. 
We also highlight the fact that, in the decoupling limit, single-clock inflation exhibits the same symmetry breaking pattern as a superfluid condensate at zero temperature. Therefore, the same effective field theory as \eqref{EFTpi} applies to the dynamics of superfluid's \textit{phonons} (see, e.g., \cite{Nicolis:2013sga,Nicolis:2013lma,Nicolis:2015sra,Joyce:2022ydd}). 

We are interested in additional weakly coupled relativistic fields during inflation (with unit sound speeds). 
In addition to the Hubble scale, there are four relevant energy scales in our setup:
\begin{itemize}
    \item The symmetry breaking scale $f_\pi=(2c_sM_\mathrm{Pl}^2|\dot{H}|)^{1/4}$ associated with the broken time translation.
    \item The strong coupling scale of the EFT, given by
\begin{align}
    \Lambda_*=\left(\frac{24\pi}{5}\right)^{\frac{1}{4}}\dfrac{c_s}{(1-c_s^2)^{\frac{1}{4}}}f_\pi\,,
\end{align}
above which the $\delta g_{00}^2$ operator in \eqref{actiondeltag} becomes strongly coupled\footnote{After performing the Stueckelberg trick, the operator $\delta g_{00}^2$ transforms to $(1+\partial_\mu(t+\pi)\partial^\mu (t+\pi))^2=(2\dot{\pi}+(\partial_\mu \pi)^2)^2$. It turns out that, within this block, the most conservative value for $\Lambda_*$ follows from the perturbative unitarity of the $(\partial_i \pi)^4$ operator.}.
\item $H/c_s$, namely the gradient energy of a relativistic field when its momentum crosses the sound horizon, i.e. $k/a=H/c_s$.
\item The mass of the new particle $m$. 
\end{itemize}
The symmetry breaking scale $f_\pi$ is tied to the amplitude of the scalar powerspectrum ($\Delta_\zeta^2\sim 2.2\times 10^{-9}$ \cite{Aghanim:2018eyx}), namely we have $f_\pi= (2\pi \Delta_\zeta)^{-\frac{1}{2}}H\approx 58 H$. The strong coupling scale $\Lambda_*$ is bounded from below because of the Planck lower limit on the speed of sound (coming from Planck's constraints on equilateral and orthogonal non-Gaussianity \cite{Planck:2019kim}):    
\begin{align}
    c_s\geq 0.021\,\,(95\% \text{CL})\qquad \Rightarrow \qquad \Lambda_*\gtrsim 2.3 H\,.
\end{align}

We are interested in the correlators of the phonon $\pi$ induced by the exchange of heavy fields in the MFS limit, namely  
\begin{align}
\label{mfs}
    H\ll m\lesssim {\cal O}(1)\,H/c_s\,. 
\end{align}
Integrating out heavy fields within this mass window yields an effective action for $\pi$ that is generally nonlocal in space but remains local in time. For two-vertex heavy graphs, the corresponding Lagrangian takes the schematic form shown in Eq.\,\eqref{MFStwovertexaction}. In contrast, integrating out heavier species produces the usual local higher-derivative operators in the EFT of inflation, while lighter particles result in fully non-local actions. While it is intriguing to describe the exchange of moderately heavy fields by adding new non-local operators to the EFT of inflation, it is not obvious how to construct a non-local EFT of inflation from a bottom-up point of view. Therefore, in all computations below, we rely on some known, local and weakly coupled UV picture. Finally, for the full consistency of our setup, we further require that the mass of the heavy field is below the cut-off of the EFT, i.e., 
\begin{align} 
m < \Lambda_*\,.
\end{align} 
Depending on the value of $c_s$, this condition might be more stringent than the upper bound in Eq.\,\eqref{mfs}. 
\begin{figure}
    \centering
    \includegraphics[scale=0.55]{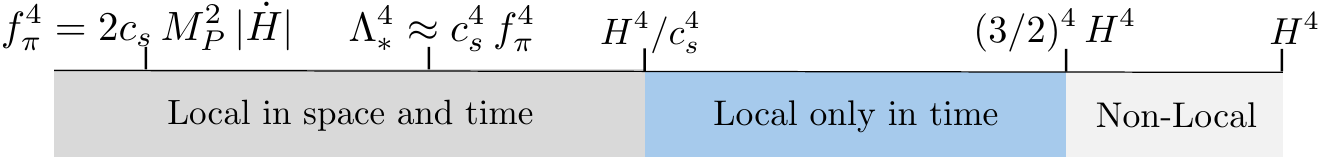}
    \caption{Characteristic energy scales in the EFT of inflation. Integrating out an additional heavy scalar results in an effective action for $\pi$, which is \textit{non-local} in space but local in time, if $m<{\cal O}(1)H/c_s$, or local in both space and time when the field is heavier. Integrating out fields in the complementary series always gives fully non-local effective actions.}
    \label{fig:enter-label}
\end{figure} 
\subsection{Coupling matter fields to the EFT of inflation}
It is straightforward to couple additional matter fields to the EFT of inflation. In this work, we study relativistic species that belong to the unitary representations of the de Sitter (dS) isometry group $\mathrm{SO}(4,1)$. While the second order Lagrangian for these matter fields preserves dS isometries, we allow their interactions with the $\pi$ sector to weakly break de Sitter boost invariance\footnote{Effective field theories can also be constructed for species that strongly break dS boosts at the level of the quadratic action. For massive scalars, this involves introducing a non-trivial speed of propagation, while for spinning fields, such breaking results in qualitatively different dynamics compared to their dS-symmetric counterparts; see, e.g., \cite{Bordin:2018pca}.}. As a result, the correlators of $\pi$ due to the exchange of these fields do not satisfy conformal Ward identities, although they remain scale-invariant. 

In this work, we do not aim to write the most general set of interactions between the matter fields and $\pi$. Instead, we only consider representative mixing terms that lead to tree-level and loop-level contributions to the $\pi$ bispectrum. Nevertheless, our techniques for the computation of these diagrams can be easily adapted to other variations of the
vertices.

In the unitary gauge, we consider the following operators (up to quadratic order in the massive field):
\begin{align}
    &\left(\delta g^{00}\right)^n\sigma\,,\left(\delta g^{00}\right)^n\sigma^2\,,
\end{align}
where $n=(1,2)$. The first operator generates tree-level processes, while the second one contributes at loop level. Notably, for charged scalars, such as a Higgs-like field in the unbroken phase, the linear term is absent and their effect on $\pi$'s correlators start at one-loop level.

We employ the Stückelberg trick to reintroduce $\pi$, which amounts to the transformation:
\begin{align}
    \delta g^{00}\to 1+\bar{g}^{\mu\nu}\partial_\mu(t+\pi)\partial_\nu(t+\pi)\,. 
\end{align}
Substituting this 
into the action and taking the decoupling limit, we obtain the following set of interaction operators: 
\begin{tcolorbox}[colframe=white,arc=0pt]
\begin{align}
S_{\text{int}}&=\int\,d\eta\,d^3x\,a^4\left({\cal L}^{\sigma^1}+{\cal L}^{\sigma^2}\right)\,,\\ \nn
    {\cal L}^{(\sigma^1)}&=\rho\, \dfrac{1}{a(\eta)}\pi'_c\sigma+\dfrac{1}{\Lambda_1}\dfrac{1}{a^2(\eta)}\pi'^2_c\sigma+{\color{NavyBlue}\dfrac{1}{\bar{\Lambda}_1}\,(\frac{1}{a}\partial_i \pi_c)^2 \sigma}+\dots\,,\\ \nn
    {\cal L}^{(\sigma^2)}&=g\,\dfrac{1}{a(\eta)}\pi'_c\sigma^2+\dfrac{1}{\Lambda_2^2}\dfrac{1}{a^2(\eta)}\pi_c'^2\sigma^2+{\color{NavyBlue}\dfrac{1}{\bar{\Lambda}_2^2}\,(\frac{1}{a}\partial_i \pi_c)^2\,\sigma^2}+\dots\,.
\end{align}
\end{tcolorbox}
\noindent Here, we have retained only the terms that contribute to tree-level and one-loop diagrams of the $\pi$ three-point function. The Lagrangian above is expressed in terms of the canonically normalized field $\pi_c=c_s^{-\frac{3}{2}}f_\pi^2\pi$. The coupling constants $\rho$, $g$ and $\Lambda_{1,2}$ are free parameters. However, due to the non-linearly realised boosts, the other scales $\bar{\Lambda}_{1,2}$ will be automatically dictated by the following relations: 
\begin{align}
\label{nonlinreal}
  \bar{\Lambda}_1\,\rho=(\bar{\Lambda}_2)^2\,g=-2 f_\pi^2\,c_s^{-\frac{3}{2}}\,.
\end{align} 
In the notation of Eq.\,\eqref{MFStwovertexaction}, we can form two scalar-exchange diagrams (hereafter SE1 and SE2) as well as two scalar-bubble diagrams (SB1 and SB2) by setting
\begin{align}
\nn
    &\left(\dfrac{1}{\Lambda_1 a^2(\eta)}\pi'^2_c,\dfrac{\rho}{a(\eta)}\pi_c'\right)\,\,\,\,\,\,(\text{SE}1)\,,\qquad \left(\dfrac{1}{\bar{\Lambda}_1}(\frac{1}{\,a}\partial_i \pi_c)^2,\dfrac{\rho}{a(\eta)}\pi_c'\right)\,\,(\text{SE}2)\,,\\ \label{diagramsofinterest}
    &\left(\dfrac{1}{\Lambda_2\, a^2(\eta)}\pi'^2_c,\dfrac{g}{a(\eta)}\pi_c'\right)\,\,\,\,(\text{SB}1)\,,
    \qquad \left(\dfrac{1}{\bar{\Lambda}_2}(\frac{1}{\,a}\partial_i \pi_c)^2,\dfrac{g}{a(\eta)}\pi_c'\right)\,\,\,\,(\text{SB}2)\,.
\end{align}
where the components in each case corresponds to ${\cal O}_L[\pi]$ and ${\cal O}_R[\pi]$. 
\\

\noindent \textbf{Unitarity bounds on the couplings.}
We work in the weak coupling regime, where the quadratic mixing $\dot{\pi}\sigma$ can be treated perturbatively. This assumption imposes an upper bound on $\rho$:  
\begin{align}
\label{weakmix}
    \rho \lesssim m\,, \quad 
\end{align}  
This bound ensures that the single-exchange diagrams correcting the propagators of $\pi$ and/or $\sigma$ remain small at energy scales of order $\omega_\pi \sim H$ \cite{Pinol:2023oux,Jazayeri:2023xcj}.  

The other free energy scales are constrained by perturbative unitarity (see \cite{Pinol:2023oux,Jazayeri:2023xcj})\footnote{A more rigorous approach to establishing perturbativity bounds on EFT coefficients involves studying the partial wave expansion of phonon amplitudes, see \cite{Grall:2020tqc}.}. For a rough estimation of the lower bounds on these scales, we neglect the mass of $\sigma$.  The bound on the $\dot{\pi}^2\sigma$ operator can be estimated by comparing the one-loop and tree-level contributions to the three-point amplitude ${\cal A}_{2\pi \to \sigma}$ at energy scales of order the Hubble scale. Requiring the one-loop correction to be small relative to the three-particle amplitude yields:  
\begin{align}
\label{lambda1bounds}
   H / \Lambda_1 \ll 2\pi\,c_s\,.
\end{align}  
Similarly, requiring the one-loop correction to ${\cal A}_{\pi \to 2\sigma}$ to remain subdominant compared to the $\dot{\pi}^2\sigma$ contact term gives:  
\begin{align}
\label{gbound}
    g \ll 2\pi\,,
\end{align}  
and, finally, perturbative unitarity of the ${\cal A}_{2\pi \to 2\sigma}$ amplitude imposes an upper bound on the size of the $\pi'^2\sigma^2$ operator:  
\begin{align}
\label{lambda2bounds}
    H / \Lambda_2 \ll 2\pi\,c_s^{1/2}\,.
\end{align}  
Similar bounds can be placed on the energy scales $\bar{\Lambda}_{1,2}$, but they are weaker than bounds derived from Eq.\,\eqref{nonlinreal}. 
\subsection{Seed correlators from Mellin transformation}
\label{seedcorrelatorssection}
\begin{figure}
    \centering
    \includegraphics[scale=0.6]{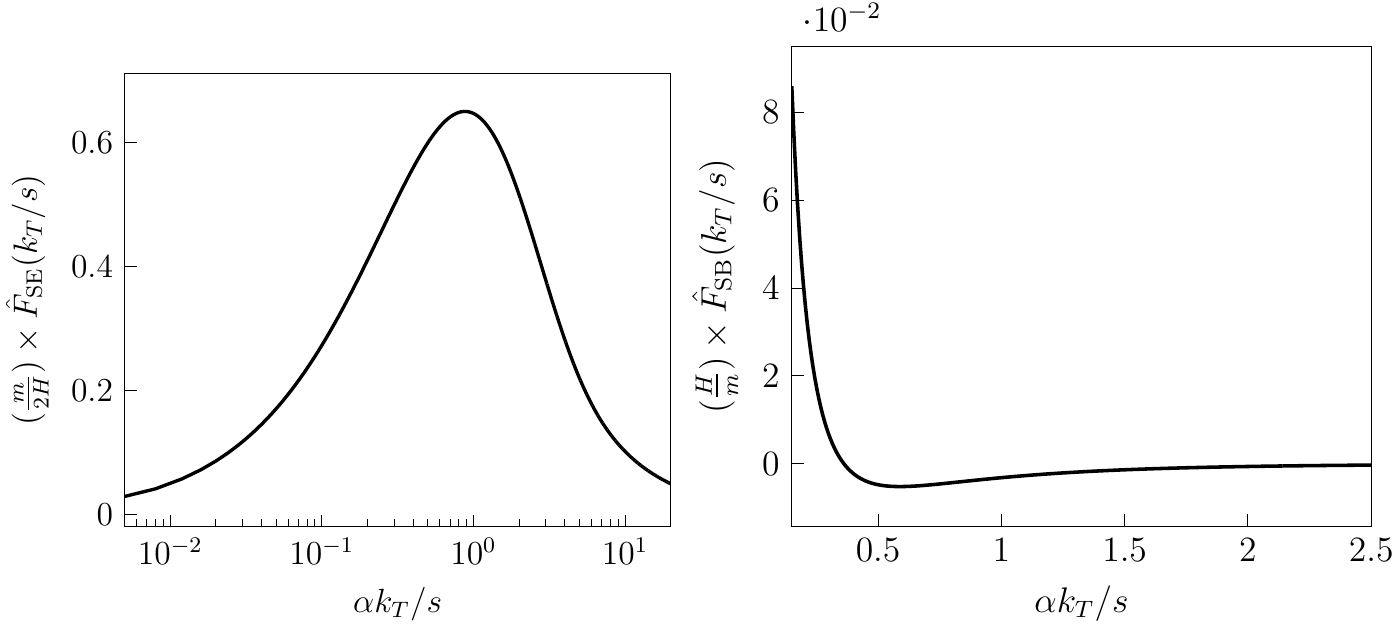}
    \caption{Plots of the tree-level and one-loop four-point functions of the conformally coupled field, scaled by the exchanged momentum $s$, i.e., $\hat{F}_{\text{SE,SB}}\equiv s\times F_{\text{SE,SB}}$.}
    \label{fhat}
\end{figure}
\textbf{Weight-shifting operators}. It is easy to show that all the four diagrams of our interest in \eqref{diagramsofinterest} can be extracted from the off-shell four-point functions of the conformally coupled field, namely $F_{\text{SE}}$ and $F_{\text{SB}}$, as defined in Eq.\,\eqref{FSE} and \eqref{FSB}, respectively. These diagrams are extracted using a set of Weight-Shifting (WS) operators, which transform the external propagators of the cc fields into massless propagators with the appropriate derivative structures. The derivation of the WS operators can be found in \cite{Jazayeri:2022kjy}; here, we only quote the final result:
\begin{align}
\nn
    \langle \pi_c(\bm{k}_1)\pi_c(\bm{k}_2)\pi_c(\bm{k}_3)\rangle'_{\text{SE1}} &=\frac{\rho}{H^3\Lambda_1}
    \left(\prod_{i=1}^3 |\pi_c(k_i,\eta_0)|^2\right)\hat{{\cal W}}_1(k_i,\dfrac{\partial}{\partial k_i}) F_{\text{SE}}(c_s k_i,s)|_{k_4\to 0}\\ \nn
    \langle \pi_c(\bm{k}_1)\pi_c(\bm{k}_2)\pi_c(\bm{k}_3)\rangle'_{\text{SB1}}&=\frac{g}{H\Lambda_2^2}\left(\prod_{i=1}^3 |\pi_c(k_i,\eta_0)|^2\right)\hat{{\cal W}}_1(k_i,\dfrac{\partial}{\partial k_i}) F_{\text{SB}}(c_s k_i,s)|_{k_4\to 0}\\ \nn
        \langle \pi_c(\bm{k}_1)\pi_c(\bm{k}_2)\pi_c(\bm{k}_3)\rangle'_{\text{SE2}} &=\frac{\rho}{H^3\bar{\Lambda}_1}
    \left(\prod_{i=1}^3 |\pi_c(k_i,\eta_0)|^2\right)\hat{{\cal W}}_2(k_i,\dfrac{\partial}{\partial k_i}) F_{\text{SE}}(c_s k_i,s)|_{k_4\to 0}\\ \label{WSbispectrum}
    \langle \pi_c(\bm{k}_1)\pi_c(\bm{k}_2)\pi_c(\bm{k}_3)\rangle'_{\text{SB2}}&=\frac{g}{H\bar{\Lambda}_2^2}\left(\prod_{i=1}^3 |\pi_c(k_i,\eta_0)|^2\right)\hat{{\cal W}}_2(k_i,\dfrac{\partial}{\partial k_i}) F_{\text{SB}}(c_s k_i,s)|_{k_4\to 0}\,,
\end{align}
where 
\begin{align}
\nn
    \hat{{\cal W}}_1 &=-c_s^4(k_1k_2k_3)^2\dfrac{\partial^2}{\partial k_{12}^2}\,,\\ 
    \hat{{\cal W}}_2 &=c_s^2 (\bm{k}_1\cdot\bm{k}_2)\, k_3^2\left(1-k_{12}\dfrac{\partial}{\partial k_{12}}+k_1 k_2 \dfrac{\partial^2}{\partial k_{12}^2}\right)\,.
\end{align}
On the RHS of the equations above, in the argument of the off-shell four-points $F_{\text{SE,SB}}$, we have set $\omega_i=c_sk_i (i=1,2,3)$ and have taken the soft limit $k_4\to 0$. Therefore, by conservation of momentum, we have $s=|\bm{k}_1+\bm{k}_2|\to k_3$. Notice that the intermediate momentum $s$ is not rescaled with the speed of sound, which reflects the fact that the exchanged fields have unit sound speeds. 

As shown in the equations above, the same WS operator $\hat{{\cal W}}_{1,2}$ generates the SE1/2 and SB1/2 diagrams from the corresponding tree-level and one-loop seeds. This is due to the identical derivative structure in the external legs of these diagrams. It is also important to note that in the equations above we have considered only one permutation in each channel, and thus the result is not manifestly symmetric under permutations of $k_1,k_2$ and $k_3$. 
The full three-point function, however, is the sum over all permutations and all channels, hence fully permutation symmetric.\\

\noindent \textbf{Seed correlators in the MFS limit.} Having established the relationship between the $\pi$ three-point function and the four-point functions of the conformally coupled field, the remaining task is to compute the latter seed correlators. In the MFS limit, these are given by
\begin{align}
\label{FSBSEfinal}
    F_{\text{SE},\text{SB}}(\omega_T,s)=2\,\text{Re}\int_{-\infty(1-i\epsilon)}^0 d\eta\, G_{\text{SE,SB}}(s^2/a^2(\eta))\exp(i\omega_T\eta)\,,
\end{align}
where $G_{\text{SE,SB}}$ are given by Eq.\,\eqref{GSEGSB}. As discussed earlier, we use the $\bar{\text{MS}}$ scheme to regulate the loop diagram. In more detail, the second term in the last line of Eq.\,\eqref{GSB} contributes as
\begin{align}
    F_{\text{SB}}\supset \dfrac{1}{4\pi^2}\left(-\dfrac{1}{(d-4)}+\dfrac{1}{2}\log(4\pi\,e^{-\gamma_E})\right)\dfrac{1}{\omega_T}\,,
\end{align}
which can be canceled by adding a $\phi^4$ counter term to the action of the conformally coupled field. From this point forward, we assume that this subtraction has been performed, and $G_\text{SB}$ denotes the remaining finite part of the loop graph, namely the first term in Eq.\,\eqref{GSB}.

In the context of the $\pi$ three-point function, the corresponding counterterms are $\dot{\pi}^3$ for the SB1 diagram and $\dot{\pi}(\partial_i\pi)^2$ for the SB2 diagram. The operator $\dot{\pi}\sigma^2$ also induces a one-loop contribution to the two-point function of $\pi$, resulting in a divergence that must be canceled by a quadratic counterterm of the form $\dot{\pi}^2$. This quadratic operator effectively modifies the speed of sound, ensuring that the coefficient of the $\dot{\pi}(\partial_i\pi)^2$ operator in the Lagrangian (Eq.\,\eqref{EFTpi}) does not change. In other words, this coefficient is protected by the underlying non-linearly realized boost, which holds to all orders in the loop expansion.
\\
\begin{figure}
    \centering
    \includegraphics[scale=0.55]{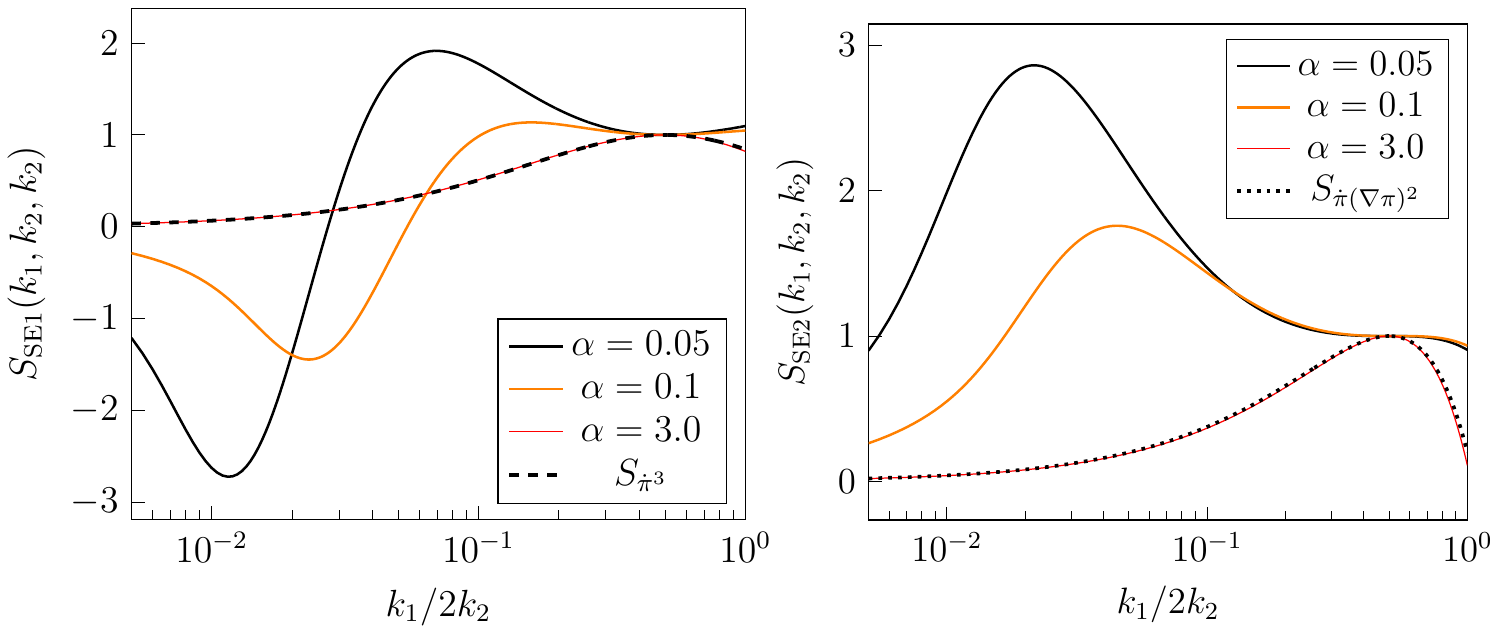}
    \caption{The shape of the bispectrum corresponding to the SE1 and SE2 diagrams, normalized to unity at the equilateral configuration ($k_1 = k_2$). For comparison, the shapes induced by the EFT operators $\dot{\pi}^3$ and $\dot{\pi}(\partial_i\pi)^2$ are also shown.}
    \label{treelevelShape}
\end{figure}

\noindent \textbf{Mellin transformation.} In order to evaluate the time integral in Eq.\,\eqref{FSBSEfinal}, it is useful to exploit the Mellin transformations of $G_{\text{SE,SB}}(\vec{p}_E^2)$ as a function of $p_E=\sqrt{\vec{p}_E^2}>0$, where $\vec{p}_E=(ip_0,p_i)$ is the $d-$ dimensional Euclidean momentum. 
The Mellin transformation of the SE diagram is well defined and is given by
\begin{align}
    \tilde{G}_{\text{SE}}(z)&=\int_0^\infty dp_E\,p_E^{z-1}\,G_{\text{SE}}(p_E^2)=\dfrac{1}{2}\dfrac{i\,\pi}{\sin(\pi z/2)}m^{z-2}\,\qquad (0<\text{Re}(z)<2)\,. 
\end{align}
Strictly speaking, the same transformation does not exist for $G_{\text{SB}}(p_E)$, because the integral over $p_E$ diverges for any $z$. Instead, we can perform the Mellin transformation after making a suitable subtraction. Specifically, we define 
\begin{align}
    \tilde{G}_{\text{SB}}(z)&=\int_0^\infty dp_E\,p_E^{z-1}\,\left[G_{\text{SB}}(p_E^2)-\dfrac{i}{16\pi^2}\log(m^2/\mu^2)\right]\\ \nn&=\dfrac{i}{16\pi^2}\dfrac{\pi}{z\,\sin(\pi z/2)}\dfrac{\Gamma^2(1-z/2)}{\Gamma(2-z)}m^z\qquad (-1<\text{Re}(z)<0)\,.
\end{align}
The inverse Mellin transformations are: 
\begin{align}
    G_{\text{SE}}(p_E^2)&=\dfrac{1}{2\pi i}\int_{c-i\infty}^{c+i\infty}dz\,p_E^{-z}\,\tilde{G}_{\text{SE}}(z)\,,\\ \nn
    G_{\text{SB}}(p_E^2)&=\dfrac{1}{2\pi i}\int_{c-i\infty}^{c+i\infty}dz\,p_E^{-z}\,\tilde{G}_{\text{SB}}(z)+\dfrac{i}{16\pi^2}\log(m^2/\mu^2)\,,
\end{align}
where $c$ is an arbitrary real number within the strip of analyticity in each case, namely $c\in (0,2)$ for the SE graph, and $c\in (-1,0)$ for the SB graph. Plugging the Mellin transformed graphs inside \eqref{FSBSEfinal} and performing the time integral yields: 
\begin{align}
\label{FSEMellinexp}
    F_{\text{SE}}(c_sk_T,s)=\dfrac{H}{2m}\dfrac{1}{s}\,\text{Re}\int_{c-i\infty}^{c+i\infty}dz\dfrac{\Gamma(1-z)}{\sin(\pi z/2)}\left(\dfrac{i\alpha k_T}{s}\right)^{z-1}\,\qquad (0<c<1)\,,
\end{align}
where $\alpha=c_sm/H$. 
It should be noted that the integral over the conformal time is IR convergent only if $\text{Re}(z)<1$. Therefore, the integration contour for the SE diagram had to be further restricted to $0<c<1$. For the one-loop diagram one finds
\begin{align}
\nn
    F_{\text{SB}}(c_sk_T,s)&=\dfrac{1}{64\pi^2}(\dfrac{m}{H})\dfrac{1}{s}\,\text{Re}\int_{c-i\infty}^{c+i\infty}dz\dfrac{1}{\sin(\pi z/2)}\dfrac{z\,\Gamma^2(-z/2)}{1-z}\left(\dfrac{i\alpha k_T}{s}\right)^{z-1}\\ \label{mellinintegrandsb}
    &+\dfrac{1}{8\pi^2}\log(m^2/\mu^2)\dfrac{s}{c_sk_T}\,\qquad (-1<c<0)\,.
\end{align}
\\

\noindent \textbf{Scalar-exchange seed.} To derive computationally useful expressions for the seed correlators, we use the Cauchy theorem to evaluate the Mellin integrals in Eq.\,\eqref{FSEMellinexp}. We first assume $\alpha k_T / s < 1$. In this case, the original contour can be closed using an infinite arc in the right complex half-plane. Within the resulting closed contour, the Mellin integrand of the SE diagram exhibits poles at $z = n \geq 1$. The Cauchy theorem then implies: 
\begin{tcolorbox}[colframe=white,arc=0pt]
\begin{align}
\label{Fevery}
    F_\text{SE}(c_sk_T,s)=\dfrac{2H}{m}\dfrac{1}{s}\sum_{n=1}^\infty \dfrac{1}{(2n-1)!}\left(\dfrac{\alpha k_T}{s}\right)^{2n-1}\,\left[-\log(\dfrac{\alpha k_T}{s})+\psi^{(0)}(2n)\right]\,,
\end{align}
\end{tcolorbox}
\noindent where $\psi^{(0)}(n)$($=-\gamma_E+\sum_{k=1}^{n-1}k^{-1}$) is the polygamma function of order zero. The power series representation of the SE diagram above matches the alternative formula for $F_{\text{SE}}$ given by Eq.\,\eqref{correl:singleExchangeMFS}. Note that after taking the real part, only the poles at $z = 2n \geq 2$ have contributed to $F_{\text{SE}}$. 
Moreover, although we initially assumed $\alpha\, k_T/s > 1$, the series expansion actually possesses an infinite radius of convergence. This implies that Eq.\,\eqref{Fevery} also applies to $\alpha k_T/s>1$.
\\

\noindent \textbf{Scalar-bubble seed.} The Mellin integrand associated with the SB diagram (Eq.\,\eqref{mellinintegrandsb}) has poles located at $z\in\lbrace 1,2n\geq 0\rbrace$. Summing over the residues of these poles we find: 
\begin{tcolorbox}[colframe=white,arc=0pt]
\begin{align}
\nn
    F_\text{SB}(c_sk_T,s)&=\dfrac{1}{16\pi^2}(\frac{m}{H})\dfrac{1}{s}\sum_{n=0}^\infty \dfrac{1}{(n!)^2}\left(\dfrac{\alpha\,k_T}{s}\right)^{2n-1}\left[X_n+Y_n\log\left(\dfrac{\alpha\,k_T}{s}\right)+Z_n\log^2\left(\dfrac{\alpha\,k_T}{s}\right)\right]\\ \nn
    &+\dfrac{1}{8\pi^2}\log(m^2/\mu^2)\dfrac{1}{c_sk_T}\,,
\end{align}
where
\begin{align}
\nn
X_n&=\dfrac{1}{(2n-1)^3}\left(4+4 n (2n-1)^2 (\psi ^{(0)}(n+1))^2\right. \\ \nn
&\qquad\qquad\qquad\qquad\left.-2 n (2 n-1)^2 \psi
   ^{(1)}(n+1)+4(2n-1) \psi ^{(0)}(n+1)\right)\,,\\ \nn
Y_n&=-\frac{4}{(2 n-1)^2}\left(2 n (2 n-1) \psi ^{(0)}(n+1)+1\right)\,,\\ 
Z_n&=\frac{4 n}{(2 n-1)}\,,
\end{align}
\end{tcolorbox}
\noindent where $\psi^{(1)}(n)$(=$\pi^2/6-\sum_{k=1}^{n-1}k^{-2}$) is the polygamma function of order one. Similar to the SE diagram, although we performed the integration for $\alpha\,k_T/s>1$, the resulting power series defines an entire function of $\alpha\,k_T/s$.  
\subsection{The bispectrum}
\label{bispectrumSection}
\begin{figure}[h]
    \centering
    \includegraphics[scale=0.8]{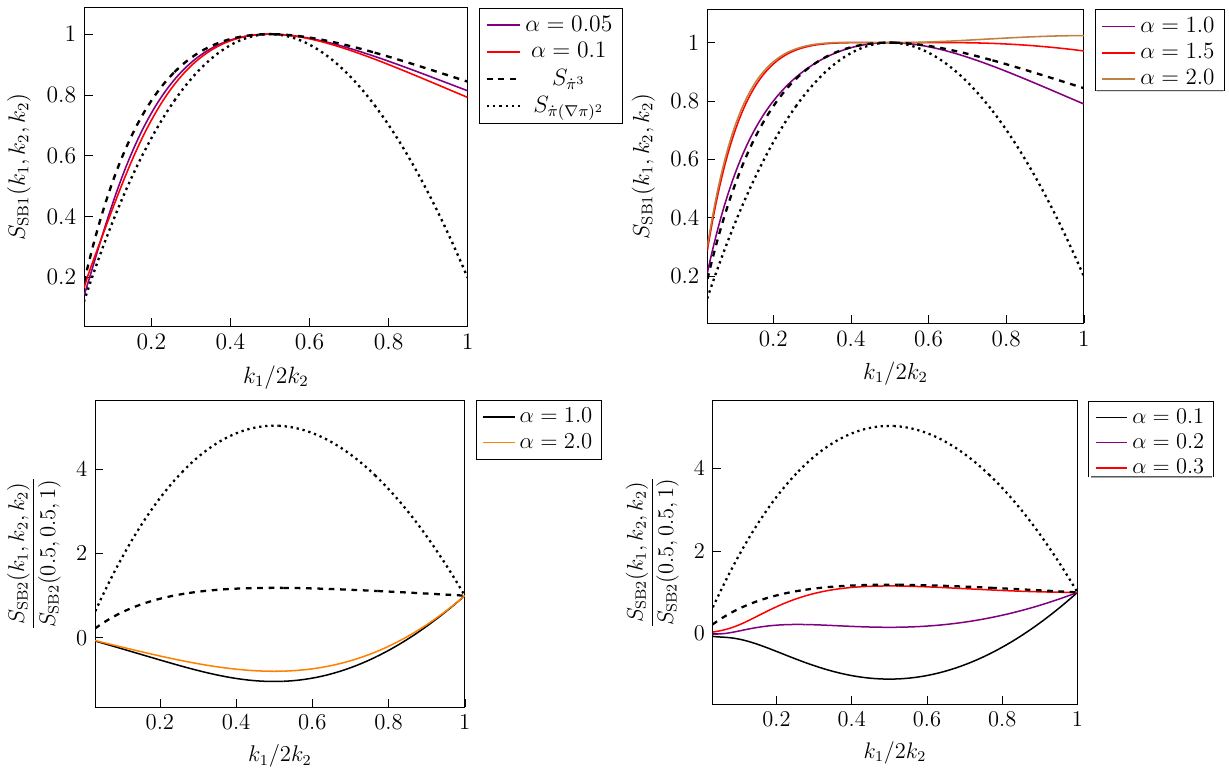}
    \caption{The shape of the bispectrum at one-loop. The plots correspond to the bispectra excluding the $\mu$-dependent contributions highlighted in the last lines of Equations \eqref{BSB1} and \eqref{BSB2}. For convenience, the shape is normalized to unity at $k_1 = k_2$ for Diagram SB1 and at $k_1 = 2k_2$ for Diagram SB2.}
    \label{looplevelshape}
\end{figure}
We now have all the ingredients to compute the bispectrum of the curvature perturbation $\zeta$, defined by 
\begin{align}
    B(k_1,k_2,k_3)=\langle \zeta(\bm{k}_1)\zeta(\bm{k}_2)\zeta(\bm{k}_3)\rangle'\,.
\end{align}
At leading order in the slow-roll parameters, $\zeta$ is given by 
\begin{align}
    \zeta\approx -H\pi=-2\pi\Delta_\zeta\,c_s^{3/2}\pi_c/H\,.
\end{align}
 We find analytical expressions for all the four diagrams, by acting with the weight-shifting operators on $F_{\text{SE,SB}}$, according to Eq.\,\eqref{WSbispectrum}. These results are summarised in the inserts below. 

The one-loop bispectrum depends on the arbitrary renormalization scale $\mu$. However, after renormalizing both the speed of sound $c_s$ and the coefficient A of the cubic operator $\dot{\pi}^3$, the complete bispectrum—which is the sum over contributions from the contact cubic terms and the one-loop diagrams—is guaranteed to be independent of $\mu$. To isolate and visualize the one-loop contribution specifically, we subtract the $\mu$-dependent parts of the bispectrum for each diagram. These subtracted components are highlighted in purple in the final lines of Equations \eqref{BSB1} and \eqref{BSB2}. Importantly, the subtracted parts are degenerate with the bispectrum induced by the local operators $\dot{\pi}^3$ (for the SB1 diagram) and $\dot{\pi}(\partial_i\pi)^2$ (for the SB2 diagram). By subtracting these terms, we isolate the genuine one-loop contributions that are not degenerate with local EFT operators. 

We first estimate the upper bounds on the size of the non-Gaussianity, which is commonly defined in terms of the bispectrum evaluated at the equilateral configuration  $k_1 = k_2 = k_3$, i.e., 
\begin{align}
    f_{\text{NL}}=\dfrac{10}{9}\dfrac{k^6 B(k,k,k)}{(2\pi\Delta_\zeta)^4}\,.
\end{align}
For each diagram, perturbative unitarity bounds on the couplings, given by \eqref{lambda1bounds}, \eqref{gbound} and \eqref{lambda2bounds}, combined with the weak mixing assumption \eqref{weakmix} (relevant for the SE1/2 graphs), impose the following upper bounds on $f_{\text{NL}}$: 
\begin{align}
\nn
    &f_{\text{NL}}^{\text{SE1}}\ll {\cal O}(1)\sqrt{c_s}\Delta_\zeta^{-1}\,,\qquad \quad f_{\text{NL}}^{\text{SE2}}\ll {\cal O}(1)/c_s^2\,,\\ 
    &f_{\text{NL}}^{\text{SB1}}\ll {\cal O}(0.01)\Delta_\zeta^{-1}/\sqrt{c_s}\,,\,\quad f_{\text{NL}}^{\text{SB2}}\ll {\cal O}(0.01)/c_s^2\,.
\end{align}
Notably, the bounds on the SE1 and SB1 diagrams, both of which involve vertices containing the $\dot{\pi}^2$ operator, are relaxed by a factor of $\Delta_\zeta^{-1}$. As a result, the bispectrum generated by the SE1/2 and SB1 diagrams can be of order 10 within the perturbative framework.
However, the observational constraints on the speed of sound $c_s$ implies $f_{\text{NL}}^{\text{SB2}} \ll {\cal O}(10)$.

In Figures \ref{treelevelShape} and \ref{looplevelshape}, we have plotted the shape of the bispectrum, defined as
\begin{align}
S(k_1,k_2,k_3)\propto(k_1k_2k_3)^2 B(k_1,k_2,k_3)\,,
\end{align}
as a function of $k_1/2k_2$, and for isosceles triangles (i.e., $k_3=k_2$). 
 The shape depends only on the parameter $\alpha$, apart from an overall amplitude, which has been normalized to unity at the equilateral configuration ($k_1 = k_2 = k_3$) for the SE1, SE2, and SB1 diagrams, and at the folded configuration ($k_1 = 2k_2$) for the SB2 diagram.

The most notable feature in the tree-level shapes is the distinct resonant behavior that emerges in the squeezed limit, near
$\frac{k_1}{2k_2}\sim {\cal O}(1)\alpha$, when $\alpha$ is small.  As $\alpha$ increases and approaches $\alpha \gtrsim 1$, the resonance fades away, and the shapes of the SE1 and SE2 diagrams reduce to those generated by the local EFT operator $\dot{\pi}^3$ and $\dot{\pi}(\partial_i\pi)^2$, respectively. 
These resonances, referred to as the low-speed collider signal, were first introduced in \cite{Jazayeri:2022kjy} and fully characterized in \cite{Jazayeri:2023xcj}. This signal has recently been employed as an observational template for identifying signatures of heavy fields in CMB and galaxy data \cite{Sohn:2024xzd,Cabass:2024wob}. 

At the one-loop level, the bispectrum shapes do not exhibit similar resonances in the squeezed limit. For $\alpha\lesssim 1$, the bispectrum corresponding to the SB1 diagram closely resembles the shape generated by the  cubic operator $\dot{\pi}^3$, indicating that the signal can be effectively captured by the equilateral template. As $\alpha$ approaches order unity, the SB1 shape develops a plateau near the equilateral configuration. In contrast, the SB2 diagram shows markedly different behavior: for $\alpha\lesssim 0.1$, the shape displays a local minimum around the equilateral configuration, which transitions into an inflection point and then a maximum as $\alpha$ increases to $\sim 0.3$. For $\alpha$ of order unity, the SB2 shape once again exhibits a local minimum near $k_1 = k_2$.

A careful analysis of the full bispectrum shape, including its overlap with the standard local, equilateral, and orthogonal templates, is deferred to future work. 
\begin{framed}
\noindent \textbf{Tree-level bispectrum.} 
\begin{align}
\nn
B_{\text{SE1}}&=4\pi^3\,c_s^{1/2}\Delta_\zeta^3\left(\dfrac{\rho}{\Lambda_1}\right)\\
    &\times \dfrac{1}{\alpha\,k_T^2 k_1k_2k_3^2}\sum_{n=1}^\infty \dfrac{1}{(2n-1)!}\left(\dfrac{\alpha\,k_T}{k_3}\right)^{2n-1}\left(\textcolor{red}{a_n}+\textcolor{red}{b_n}\log(\alpha\,k_T/k_3)\right)+\text{2 perms}\,,\\ \nn
\end{align}
\begin{align}
&B_{\text{SE2}} =4\pi^4 \Delta_\zeta^4\left(\dfrac{\rho}{H}\right)^2\dfrac{(\bm{k}_1\cdot\bm{k}_2)}{\alpha k_1^3 k_2^3\, k_3^2k_T^2}\sum_{n=1}^\infty \dfrac{1}{(2n-1)!}\left(\dfrac{\alpha\,k_T}{k_3}\right)^{2n-1}\\ \nn
&\times \left[\psi^{(0)}(n)\,k_T^2+\textcolor{Blue}{c_n}\,k_{12}k_T+\textcolor{Blue}{d_n}\,k_1 k_2+\log(\alpha\,k_T/k_3)\left(\textcolor{Blue}{e_n} k_1k_2+2n k^2_{12}+(1+2n)k_3 k_{12}+k_3^2\right)\right]\,\\ 
&+\text{2 perms}\,.
\end{align}
where
\begin{align}
\nn
    &\textcolor{red}{a_n}=(4n^2-6n+2)\psi^{(0)}(2n)+3-4n\,,\qquad \textcolor{red}{b_n}=-4n^2+6n-2\,,\\ 
    &\textcolor{Blue}{c_n}=1+\psi^{(0)}(2n)(1-2n)\,,\quad \textcolor{Blue}{d_n}=\psi^{(0)}(2n)(4n^2-6n+2)-4n+3\,,\quad \textcolor{Blue}{e_n}=-4n^2+6n-2\,.
\end{align}
\end{framed}
\begin{framed}
\noindent \textbf{Loop-level bispectrum.} 
\begin{align}
\nn
    &B_{\text{SB1}}=\dfrac{\pi}{8}c_s^{-3/2}\Delta_\zeta^3\left(\dfrac{g H^2}{\Lambda_2^2}\right)\\ \nn
    &\times \dfrac{\alpha}{k_T^2(k_1 k_2 k_3^2)}\sum_{n=0}^\infty \dfrac{1}{(n!)^2}\left(\dfrac{\alpha k_T}{k_3}\right)^{2n-1}\left[\textcolor{red}{A_n}+\textcolor{red}{B_n}\log\left(\dfrac{\alpha k_T}{k_3}\right)+8n(n-1) \log^2\left(\dfrac{\alpha k_T}{k_3}\right)\right]\,\\ \label{BSB1}
    &+\text{2 perms}+\textcolor{Blue}{\dfrac{3\pi}{2}c_s^{-3/2}\Delta_\zeta^3 \left(\dfrac{g H^2}{\Lambda_2^2}\right)\log(m^2/\mu^2)\dfrac{1}{k_1k_2k_3k_T^3}}\,,\\ \nn
    \\ \nn
    B_{\text{SB2}}&=-\dfrac{\pi^2}{16}c_s^{-2}\,\Delta_\zeta^4\,g^2 \times \dfrac{\alpha\,(k_3^2-k_1^2-k_2^2)}{k_1^3 k_2^3 k_3^2 k_T^2}\sum_{n=0}^\infty \dfrac{-1}{(2n-1)^3(n!)^2}\left(\dfrac{\alpha k_T}{k_3}\right)^{2n-1}\\ \nn
    &\left[\textcolor{Green}{C_n} k_{12}k_T+\textcolor{Green}{D_n} k_T^2+\textcolor{Green}{E_n} k_1 k_2 +\log\left(\dfrac{\alpha k_T}{k_3}\right)\left(\textcolor{Brown}{F_n} k_{12}k_T+\textcolor{Brown}{G_n} k_T^2+\textcolor{Brown}{H_n} k_1 k_2 \right)\right.\\ \nn
    &\left.+\log^2\left(\dfrac{\alpha k_T}{k_3}\right)\left(\textcolor{purple}{I_n} k_{12}k_T+\textcolor{purple}{J_n} k_T^2+\textcolor{purple}{K_n} k_1 k_2 \right) +\text{2 perms}\right]\,\\ \label{BSB2}
    &+\textcolor{Blue}{\dfrac{\pi^2}{4}c_s^{-1}\Delta_\zeta^4\,g^2\log(m^2/\mu^2)\left[\dfrac{1}{2k_1^3k_2^3k_3k_T^3}(k_3^2-k_1^2-k_2^2)(2k_1k_2+k_{12}k_T+k_T^2)+2\,\text{perms}\right]}\,,
\end{align}
where 
\begin{align}
\nn
    \textcolor{red}{A_n}&=4+8 n(n-1)\, \psi ^{(0)}(n+1)^2+4(-4 n+2)
   \psi ^{(0)}(n+1)-4 n(n-1)\, \psi ^{(1)}(n+1)\,,\\ \nn
   \textcolor{red}{B_n} &=16n-8-16n(n-1)\psi^{(0)}(n+1)\,,\\ \nn
   \textcolor{Green}{C_n} &=2(2 n-1)^3 \left(-n \psi ^{(1)}(n+1)+2 \psi ^{(0)}(n+1) (n \psi
   ^{(0)}(n+1)-1)\right)\,, \\ \nn
   \textcolor{Green}{D_n} &=-4+2 (2 n-1) \left(-2 n(2 n-1) \psi ^{(0)}(n+1)^2-2 \psi
   ^{(0)}(n+1)+n (2 n-1) \psi ^{(1)}(n+1)\right)\,,\\ \nn
   \textcolor{Green}{E_n} &=-4 (2 n-1)^3 \\ \nn
   &\times \left(1+2n(n-1)\psi ^{(0)}(n+1)^2+(2-4 n) \psi
   ^{(0)}(n+1)-n(n-1)\psi ^{(1)}(n+1)\right)\,,\\ \nn
   \textcolor{Brown}{F_n} &=4 (2 n-1)^3 (1- n \psi ^{(0)}(n+1))\,,\qquad \textcolor{Brown}{G_n} =4 (2 n-1) (2 n (2 n-1) \psi ^{(0)}(n+1)+1)\,,\\ \nn
   \textcolor{Brown}{H_n} &=8 (2 n-1)^3 (1-2 n+2n(n-1)\psi ^{(0)}(n+1))\,,\\ 
   \textcolor{purple}{I_n}&=4 n (2 n-1)^3\,,\qquad \textcolor{purple}{J_n}=-4 (2 n-1)^2 n\,,\qquad \textcolor{purple}{K_n} =-8 n(n-1)(2 n-1)^3\,.
\end{align}
\end{framed}
\section{Summary and Outlook}
The established relationship between de Sitter correlators and the flat-space S-matrix, emerging at the total energy singularity of perturbative de Sitter graphs, has proven to be a valuable input in the cosmological bootstrap program. In this paper, we made further progress on linking perturbative processes in de Sitter and flat space by developing a novel flat-space limit for a subset of dS correlators, which are characterized by a mass gap between their external and internal legs.

We demonstrated that, in a double-scaling limit—where the external energies approach zero inversely proportional to the internal masses—the correlators undergo significant simplifications. In this limit, we derived a reduction formula that expresses the original in-in diagram in terms of the corresponding amputated diagram in flat space. Importantly, the internal lines in this limit remain massive, contrasting sharply with the conventional amplitude limit, where propagators are massless. Our results illustrate the connection between the rich structure of massive Feynman integrals and that of massive exchange processes in de Sitter.

We applied our MFS limit to the phenomenology of inflation by calculating the massive tree-level and one-loop exchange diagrams for the three-point function of curvature perturbations, assuming a small sound speed. We showed that the exchange of heavy fields in an intermediate mass range, specified by $H \ll m \lesssim \mathcal{O}(1)H/c_s$, leads to intriguing non-Gaussian signatures. We termed this framework the "cosmological phonon collider". The non-Gaussian shapes we identified in this setup are distinct from the cosmological collider oscillations extensively studied in the literature. We observed that the bispectrum can exhibit: ($i$) resonances in the mildly squeezed regime, referred to as the "low-speed collider signal"\cite{Jazayeri:2023xcj,Jazayeri:2022kjy}, arising at tree level and for $\alpha=c_s m/H\ll 1$, or, depending on the vertex type and the parameter $\alpha$, ($ii$) a plateau or a local minimum near the equilateral configuration, at one-loop level.

There are several directions that remain to be explored. 
\begin{itemize}
    \item \textbf{On-shell massive flat-space limit.} The MFS limit proposed in this work establishes a connection between off-shell correlators and off-shell flat-space Feynman diagrams. A significant future avenue involves obtaining an on-shell massive flat-space limit applicable directly to the original in-in diagram (without resorting to the additional energy variables \(\omega_i\)). A natural proposal could be a double-scaling limit where \(k_T \to 0\) and \(m \to \infty\), while maintaining \(k_T \times m < \infty\). One concern about such a limit is that contributions from particle production—negligible in the MFS limit due to their Boltzmann-suppressed nature—may become important. This difference arises from branch cuts introduced by particle production effects in the final correlator, leading to large exponential factors, such as \((-1)^{-im/H}\), when some of the external energies \(k_i\) take negative values. This is an unavoidable situation when \(k_T \to 0\); however, it was circumvented in the MFS limit by assuming that all $\omega_i$'s are positive real numbers. As a result of this technical difference, the structure of the MFS reduction formula might qualitatively change in an on-shell flat-space limit. We leave this direction to future work. 
    \item \textbf{Connections with the flat-space limit of AdS/CFT.} Several frameworks have been constructed in recent years to study the flat-space limit of AdS/CFT, employing different representations of the boundary CFT, e.g., position space, Mellin space, or the conformal particle-wave expansion (see \cite{Polchinski:1999ry, Giddings:1999jq, Gary:2009ae, Heemskerk:2009pn, Fitzpatrick:2010zm, Maldacena:2015iua, Komatsu:2020sag} for an incomplete list of references and \cite{ Li:2021snj} for a review). It would be interesting to find concrete connections between our MFS limit and such constructions in AdS and explore other similar flat-space limits of dS.
    \item \textbf{Cosmological Phonon Collider.} On the phenomenological side, our results can be easily extended to include other species, such as fermions and massive vector fields. It would be interesting to explore the observational signatures of such particles especially in situations where their contributions can be enhanced, e.g. by color factors, when multiple species circle in the loop diagrams. 
    Additionally, it would be interesting to explore whether triangle and box diagrams exhibit low-speed collider bumps, which were absent in the bubble graphs computed in this paper.
\end{itemize}

\section*{Acknowledgements}
SJ acknowledges the use of ChatGPT‑4o, particularly for (1) significant rewrites/reformulation of nearly all sentences (including enhancements to the scientific presentations), some of which were iterative, (2) iterative development and reformulation of several paragraphs particularly in the abstract and introduction, and (3) adding/editing domain-specific phrasings. The core scientific content belongs solely to the authors. The authors have reviewed and take full responsibility for the published version.  
We thank Chandramouli Chowdhury, Paolo Creminelli, Claudia de Rham, Carlos Duaso Pueyo, Viola Gattus, Shota Komatsu, Arthur Lipstein, Scott Melville, Enrico Pajer, Zhehan Qin, Luca Santoni, Kostas Skenderis, David Stefanyszyn, Massimo Taronna, Andrew Tolley, Xi Tong, Toby Wiseman and Yuhang Zhu for stimulating discussions. SJ would like to thank Sébastien Renaux-Petel and Denis Werth for earlier collaborations on related topics. SJ would also like to thank the Astroparticle and Cosmology laboratory (APC) in Paris for its hospitality when parts of this work was under progress. SC would like to thank the theory group at CERN where part of this work took place. The work of SJ is supported by a Simons Investigator award 690508. SC is supported in part by the SSTFC Consolidated Grants ST/T000791/1 and ST/X000575/1 and by a Simons Investigator award 690508.
\appendix
\section{$\pi$ Effective Action from Integrating Out The Heavy Field}
\label{EFTintout}
In this appendix we integrate out the heavy field at the level of the action and obtain a spatially non-local EFT for $\pi$ (see also \cite{Gwyn:2012mw,Gwyn:2014doa,Jazayeri:2022kjy,Jazayeri:2023xcj}). We consider the simplest example in which the interactions are linear in the heavy field, namely we take 
\begin{align}
    S_\sigma=\int d^4x\,\sqrt{-g}\left(-\dfrac{1}{2}(\partial \sigma)^2-\dfrac{1}{2}m^2\sigma^2-g\,{\cal O}[\pi(x)]\sigma\right)\,.
\end{align}
Integrating out $\sigma$ induces the following effective action: 
\begin{align}
    S_{\text{eff}}=-\dfrac{g^2}{2}\int d\eta\,d^3\bm{x}\dfrac{1}{H^4\eta^4}\,{\cal O}[\pi(\eta,\bfx)]\dfrac{1}{\Box-m^2-i\epsilon}{\cal O}[\pi(\eta,\bfx)]\,.
\end{align}
The $i\epsilon$ prescription above matches with the Feynman in-out propagator. The above action is clearly non-local in both space and time. However, in the MFS limit, time derivatives acting on ${\cal O}$ are suppressed compared to spatial derivatives, therefore one can expand in 
\begin{align}
    \nn
    \hat{\delta}=-\Box+\nabla^2/a^2=\eta^2\partial_\eta^2-2\eta\partial_\eta\,,
\end{align}
and re-write $S_{\text{eff}}$ as\cite{Jazayeri:2022kjy}: 
\begin{align}
    S_{\text{eff}}=-\dfrac{1}{2}\int d\eta\,d^3\bm{x}\dfrac{1}{H^4\eta^4}\,{\cal O}[\pi(\eta,\bfx)]\sum_{n=0}^\infty (-1)^n\,[G(\nabla^2/a^2(\eta))\hat{\delta}]^n\,G(\nabla^2/a^2(\eta)) {\cal O}[\pi(\eta,\bfx)]\,,
\end{align}
where $G(\nabla^2/a^2)=1/(-\nabla^2/a^2+m^2)$. The effective action above is local in time and it systematically corrects the leading order MFS effective action in Eq.\,\eqref{MFStwovertexaction}. 

In more detail, the typical size of $\hat{\delta}$ at the sound horizon crossing of the phonon modes, namely when $c_s|\bm{k}|\eta\sim H$, is of order $H^2$, whereas $G$ at the same time is of order $1/m^2$ (notice that in the MFS limit, $c_s\sim {\cal O}(1)H/m$). This scaling implies that the above action is organized in increasing powers of $H/m$.

\section{Spectral Decomposition}
\label{app:spectral}
To examine the MFS limit of the bubble graph, we can resort to the spectral decomposition of the composite operator $\sigma^2$. This decomposition allows one to write the one-loop graph given by Eq.\,\eqref{bubblegraph} as a spectral integral over tree-level exchange diagrams with varied intermediate masses \cite{Marolf:2010zp, Xianyu:2022jwk}\footnote{The Källén–Lehmann decomposition assumes de Sitter isometries. For the example we are considering the only relevant sector to the one-loop, spectral decomposition is the free-field Lagrangian for the massive field $\sigma$ which is indeed de Sitter invariant. Consequently, the spectral decomposition of the one-loop graph is applicable even if the external legs break dS isometries, as for example in the EFT of inflation. For more generic cases, such as when the $\sigma$ propagator itself break dS symmetries, the spectral representation of the one-loop diagrams is not applicable. Nevertheless, our effective action reasoning in Section \ref{MFSfromeffectiveaction} ensures that the MFS reduction formulae remains valid in all those cases.}. 

Once translated to momentum space, the Källén–Lehmann decomposition of the scalar operator $\sigma^2(\eta,\bfx)$ implies the following identity \cite{Xianyu:2022jwk}: 
\begin{align}
    \int \dfrac{d^dq}{(2\pi)^d}\,G_{\pm\pm}(|\bm{q}|,\eta_1,\eta_2;\mu_\sigma)G_{\pm\pm}(|\bm{q}-\bfk|,\eta_1,\eta_2;\mu_\sigma)=\int_{0}^{
    \infty}
     d\mu\ \rho_\sigma(\mu)\,G_{\pm\pm}(|\bfk|,\eta_1,\eta_2;\mu)\,,
     \label{squareprop}
\end{align}
where we have included $\mu_\sigma=\sqrt{\frac{m^2}{H^2}-\frac{d^2}{4}}$ in the propagator's arguments. $\mu$ is a similar parameter characterizing the mass spectrum appearing on the RHS above. Notice that for heavy fields the spectral integral goes over the principal series only, for which $\mu>0$ \cite{Hogervorst:2021uvp}. The spectral density~$\rho_\sigma(\mu)$ is given by the following formula: 
\begin{align}
    \rho_\sigma(\mu)&=\frac{\mu\sinh(\pi\mu)}{2^3\pi^{d/2+2}\Gamma(d/2)}\frac{\Gamma^2(\frac{\Delta}{2})\Gamma^2(\frac{d-\Delta}{2})}{\Gamma(\Delta)\Gamma(d-\Delta)}\nonumber\\
&\Gamma\left(\frac{2\Delta_\sigma+\Delta-2d}{2}\right)\Gamma\left(\frac{2\Delta_\sigma-\Delta}{2}\right)\Gamma\left(\frac{d-2\Delta_\sigma+\Delta}{2}\right)\Gamma\left(\frac{2d-2\Delta_\sigma+\Delta}{2}\right)\,,
\label{spectral_density}
\end{align}
where $\Delta=\frac{3}{2}+i\mu$ and $\Delta_\sigma=\frac{3}{2}+i\mu_\sigma$.  
The spectral density, written in this form, is manifestly positive and explicitly invariant under the shadow transformation $\Delta \to d - \Delta$. 

We will now show that, in the flat-space limit, the combination $\rho_\sigma/\mu$ asymptotically approaches the spectral density appearing in the Källén–Lehmann decomposition of the $\sigma^2$ operator in Minkowski, namely
\begin{align}
    \rho_\sigma^{\mathrm{mink}}(m'^2)=\dfrac{1}{32\pi^2}\left(1-\dfrac{4m^2}{m'^2}\right)^{1/2} \theta(m'^2-4m^2)\,,
    \label{minkowski_spectral}
\end{align}
where $m'$ is the mass parameter appearing in $\mu$.
To see this, we first note that there are three asymptotic cases in \eqref{spectral_density} depending on the relative size of $2\Delta_\sigma-\Delta$. For each case, we will expand the $\Gamma$ functions for large or small parameters. It will come in handy to use the following approximation 
\begin{align}
\Gamma(A+i\mu/2)\approx\sqrt{2\pi}\left(\vert\mu\vert/2\right)^{A-1/2} e^{-\pi\vert\mu\vert/4+\frac{i}{4}(\mathrm{sign}(\mu)(1-2A)\pi+2\mu(1-\log(\mu/2)))+\mathcal{O}(1/\mu)}\,,
   \label{complex_Gamma_asym}
\end{align}
which is valid for $\mu\gg 1$ and $A>0$. Notice that the real part in the exponential depends only on the absolute value of the imaginary part of the argument of the $\Gamma$ function. 

The first case is when $\mu\ll\mu_\sigma$. Given that $\mu_\sigma\gg 1 $, we can use the formula in Eq.\,\eqref{complex_Gamma_asym} for all the Gamma functions and obtain: 
\begin{align}
    \frac{\rho_\sigma}{\mu} \sim e^{-\frac{2\mu_\sigma}{\pi}+\mathcal{O}(1/\mu_\sigma)},\qquad\vert\mu\vert\ll 1.
\end{align}
This effect is exponentially suppressed but remains nonzero. This is not surprising, as particle creation exists in de Sitter space even below the naive particle production threshold in flat space  (i.e., $m'<2m$).
The expression above, however, ceases to be valid when  $\vert\mu-\mu_\sigma\vert \leq 1$. In this regime, we can approximate the Gamma function for small real $\delta$,
\begin{align}
    \Gamma(3/4+i\delta/2)\approx\Gamma(3/4)\left(1-\frac{\delta}{4}(2\gamma_E-\pi-\log(64))+\mathcal{O}(\delta^2)\right)
\end{align}
where this form is applied to the Gamma functions with argument $\vert\Delta-\Delta_\sigma\vert$. For the remaining terms, we use the asymptotic expression for the Gamma function from \eqref{complex_Gamma_asym}, which yields the following result,
\begin{align}
      & \frac{ \rho_\sigma(\mu)}{\mu}=\frac{\mu  \Gamma \left(\frac{3}{4}\right)^2 e^{-\frac{3 \pi  \mu }{2}-\pi  \mu_\sigma+\mathcal{O}(\mu-\mu_\sigma)^{-2}} \sqrt{\frac{\mu }{2}+\mu_\sigma} }{8 \pi ^2 \left(4 \mu ^2+1\right)}\nonumber\\
      &\times\sinh (2 \pi  \mu ) \left((2 \gamma_E -\pi
   +\log (64))^2 (\mu -2 \mu_\sigma)^2+16+\mathcal{O}((\mu-2\mu_\sigma)^3)\right), \qquad \mathrm{for} \ \ \vert\mu- 2\mu_\sigma\vert\leq 1.
   \label{approx2}
\end{align}
We observe that as  $\mu$ increases, the exponential suppression, coming from the factors of $\mu_\sigma$from the first limit, diminishes until it is eventually overcome. Additionally, notice that when $\mu=2\mu_\sigma$ the spectral density is non-zero, again because the spectral density has  support for particle creation below the threshold. 
Finally, when $\vert \mu-2\mu_\sigma\vert\geq 1$, we have
\begin{align}
   \frac{ \rho_\sigma(\mu)}{\mu}= 
			\frac{2\mu^2}{\pi(1+4\mu^2)}\left\vert 1-\frac{4\mu_\sigma^2}{4\mu^2}\right\vert^{1/2} \sinh(\pi\mu)e^{-\frac{\pi}{2}\left(3\mu+2\mu_\sigma+\vert\mu-2\mu_\sigma\vert\right)+\mathcal{O}(\mu^{-2})} \qquad  (\vert\mu- 2\mu_\sigma\vert\geq 1)\,.
   \label{approx1}
\end{align}
In this case, we observe that the spectral density increases monotonically until $\mu$ becomes significantly larger than $\mu_\sigma$, at which point it asymptotes to a constant value, resembling the behavior in Minkowski space.

The spectral density in de Sitter space differs from its Minkowski counterpart in several key aspects. First, spontaneous particle creation in de Sitter leads to a nonzero spectral density even for particles below the energy threshold, albeit with an exponentially small contribution. This contrasts with Minkowski space, where the spectral density is strictly zero below threshold, reflecting the absence of spontaneous particle creation in flat spacetime.
Around the point $\mu = 2\mu_\sigma$, there is a subtle difference in de Sitter, attributed to particle creation, though this effect is minor. For $\mu \geq 2\mu_\sigma$, however, the spectral density behaves as
\begin{align}
   \frac{ \rho_\sigma(\mu)}{\mu}\sim \frac{1}{4\pi}\left( 1-\frac{4\mu_\sigma^2}{\mu^2}\right)^{1/2}\left(1+\mathcal{O}(e^{-2\pi\mu})\right), 
\end{align}
very similarly as the Minkowski spectral density. 
Finally we show that in the MFS limit $\rho_\sigma(\mu)/\mu\to\ 8\pi \rho_\sigma^{\mathrm{mink}}$. To see this, we reintroduce $H$ on the expressions above, and we considerthree cases previously discussed. First, in the regime where  $2\mu_\sigma\ll \mu$ the spectral density scales as
\begin{align}
    \rho_{\sigma}(\mu)/\mu\sim \frac{1}{4\pi}\left(\frac{4m^2}{m'}-1\right)^{1/2}e^{-\frac{\pi}{H}(2m'-m)}\to 0
\end{align}
In the intermediate regime, where $\vert 2\mu_\phi-\mu\vert\leq 1$ we get that
\begin{align}
        \rho_\sigma(\mu)/\mu\sim e^{-\frac{\pi}{2H}(3m+m')}\sqrt{\frac{m}{H}}\to 0
\end{align}
which is also exponentially suppressed when taking the flat-space limit. Finally  when  $|\mu-2\mu_\sigma|\geq 1$ we have that
\begin{align}
        \rho_\sigma(\mu)/\mu\sim  \frac{1}{4\pi}\left(1-\frac{4m^2}{m'}\right)^{1/2}(1-\mathcal{O}(H^2/m^2)),
\end{align}
in agreement with \cite{Loparco:2023rug}. We can now discuss the relation between the flat space limit of the spectral density and the  MFS reduction formula. Starting from the heavy graph described in \eqref{FSB}, the spectral decomposition replaces the bubble diagram with the simplified expression in \eqref{squareprop}. In the massive flat-space limit, where $m/H \to \infty$, the heavy graph simplifies to the standard MFS reduction formula, as shown in \eqref{flatspace}. In this formulation, the amputated propagator becomes:

\begin{align}
G_n(p_1(\eta), \dots, p_n(\eta)) = (i g_L)(i g_R) \int \frac{d^4 \bar{q}}{(2\pi)^4} \frac{-i}{\bar{q}^2 + m^2 + i\epsilon} \frac{-i}{(\bar{q} - \sum_{i=1}^{n_L} p_i(\eta))^2 + m^2 + i\epsilon}.
\end{align}
Using the spectral decomposition, this can be equivalently expressed as:
\begin{align}
G_n(p_1(\eta), \dots, p_n(\eta)) = (i g_L)(i g_R) \int d m' \, m' \, \rho_\sigma^{\mathrm{mink}}(m'^2) \frac{i}{(p_1(\eta) + \dots + p_{n_L}(\eta))^2 + m'^2}.
\end{align}
Here, the spectral density  $\rho_\sigma^{\mathrm{mink}}(m'^2)$  approximates the spectral density of the one-loop graph in de Sitter, in the limit $m/H \to \infty$.

Finally, we note that the UV divergence of the spectral integral can be regularized by introducing either a hard cutoff or using dim-reg, although the specifics of this regularization procedure are not central to the analysis presented above.

\bibliographystyle{JHEP}
\bibliography{refs}

\providecommand{\href}[2]{#2}\begingroup\raggedright\begin{thebibliography}{100}

\bibitem{Maldacena:2002vr}
J.M.~Maldacena, \emph{{Non-Gaussian features of primordial fluctuations in
  single field inflationary models}},
  \href{https://doi.org/10.1088/1126-6708/2003/05/013}{\emph{JHEP} {\bfseries
  05} (2003) 013} [\href{https://arxiv.org/abs/astro-ph/0210603}{{\ttfamily
  astro-ph/0210603}}].

\bibitem{Achucarro:2022qrl}
A.~Ach\'ucarro et~al., \emph{{Inflation: Theory and Observations}},
  \href{https://arxiv.org/abs/2203.08128}{{\ttfamily 2203.08128}}.

\bibitem{Maldacena:2011nz}
J.M.~Maldacena and G.L.~Pimentel, \emph{{On graviton non-Gaussianities during
  inflation}}, \href{https://doi.org/10.1007/JHEP09(2011)045}{\emph{JHEP}
  {\bfseries 09} (2011) 045} [\href{https://arxiv.org/abs/1104.2846}{{\ttfamily
  1104.2846}}].

\bibitem{Bzowski:2011ab}
A.~Bzowski, P.~McFadden and K.~Skenderis, \emph{{Holographic predictions for
  cosmological 3-point functions}},
  \href{https://doi.org/10.1007/JHEP03(2012)091}{\emph{JHEP} {\bfseries 03}
  (2012) 091} [\href{https://arxiv.org/abs/1112.1967}{{\ttfamily 1112.1967}}].

\bibitem{Creminelli:2011mw}
P.~Creminelli, \emph{{Conformal invariance of scalar perturbations in
  inflation}}, \href{https://doi.org/10.1103/PhysRevD.85.041302}{\emph{Phys.
  Rev. D} {\bfseries 85} (2012) 041302}
  [\href{https://arxiv.org/abs/1108.0874}{{\ttfamily 1108.0874}}].

\bibitem{Mata:2012bx}
I.~Mata, S.~Raju and S.~Trivedi, \emph{{CMB from CFT}},
  \href{https://doi.org/10.1007/JHEP07(2013)015}{\emph{JHEP} {\bfseries 07}
  (2013) 015} [\href{https://arxiv.org/abs/1211.5482}{{\ttfamily 1211.5482}}].

\bibitem{Bzowski:2012ih}
A.~Bzowski, P.~McFadden and K.~Skenderis, \emph{{Holography for inflation using
  conformal perturbation theory}},
  \href{https://doi.org/10.1007/JHEP04(2013)047}{\emph{JHEP} {\bfseries 04}
  (2013) 047} [\href{https://arxiv.org/abs/1211.4550}{{\ttfamily 1211.4550}}].

\bibitem{Bzowski:2013sza}
A.~Bzowski, P.~McFadden and K.~Skenderis, \emph{{Implications of conformal
  invariance in momentum space}},
  \href{https://doi.org/10.1007/JHEP03(2014)111}{\emph{JHEP} {\bfseries 03}
  (2014) 111} [\href{https://arxiv.org/abs/1304.7760}{{\ttfamily 1304.7760}}].

\bibitem{Arkani-Hamed:2015bza}
N.~Arkani-Hamed and J.~Maldacena, \emph{{Cosmological Collider Physics}},
  \href{https://arxiv.org/abs/1503.08043}{{\ttfamily 1503.08043}}.

\bibitem{Arkani-Hamed:2017fdk}
N.~Arkani-Hamed, P.~Benincasa and A.~Postnikov, \emph{{Cosmological Polytopes
  and the Wavefunction of the Universe}},
  \href{https://arxiv.org/abs/1709.02813}{{\ttfamily 1709.02813}}.

\bibitem{Arkani-Hamed:2018kmz}
N.~Arkani-Hamed, D.~Baumann, H.~Lee and G.L.~Pimentel, \emph{{The Cosmological
  Bootstrap: Inflationary Correlators from Symmetries and Singularities}},
  \href{https://doi.org/10.1007/JHEP04(2020)105}{\emph{JHEP} {\bfseries 04}
  (2020) 105} [\href{https://arxiv.org/abs/1811.00024}{{\ttfamily
  1811.00024}}].

\bibitem{Arkani-Hamed:2018bjr}
N.~Arkani-Hamed and P.~Benincasa, \emph{{On the Emergence of Lorentz Invariance
  and Unitarity from the Scattering Facet of Cosmological Polytopes}},
  \href{https://arxiv.org/abs/1811.01125}{{\ttfamily 1811.01125}}.

\bibitem{Baumann:2019oyu}
D.~Baumann, C.~Duaso~Pueyo, A.~Joyce, H.~Lee and G.L.~Pimentel, \emph{{The
  cosmological bootstrap: weight-shifting operators and scalar seeds}},
  \href{https://doi.org/10.1007/JHEP12(2020)204}{\emph{JHEP} {\bfseries 12}
  (2020) 204} [\href{https://arxiv.org/abs/1910.14051}{{\ttfamily
  1910.14051}}].

\bibitem{Benincasa:2019vqr}
P.~Benincasa, \emph{{Cosmological Polytopes and the Wavefuncton of the Universe
  for Light States}},  \href{https://arxiv.org/abs/1909.02517}{{\ttfamily
  1909.02517}}.

\bibitem{COT}
H.~Goodhew, S.~Jazayeri and E.~Pajer, \emph{{The Cosmological Optical
  Theorem}}, \href{https://doi.org/10.1088/1475-7516/2021/04/021}{\emph{JCAP}
  {\bfseries 04} (2021) 021}
  [\href{https://arxiv.org/abs/2009.02898}{{\ttfamily 2009.02898}}].

\bibitem{Cespedes:2020xqq}
S.~C\'espedes, A.-C.~Davis and S.~Melville, \emph{{On the time evolution of
  cosmological correlators}},
  \href{https://doi.org/10.1007/JHEP02(2021)012}{\emph{JHEP} {\bfseries 02}
  (2021) 012} [\href{https://arxiv.org/abs/2009.07874}{{\ttfamily
  2009.07874}}].

\bibitem{Baumann:2020dch}
D.~Baumann, C.~Duaso~Pueyo, A.~Joyce, H.~Lee and G.L.~Pimentel, \emph{{The
  Cosmological Bootstrap: Spinning Correlators from Symmetries and
  Factorization}},
  \href{https://doi.org/10.21468/SciPostPhys.11.3.071}{\emph{SciPost Phys.}
  {\bfseries 11} (2021) 071}
  [\href{https://arxiv.org/abs/2005.04234}{{\ttfamily 2005.04234}}].

\bibitem{Benincasa:2020aoj}
P.~Benincasa, A.J.~McLeod and C.~Vergu, \emph{{Steinmann Relations and the
  Wavefunction of the Universe}},
  \href{https://doi.org/10.1103/PhysRevD.102.125004}{\emph{Phys. Rev. D}
  {\bfseries 102} (2020) 125004}
  [\href{https://arxiv.org/abs/2009.03047}{{\ttfamily 2009.03047}}].

\bibitem{Pajer:2020wxk}
E.~Pajer, \emph{{Building a Boostless Bootstrap for the Bispectrum}},
  \href{https://doi.org/10.1088/1475-7516/2021/01/023}{\emph{JCAP} {\bfseries
  01} (2021) 023} [\href{https://arxiv.org/abs/2010.12818}{{\ttfamily
  2010.12818}}].

\bibitem{Jazayeri:2021fvk}
S.~Jazayeri, E.~Pajer and D.~Stefanyszyn, \emph{{From locality and unitarity to
  cosmological correlators}},
  \href{https://doi.org/10.1007/JHEP10(2021)065}{\emph{JHEP} {\bfseries 10}
  (2021) 065} [\href{https://arxiv.org/abs/2103.08649}{{\ttfamily
  2103.08649}}].

\bibitem{Baumann:2021fxj}
D.~Baumann, W.-M.~Chen, C.~Duaso~Pueyo, A.~Joyce, H.~Lee and G.L.~Pimentel,
  \emph{{Linking the Singularities of Cosmological Correlators}},
  \href{https://arxiv.org/abs/2106.05294}{{\ttfamily 2106.05294}}.

\bibitem{Goodhew:2021oqg}
H.~Goodhew, S.~Jazayeri, M.H.~Gordon~Lee and E.~Pajer, \emph{{Cutting
  cosmological correlators}},
  \href{https://doi.org/10.1088/1475-7516/2021/08/003}{\emph{JCAP} {\bfseries
  08} (2021) 003} [\href{https://arxiv.org/abs/2104.06587}{{\ttfamily
  2104.06587}}].

\bibitem{Sleight:2021plv}
C.~Sleight and M.~Taronna, \emph{{From dS to AdS and back}},
  \href{https://doi.org/10.1007/JHEP12(2021)074}{\emph{JHEP} {\bfseries 12}
  (2021) 074} [\href{https://arxiv.org/abs/2109.02725}{{\ttfamily
  2109.02725}}].

\bibitem{Bonifacio:2021azc}
J.~Bonifacio, E.~Pajer and D.-G.~Wang, \emph{{From amplitudes to contact
  cosmological correlators}},
  \href{https://doi.org/10.1007/JHEP10(2021)001}{\emph{JHEP} {\bfseries 10}
  (2021) 001} [\href{https://arxiv.org/abs/2106.15468}{{\ttfamily
  2106.15468}}].

\bibitem{Benincasa:2022gtd}
P.~Benincasa, \emph{{Amplitudes meet Cosmology: A (Scalar) Primer}},
  \href{https://arxiv.org/abs/2203.15330}{{\ttfamily 2203.15330}}.

\bibitem{Penedones:2023uqc}
J.~Penedones, K.~Salehi~Vaziri and Z.~Sun, \emph{{Hilbert space of Quantum
  Field Theory in de Sitter spacetime}},
  \href{https://arxiv.org/abs/2301.04146}{{\ttfamily 2301.04146}}.

\bibitem{AguiSalcedo:2023nds}
S.~Agui~Salcedo and S.~Melville, \emph{{The cosmological tree theorem}},
  \href{https://doi.org/10.1007/JHEP12(2023)076}{\emph{JHEP} {\bfseries 12}
  (2023) 076} [\href{https://arxiv.org/abs/2308.00680}{{\ttfamily
  2308.00680}}].

\bibitem{Albayrak:2023hie}
S.~Albayrak, P.~Benincasa and C.~Duaso~Pueyo, \emph{{Perturbative unitarity and
  the wavefunction of the Universe}},
  \href{https://doi.org/10.21468/SciPostPhys.16.6.157}{\emph{SciPost Phys.}
  {\bfseries 16} (2024) 157}
  [\href{https://arxiv.org/abs/2305.19686}{{\ttfamily 2305.19686}}].

\bibitem{Loparco:2023rug}
M.~Loparco, J.~Penedones, K.~Salehi~Vaziri and Z.~Sun, \emph{{The
  K\"all\'en-Lehmann representation in de Sitter spacetime}},
  \href{https://doi.org/10.1007/JHEP12(2023)159}{\emph{JHEP} {\bfseries 12}
  (2023) 159} [\href{https://arxiv.org/abs/2306.00090}{{\ttfamily
  2306.00090}}].

\bibitem{Lee:2024sks}
M.H.G.~Lee, E.~Pajer, M.~Giroux, H.S.~Hannesdottir, S.~Mizera and
  C.~Pasiecznik, \emph{{Records from the S-Matrix Marathon: A Timeless History
  of Time}},  9, 2024 [\href{https://arxiv.org/abs/2410.00227}{{\ttfamily
  2410.00227}}].

\bibitem{SalehiVaziri:2024joi}
K.~Salehi~Vaziri, \emph{{A non-perturbative construction of the de Sitter
  late-time boundary}},  \href{https://arxiv.org/abs/2412.00183}{{\ttfamily
  2412.00183}}.

\bibitem{Goodhew:2024eup}
H.~Goodhew, A.~Thavanesan and A.C.~Wall, \emph{{The Cosmological CPT Theorem}},
   \href{https://arxiv.org/abs/2408.17406}{{\ttfamily 2408.17406}}.

\bibitem{Stefanyszyn:2024msm}
D.~Stefanyszyn, X.~Tong and Y.~Zhu, \emph{{There and Back Again: Mapping and
  Factorizing Cosmological Observables}},
  \href{https://doi.org/10.1103/PhysRevLett.133.221501}{\emph{Phys. Rev. Lett.}
  {\bfseries 133} (2024) 221501}
  [\href{https://arxiv.org/abs/2406.00099}{{\ttfamily 2406.00099}}].

\bibitem{Stefanyszyn:2023qov}
D.~Stefanyszyn, X.~Tong and Y.~Zhu, \emph{{Cosmological correlators through the
  looking glass: reality, parity, and factorisation}},
  \href{https://doi.org/10.1007/JHEP05(2024)196}{\emph{JHEP} {\bfseries 05}
  (2024) 196} [\href{https://arxiv.org/abs/2309.07769}{{\ttfamily
  2309.07769}}].

\bibitem{Melville:2024ove}
S.~Melville and G.L.~Pimentel, \emph{{A de Sitter S-matrix from amputated
  cosmological correlators}},
  \href{https://doi.org/10.1007/JHEP08(2024)211}{\emph{JHEP} {\bfseries 08}
  (2024) 211} [\href{https://arxiv.org/abs/2404.05712}{{\ttfamily
  2404.05712}}].

\bibitem{Melville:2023kgd}
S.~Melville and G.L.~Pimentel, \emph{{de Sitter S matrix for the masses}},
  \href{https://doi.org/10.1103/PhysRevD.110.103530}{\emph{Phys. Rev. D}
  {\bfseries 110} (2024) 103530}
  [\href{https://arxiv.org/abs/2309.07092}{{\ttfamily 2309.07092}}].

\bibitem{Baumann:2022jpr}
D.~Baumann, D.~Green, A.~Joyce, E.~Pajer, G.L.~Pimentel, C.~Sleight et~al.,
  \emph{{Snowmass White Paper: The Cosmological Bootstrap}},  in \emph{{2022
  Snowmass Summer Study}}, 3, 2022
  [\href{https://arxiv.org/abs/2203.08121}{{\ttfamily 2203.08121}}].

\bibitem{Gomez:2021qfd}
H.~Gomez, R.L.~Jusinskas and A.~Lipstein, \emph{{Cosmological Scattering
  Equations}},
  \href{https://doi.org/10.1103/PhysRevLett.127.251604}{\emph{Phys. Rev. Lett.}
  {\bfseries 127} (2021) 251604}
  [\href{https://arxiv.org/abs/2106.11903}{{\ttfamily 2106.11903}}].

\bibitem{Gomez:2021ujt}
H.~Gomez, R.L.~Jusinskas and A.~Lipstein, \emph{{Cosmological Scattering
  Equations at Tree-level and One-loop}},
  \href{https://arxiv.org/abs/2112.12695}{{\ttfamily 2112.12695}}.

\bibitem{Arkani-Hamed:2023bsv}
N.~Arkani-Hamed, D.~Baumann, A.~Hillman, A.~Joyce, H.~Lee and G.L.~Pimentel,
  \emph{{Kinematic Flow and the Emergence of Time}},
  \href{https://arxiv.org/abs/2312.05300}{{\ttfamily 2312.05300}}.

\bibitem{Arkani-Hamed:2023kig}
N.~Arkani-Hamed, D.~Baumann, A.~Hillman, A.~Joyce, H.~Lee and G.L.~Pimentel,
  \emph{{Differential Equations for Cosmological Correlators}},
  \href{https://arxiv.org/abs/2312.05303}{{\ttfamily 2312.05303}}.

\bibitem{Chen:2023iix}
J.~Chen and B.~Feng, \emph{{Towards systematic evaluation of de Sitter
  correlators via Generalized Integration-By-Parts relations}},
  \href{https://doi.org/10.1007/JHEP06(2024)199}{\emph{JHEP} {\bfseries 06}
  (2024) 199} [\href{https://arxiv.org/abs/2401.00129}{{\ttfamily
  2401.00129}}].

\bibitem{De:2023xue}
S.~De and A.~Pokraka, \emph{{Cosmology meets cohomology}},
  \href{https://doi.org/10.1007/JHEP03(2024)156}{\emph{JHEP} {\bfseries 03}
  (2024) 156} [\href{https://arxiv.org/abs/2308.03753}{{\ttfamily
  2308.03753}}].

\bibitem{Hang:2024xas}
Y.~Hang and C.~Shen, \emph{{A Note on Kinematic Flow and Differential Equations
  for Two-Site One-Loop Graph in FRW Spacetime}},
  \href{https://arxiv.org/abs/2410.17192}{{\ttfamily 2410.17192}}.

\bibitem{Baumann:2024mvm}
D.~Baumann, H.~Goodhew and H.~Lee, \emph{{Kinematic Flow for Cosmological Loop
  Integrands}},  \href{https://arxiv.org/abs/2410.17994}{{\ttfamily
  2410.17994}}.

\bibitem{Benincasa:2024ptf}
P.~Benincasa, G.~Brunello, M.K.~Mandal, P.~Mastrolia and F.~Vaz\~ao, \emph{{On
  one-loop corrections to the Bunch-Davies wavefunction of the universe}},
  \href{https://arxiv.org/abs/2408.16386}{{\ttfamily 2408.16386}}.

\bibitem{Grimm:2024mbw}
T.W.~Grimm, A.~Hoefnagels and M.~van Vliet, \emph{{Structure and complexity of
  cosmological correlators}},
  \href{https://doi.org/10.1103/PhysRevD.110.123531}{\emph{Phys. Rev. D}
  {\bfseries 110} (2024) 123531}
  [\href{https://arxiv.org/abs/2404.03716}{{\ttfamily 2404.03716}}].

\bibitem{De:2024zic}
S.~De and A.~Pokraka, \emph{{A physical basis for cosmological correlators from
  cuts}},  \href{https://arxiv.org/abs/2411.09695}{{\ttfamily 2411.09695}}.

\bibitem{Chen:2024glu}
J.~Chen, B.~Feng and Y.-X.~Tao, \emph{{Multivariate hypergeometric solutions of
  cosmological (dS) correlators by $\text{d} \log$-form differential
  equations}},  \href{https://arxiv.org/abs/2411.03088}{{\ttfamily
  2411.03088}}.

\bibitem{Melville:2021lst}
S.~Melville and E.~Pajer, \emph{{Cosmological Cutting Rules}},
  \href{https://doi.org/10.1007/JHEP05(2021)249}{\emph{JHEP} {\bfseries 05}
  (2021) 249} [\href{https://arxiv.org/abs/2103.09832}{{\ttfamily
  2103.09832}}].

\bibitem{Tong:2021wai}
X.~Tong, Y.~Wang and Y.~Zhu, \emph{{Cutting Rule for Cosmological Collider
  Signals: A Bulk Evolution Perspective}},
  \href{https://arxiv.org/abs/2112.03448}{{\ttfamily 2112.03448}}.

\bibitem{Qin:2023bjk}
Z.~Qin and Z.-Z.~Xianyu, \emph{{Inflation correlators at the one-loop order:
  nonanalyticity, factorization, cutting rule, and OPE}},
  \href{https://doi.org/10.1007/JHEP09(2023)116}{\emph{JHEP} {\bfseries 09}
  (2023) 116} [\href{https://arxiv.org/abs/2304.13295}{{\ttfamily
  2304.13295}}].

\bibitem{Ghosh:2024aqd}
D.~Ghosh, E.~Pajer and F.~Ullah, \emph{{Cosmological cutting rules for
  Bogoliubov initial states}},
  \href{https://arxiv.org/abs/2407.06258}{{\ttfamily 2407.06258}}.

\bibitem{Sleight:2019hfp}
C.~Sleight and M.~Taronna, \emph{{Bootstrapping Inflationary Correlators in
  Mellin Space}}, \href{https://doi.org/10.1007/JHEP02(2020)098}{\emph{JHEP}
  {\bfseries 02} (2020) 098}
  [\href{https://arxiv.org/abs/1907.01143}{{\ttfamily 1907.01143}}].

\bibitem{Sleight:2019mgd}
C.~Sleight, \emph{{A Mellin Space Approach to Cosmological Correlators}},
  \href{https://doi.org/10.1007/JHEP01(2020)090}{\emph{JHEP} {\bfseries 01}
  (2020) 090} [\href{https://arxiv.org/abs/1906.12302}{{\ttfamily
  1906.12302}}].

\bibitem{Sleight:2020obc}
C.~Sleight and M.~Taronna, \emph{{From AdS to dS Exchanges: Spectral
  Representation, Mellin Amplitudes and Crossing}},
  \href{https://arxiv.org/abs/2007.09993}{{\ttfamily 2007.09993}}.

\bibitem{Chopping:2024oiu}
A.J.~Chopping, C.~Sleight and M.~Taronna, \emph{{Cosmological correlators for
  Bogoliubov initial states}},
  \href{https://doi.org/10.1007/JHEP09(2024)152}{\emph{JHEP} {\bfseries 09}
  (2024) 152} [\href{https://arxiv.org/abs/2407.16652}{{\ttfamily
  2407.16652}}].

\bibitem{Meltzer:2021bmb}
D.~Meltzer, \emph{{Dispersion Formulas in QFTs, CFTs, and Holography}},
  \href{https://doi.org/10.1007/JHEP05(2021)098}{\emph{JHEP} {\bfseries 05}
  (2021) 098} [\href{https://arxiv.org/abs/2103.15839}{{\ttfamily
  2103.15839}}].

\bibitem{Meltzer:2021zin}
D.~Meltzer, \emph{{The inflationary wavefunction from analyticity and
  factorization}},
  \href{https://doi.org/10.1088/1475-7516/2021/12/018}{\emph{JCAP} {\bfseries
  12} (2021) 018} [\href{https://arxiv.org/abs/2107.10266}{{\ttfamily
  2107.10266}}].

\bibitem{Salcedo:2022aal}
S.A.~Salcedo, M.H.G.~Lee, S.~Melville and E.~Pajer, \emph{{The Analytic
  Wavefunction}}, \href{https://doi.org/10.1007/JHEP06(2023)020}{\emph{JHEP}
  {\bfseries 06} (2023) 020}
  [\href{https://arxiv.org/abs/2212.08009}{{\ttfamily 2212.08009}}].

\bibitem{Liu:2024xyi}
H.~Liu, Z.~Qin and Z.-Z.~Xianyu, \emph{{Dispersive Bootstrap of Massive
  Inflation Correlators}},  \href{https://arxiv.org/abs/2407.12299}{{\ttfamily
  2407.12299}}.

\bibitem{Hogervorst:2021uvp}
M.~Hogervorst, J.a.~Penedones and K.S.~Vaziri, \emph{{Towards the
  non-perturbative cosmological bootstrap}},
  \href{https://doi.org/10.1007/JHEP02(2023)162}{\emph{JHEP} {\bfseries 02}
  (2023) 162} [\href{https://arxiv.org/abs/2107.13871}{{\ttfamily
  2107.13871}}].

\bibitem{DiPietro:2021sjt}
L.~Di~Pietro, V.~Gorbenko and S.~Komatsu, \emph{{Analyticity and Unitarity for
  Cosmological Correlators}},
  \href{https://arxiv.org/abs/2108.01695}{{\ttfamily 2108.01695}}.

\bibitem{Xianyu:2022jwk}
Z.-Z.~Xianyu and H.~Zhang, \emph{{Bootstrapping one-loop inflation correlators
  with the spectral decomposition}},
  \href{https://doi.org/10.1007/JHEP04(2023)103}{\emph{JHEP} {\bfseries 04}
  (2023) 103} [\href{https://arxiv.org/abs/2211.03810}{{\ttfamily
  2211.03810}}].

\bibitem{DiPietro:2023inn}
L.~Di~Pietro, V.~Gorbenko and S.~Komatsu, \emph{{Cosmological Correlators at
  Finite Coupling}},  \href{https://arxiv.org/abs/2312.17195}{{\ttfamily
  2312.17195}}.

\bibitem{Werth:2024mjg}
D.~Werth, \emph{{Spectral representation of cosmological correlators}},
  \href{https://doi.org/10.1007/JHEP12(2024)017}{\emph{JHEP} {\bfseries 12}
  (2024) 017} [\href{https://arxiv.org/abs/2409.02072}{{\ttfamily
  2409.02072}}].

\bibitem{Cabass:2021fnw}
G.~Cabass, E.~Pajer, D.~Stefanyszyn and J.~Supe\l{}, \emph{{Bootstrapping large
  graviton non-Gaussianities}},
  \href{https://doi.org/10.1007/JHEP05(2022)077}{\emph{JHEP} {\bfseries 05}
  (2022) 077} [\href{https://arxiv.org/abs/2109.10189}{{\ttfamily
  2109.10189}}].

\bibitem{Jazayeri:2022kjy}
S.~Jazayeri and S.~Renaux-Petel, \emph{{Cosmological bootstrap in slow
  motion}}, \href{https://doi.org/10.1007/JHEP12(2022)137}{\emph{JHEP}
  {\bfseries 12} (2022) 137}
  [\href{https://arxiv.org/abs/2205.10340}{{\ttfamily 2205.10340}}].

\bibitem{Pimentel:2022fsc}
G.L.~Pimentel and D.-G.~Wang, \emph{{Boostless Cosmological Collider
  Bootstrap}},  \href{https://arxiv.org/abs/2205.00013}{{\ttfamily
  2205.00013}}.

\bibitem{Qin:2022fbv}
Z.~Qin and Z.-Z.~Xianyu, \emph{{Helical inflation correlators: partial
  Mellin-Barnes and bootstrap equations}},
  \href{https://doi.org/10.1007/JHEP04(2023)059}{\emph{JHEP} {\bfseries 04}
  (2023) 059} [\href{https://arxiv.org/abs/2208.13790}{{\ttfamily
  2208.13790}}].

\bibitem{Cabass:2022jda}
G.~Cabass, D.~Stefanyszyn, J.~Supe\l{} and A.~Thavanesan, \emph{{On graviton
  non-Gaussianities in the Effective Field Theory of Inflation}},
  \href{https://doi.org/10.1007/JHEP10(2022)154}{\emph{JHEP} {\bfseries 10}
  (2022) 154} [\href{https://arxiv.org/abs/2209.00677}{{\ttfamily
  2209.00677}}].

\bibitem{Bonifacio:2022vwa}
J.~Bonifacio, H.~Goodhew, A.~Joyce, E.~Pajer and D.~Stefanyszyn, \emph{{The
  graviton four-point function in de Sitter space}},
  \href{https://doi.org/10.1007/JHEP06(2023)212}{\emph{JHEP} {\bfseries 06}
  (2023) 212} [\href{https://arxiv.org/abs/2212.07370}{{\ttfamily
  2212.07370}}].

\bibitem{Wang:2022eop}
D.-G.~Wang, G.L.~Pimentel and A.~Ach\'ucarro, \emph{{Bootstrapping multi-field
  inflation: non-Gaussianities from light scalars revisited}},
  \href{https://doi.org/10.1088/1475-7516/2023/05/043}{\emph{JCAP} {\bfseries
  05} (2023) 043} [\href{https://arxiv.org/abs/2212.14035}{{\ttfamily
  2212.14035}}].

\bibitem{Armstrong:2023phb}
C.~Armstrong, H.~Goodhew, A.~Lipstein and J.~Mei, \emph{{Graviton trispectrum
  from gluons}}, \href{https://doi.org/10.1007/JHEP08(2023)206}{\emph{JHEP}
  {\bfseries 08} (2023) 206}
  [\href{https://arxiv.org/abs/2304.07206}{{\ttfamily 2304.07206}}].

\bibitem{Qin:2023ejc}
Z.~Qin and Z.-Z.~Xianyu, \emph{{Closed-form formulae for inflation
  correlators}}, \href{https://doi.org/10.1007/JHEP07(2023)001}{\emph{JHEP}
  {\bfseries 07} (2023) 001}
  [\href{https://arxiv.org/abs/2301.07047}{{\ttfamily 2301.07047}}].

\bibitem{Jazayeri:2023xcj}
S.~Jazayeri, S.~Renaux-Petel and D.~Werth, \emph{{Shapes of the cosmological
  low-speed collider}},
  \href{https://doi.org/10.1088/1475-7516/2023/12/035}{\emph{JCAP} {\bfseries
  12} (2023) 035} [\href{https://arxiv.org/abs/2307.01751}{{\ttfamily
  2307.01751}}].

\bibitem{Xianyu:2023ytd}
Z.-Z.~Xianyu and J.~Zang, \emph{{Inflation correlators with multiple massive
  exchanges}}, \href{https://doi.org/10.1007/JHEP03(2024)070}{\emph{JHEP}
  {\bfseries 03} (2024) 070}
  [\href{https://arxiv.org/abs/2309.10849}{{\ttfamily 2309.10849}}].

\bibitem{DuasoPueyo:2023kyh}
C.~Duaso~Pueyo and E.~Pajer, \emph{{A cosmological bootstrap for resonant
  non-Gaussianity}}, \href{https://doi.org/10.1007/JHEP03(2024)098}{\emph{JHEP}
  {\bfseries 03} (2024) 098}
  [\href{https://arxiv.org/abs/2311.01395}{{\ttfamily 2311.01395}}].

\bibitem{Chakraborty:2023qbp}
P.~Chakraborty and J.~Stout, \emph{{Light scalars at the cosmological
  collider}}, \href{https://doi.org/10.1007/JHEP02(2024)021}{\emph{JHEP}
  {\bfseries 02} (2024) 021}
  [\href{https://arxiv.org/abs/2310.01494}{{\ttfamily 2310.01494}}].

\bibitem{Chakraborty:2023eoq}
P.~Chakraborty and J.~Stout, \emph{{Compact scalars at the cosmological
  collider}}, \href{https://doi.org/10.1007/JHEP03(2024)149}{\emph{JHEP}
  {\bfseries 03} (2024) 149}
  [\href{https://arxiv.org/abs/2311.09219}{{\ttfamily 2311.09219}}].

\bibitem{Chowdhury:2023khl}
C.~Chowdhury and K.~Singh, \emph{{Analytic results for loop-level momentum
  space Witten diagrams}},
  \href{https://doi.org/10.1007/JHEP12(2023)109}{\emph{JHEP} {\bfseries 12}
  (2023) 109} [\href{https://arxiv.org/abs/2305.18529}{{\ttfamily
  2305.18529}}].

\bibitem{Aoki:2024uyi}
S.~Aoki, L.~Pinol, F.~Sano, M.~Yamaguchi and Y.~Zhu, \emph{{Cosmological
  correlators with double massive exchanges: bootstrap equation and
  phenomenology}}, \href{https://doi.org/10.1007/JHEP09(2024)176}{\emph{JHEP}
  {\bfseries 09} (2024) 176}
  [\href{https://arxiv.org/abs/2404.09547}{{\ttfamily 2404.09547}}].

\bibitem{Qin:2024gtr}
Z.~Qin, \emph{{Cosmological Correlators at the Loop Level}},
  \href{https://arxiv.org/abs/2411.13636}{{\ttfamily 2411.13636}}.

\bibitem{Liu:2024str}
H.~Liu and Z.-Z.~Xianyu, \emph{{Massive Inflationary Amplitudes: Differential
  Equations and Complete Solutions for General Trees}},
  \href{https://arxiv.org/abs/2412.07843}{{\ttfamily 2412.07843}}.

\bibitem{Chowdhury:2023arc}
C.~Chowdhury, A.~Lipstein, J.~Mei, I.~Sachs and P.~Vanhove, \emph{{The Subtle
  Simplicity of Cosmological Correlators}},
  \href{https://arxiv.org/abs/2312.13803}{{\ttfamily 2312.13803}}.

\bibitem{Chowdhury:2024snc}
C.~Chowdhury, P.~Chowdhury, R.N.~Moga and K.~Singh, \emph{{Loops, recursions,
  and soft limits for fermionic correlators in (A)dS}},
  \href{https://doi.org/10.1007/JHEP10(2024)202}{\emph{JHEP} {\bfseries 10}
  (2024) 202} [\href{https://arxiv.org/abs/2408.00074}{{\ttfamily
  2408.00074}}].

\bibitem{Anninos:2024fty}
D.~Anninos, T.~Anous and A.~Rios~Fukelman, \emph{{De Sitter at all loops: the
  story of the Schwinger model}},
  \href{https://doi.org/10.1007/JHEP08(2024)155}{\emph{JHEP} {\bfseries 08}
  (2024) 155} [\href{https://arxiv.org/abs/2403.16166}{{\ttfamily
  2403.16166}}].

\bibitem{Raju:2012zr}
S.~Raju, \emph{{New Recursion Relations and a Flat Space Limit for AdS/CFT
  Correlators}}, \href{https://doi.org/10.1103/PhysRevD.85.126009}{\emph{Phys.
  Rev. D} {\bfseries 85} (2012) 126009}
  [\href{https://arxiv.org/abs/1201.6449}{{\ttfamily 1201.6449}}].

\bibitem{Pimentel:2012tw}
G.L.~Pimentel, L.~Senatore and M.~Zaldarriaga, \emph{{On Loops in Inflation
  III: Time Independence of zeta in Single Clock Inflation}},
  \href{https://doi.org/10.1007/JHEP07(2012)166}{\emph{JHEP} {\bfseries 07}
  (2012) 166} [\href{https://arxiv.org/abs/1203.6651}{{\ttfamily 1203.6651}}].

\bibitem{Marotta:2024sce}
R.~Marotta, K.~Skenderis and M.~Verma, \emph{{Flat space spinning massive
  amplitudes from momentum space CFT}},
  \href{https://doi.org/10.1007/JHEP08(2024)226}{\emph{JHEP} {\bfseries 08}
  (2024) 226} [\href{https://arxiv.org/abs/2406.06447}{{\ttfamily
  2406.06447}}].

\bibitem{Weinzierl:2022eaz}
S.~Weinzierl, \emph{{Feynman Integrals. A Comprehensive Treatment for Students
  and Researchers}}, UNITEXT for Physics, Springer (2022),
  \href{https://doi.org/10.1007/978-3-030-99558-4}{10.1007/978-3-030-99558-4},
  [\href{https://arxiv.org/abs/2201.03593}{{\ttfamily 2201.03593}}].

\bibitem{Boos:1990rg}
E.E.~Boos and A.I.~Davydychev, \emph{{A Method of evaluating massive Feynman
  integrals}}, \href{https://doi.org/10.1007/BF01016805}{\emph{Theor. Math.
  Phys.} {\bfseries 89} (1991) 1052}.

\bibitem{vanOldenborgh:1990yc}
G.J.~van Oldenborgh, \emph{{FF: A Package to evaluate one loop Feynman
  diagrams}}, \href{https://doi.org/10.1016/0010-4655(91)90002-3}{\emph{Comput.
  Phys. Commun.} {\bfseries 66} (1991) 1}.

\bibitem{tHooft:1978jhc}
G.~'t~Hooft and M.J.G.~Veltman, \emph{{Scalar One Loop Integrals}},
  \href{https://doi.org/10.1016/0550-3213(79)90605-9}{\emph{Nucl. Phys. B}
  {\bfseries 153} (1979) 365}.

\bibitem{Weinberg:2005vy}
S.~Weinberg, \emph{{Quantum contributions to cosmological correlations}},
  \href{https://doi.org/10.1103/PhysRevD.72.043514}{\emph{Phys. Rev. D}
  {\bfseries 72} (2005) 043514}
  [\href{https://arxiv.org/abs/hep-th/0506236}{{\ttfamily hep-th/0506236}}].

\bibitem{Senatore:2009cf}
L.~Senatore and M.~Zaldarriaga, \emph{{On Loops in Inflation}},
  \href{https://doi.org/10.1007/JHEP12(2010)008}{\emph{JHEP} {\bfseries 12}
  (2010) 008} [\href{https://arxiv.org/abs/0912.2734}{{\ttfamily 0912.2734}}].

\bibitem{Baumgart:2019clc}
M.~Baumgart and R.~Sundrum, \emph{{De Sitter Diagrammar and the Resummation of
  Time}}, \href{https://doi.org/10.1007/JHEP07(2020)119}{\emph{JHEP} {\bfseries
  07} (2020) 119} [\href{https://arxiv.org/abs/1912.09502}{{\ttfamily
  1912.09502}}].

\bibitem{Gorbenko:2019rza}
V.~Gorbenko and L.~Senatore, \emph{{$\lambda \phi^4$ in dS}},
  \href{https://arxiv.org/abs/1911.00022}{{\ttfamily 1911.00022}}.

\bibitem{Green:2020txs}
D.~Green and A.~Premkumar, \emph{{Dynamical RG and Critical Phenomena in de
  Sitter Space}}, \href{https://doi.org/10.1007/JHEP04(2020)064}{\emph{JHEP}
  {\bfseries 04} (2020) 064}
  [\href{https://arxiv.org/abs/2001.05974}{{\ttfamily 2001.05974}}].

\bibitem{Wang:2021qez}
L.-T.~Wang, Z.-Z.~Xianyu and Y.-M.~Zhong, \emph{{Precision calculation of
  inflation correlators at one loop}},
  \href{https://doi.org/10.1007/JHEP02(2022)085}{\emph{JHEP} {\bfseries 02}
  (2022) 085} [\href{https://arxiv.org/abs/2109.14635}{{\ttfamily
  2109.14635}}].

\bibitem{Lee:2023jby}
M.H.G.~Lee, C.~McCulloch and E.~Pajer, \emph{{Leading loops in cosmological
  correlators}}, \href{https://doi.org/10.1007/JHEP11(2023)038}{\emph{JHEP}
  {\bfseries 11} (2023) 038}
  [\href{https://arxiv.org/abs/2305.11228}{{\ttfamily 2305.11228}}].

\bibitem{Cespedes:2023aal}
S.~C\'espedes, A.-C.~Davis and D.-G.~Wang, \emph{{On the IR divergences in de
  Sitter space: loops, resummation and the semi-classical wavefunction}},
  \href{https://doi.org/10.1007/JHEP04(2024)004}{\emph{JHEP} {\bfseries 04}
  (2024) 004} [\href{https://arxiv.org/abs/2311.17990}{{\ttfamily
  2311.17990}}].

\bibitem{Beneke:2023wmt}
M.~Beneke, P.~Hager and A.F.~Sanfilippo, \emph{{Cosmological correlators in
  massless \ensuremath{\phi}$^{4}$-theory and the method of regions}},
  \href{https://doi.org/10.1007/JHEP04(2024)006}{\emph{JHEP} {\bfseries 04}
  (2024) 006} [\href{https://arxiv.org/abs/2312.06766}{{\ttfamily
  2312.06766}}].

\bibitem{Ballesteros:2024qqx}
G.~Ballesteros, J.~Gamb\'\i{}n~Egea and F.~Riccardi, \emph{{Finite parts of
  inflationary loops}},  \href{https://arxiv.org/abs/2411.19674}{{\ttfamily
  2411.19674}}.

\bibitem{Chen:2015lza}
X.~Chen, M.H.~Namjoo and Y.~Wang, \emph{{Quantum Primordial Standard Clocks}},
  \href{https://doi.org/10.1088/1475-7516/2016/02/013}{\emph{JCAP} {\bfseries
  02} (2016) 013} [\href{https://arxiv.org/abs/1509.03930}{{\ttfamily
  1509.03930}}].

\bibitem{Chen:2016cbe}
X.~Chen, M.H.~Namjoo and Y.~Wang, \emph{{Probing the Primordial Universe using
  Massive Fields}}, \href{https://doi.org/10.1142/S0218271817400041}{\emph{Int.
  J. Mod. Phys. D} {\bfseries 26} (2016) 1740004}
  [\href{https://arxiv.org/abs/1601.06228}{{\ttfamily 1601.06228}}].

\bibitem{Lee:2016vti}
H.~Lee, D.~Baumann and G.L.~Pimentel, \emph{{Non-Gaussianity as a Particle
  Detector}}, \href{https://doi.org/10.1007/JHEP12(2016)040}{\emph{JHEP}
  {\bfseries 12} (2016) 040}
  [\href{https://arxiv.org/abs/1607.03735}{{\ttfamily 1607.03735}}].

\bibitem{Chen:2016uwp}
X.~Chen, Y.~Wang and Z.-Z.~Xianyu, \emph{{Standard Model Background of the
  Cosmological Collider}},
  \href{https://doi.org/10.1103/PhysRevLett.118.261302}{\emph{Phys. Rev. Lett.}
  {\bfseries 118} (2017) 261302}
  [\href{https://arxiv.org/abs/1610.06597}{{\ttfamily 1610.06597}}].

\bibitem{Chen:2016hrz}
X.~Chen, Y.~Wang and Z.-Z.~Xianyu, \emph{{Standard Model Mass Spectrum in
  Inflationary Universe}},
  \href{https://doi.org/10.1007/JHEP04(2017)058}{\emph{JHEP} {\bfseries 04}
  (2017) 058} [\href{https://arxiv.org/abs/1612.08122}{{\ttfamily
  1612.08122}}].

\bibitem{Kehagias:2017cym}
A.~Kehagias and A.~Riotto, \emph{{On the Inflationary Perturbations of Massive
  Higher-Spin Fields}},
  \href{https://doi.org/10.1088/1475-7516/2017/07/046}{\emph{JCAP} {\bfseries
  07} (2017) 046} [\href{https://arxiv.org/abs/1705.05834}{{\ttfamily
  1705.05834}}].

\bibitem{Chen:2018sce}
X.~Chen, W.Z.~Chua, Y.~Guo, Y.~Wang, Z.-Z.~Xianyu and T.~Xie, \emph{{Quantum
  Standard Clocks in the Primordial Trispectrum}},
  \href{https://doi.org/10.1088/1475-7516/2018/05/049}{\emph{JCAP} {\bfseries
  05} (2018) 049} [\href{https://arxiv.org/abs/1803.04412}{{\ttfamily
  1803.04412}}].

\bibitem{Kumar:2019ebj}
S.~Kumar and R.~Sundrum, \emph{{Cosmological Collider Physics and the
  Curvaton}}, \href{https://doi.org/10.1007/JHEP04(2020)077}{\emph{JHEP}
  {\bfseries 04} (2020) 077}
  [\href{https://arxiv.org/abs/1908.11378}{{\ttfamily 1908.11378}}].

\bibitem{Liu:2019fag}
T.~Liu, X.~Tong, Y.~Wang and Z.-Z.~Xianyu, \emph{{Probing P and CP Violations
  on the Cosmological Collider}},
  \href{https://doi.org/10.1007/JHEP04(2020)189}{\emph{JHEP} {\bfseries 04}
  (2020) 189} [\href{https://arxiv.org/abs/1909.01819}{{\ttfamily
  1909.01819}}].

\bibitem{Wang:2020ioa}
L.-T.~Wang and Z.-Z.~Xianyu, \emph{{Gauge Boson Signals at the Cosmological
  Collider}}, \href{https://doi.org/10.1007/JHEP11(2020)082}{\emph{JHEP}
  {\bfseries 11} (2020) 082}
  [\href{https://arxiv.org/abs/2004.02887}{{\ttfamily 2004.02887}}].

\bibitem{Sou:2021juh}
C.M.~Sou, X.~Tong and Y.~Wang, \emph{{Chemical-potential-assisted particle
  production in FRW spacetimes}},
  \href{https://doi.org/10.1007/JHEP06(2021)129}{\emph{JHEP} {\bfseries 06}
  (2021) 129} [\href{https://arxiv.org/abs/2104.08772}{{\ttfamily
  2104.08772}}].

\bibitem{Li:2020xwr}
L.~Li, S.~Lu, Y.~Wang and S.~Zhou, \emph{{Cosmological Signatures of Superheavy
  Dark Matter}}, \href{https://doi.org/10.1007/JHEP07(2020)231}{\emph{JHEP}
  {\bfseries 07} (2020) 231}
  [\href{https://arxiv.org/abs/2002.01131}{{\ttfamily 2002.01131}}].

\bibitem{Lu:2019tjj}
S.~Lu, Y.~Wang and Z.-Z.~Xianyu, \emph{{A Cosmological Higgs Collider}},
  \href{https://doi.org/10.1007/JHEP02(2020)011}{\emph{JHEP} {\bfseries 02}
  (2020) 011} [\href{https://arxiv.org/abs/1907.07390}{{\ttfamily
  1907.07390}}].

\bibitem{Lu:2021wxu}
Q.~Lu, M.~Reece and Z.-Z.~Xianyu, \emph{{Missing scalars at the cosmological
  collider}}, \href{https://doi.org/10.1007/JHEP12(2021)098}{\emph{JHEP}
  {\bfseries 12} (2021) 098}
  [\href{https://arxiv.org/abs/2108.11385}{{\ttfamily 2108.11385}}].

\bibitem{Pinol:2021aun}
L.~Pinol, S.~Aoki, S.~Renaux-Petel and M.~Yamaguchi, \emph{{Inflationary flavor
  oscillations and the cosmic spectroscopy}},
  \href{https://doi.org/10.1103/PhysRevD.107.L021301}{\emph{Phys. Rev. D}
  {\bfseries 107} (2023) L021301}
  [\href{https://arxiv.org/abs/2112.05710}{{\ttfamily 2112.05710}}].

\bibitem{Cui:2021iie}
Y.~Cui and Z.-Z.~Xianyu, \emph{{Probing Leptogenesis with the Cosmological
  Collider}},  \href{https://arxiv.org/abs/2112.10793}{{\ttfamily 2112.10793}}.

\bibitem{Tong:2022cdz}
X.~Tong and Z.-Z.~Xianyu, \emph{{Large spin-2 signals at the cosmological
  collider}}, \href{https://doi.org/10.1007/JHEP10(2022)194}{\emph{JHEP}
  {\bfseries 10} (2022) 194}
  [\href{https://arxiv.org/abs/2203.06349}{{\ttfamily 2203.06349}}].

\bibitem{Reece:2022soh}
M.~Reece, L.-T.~Wang and Z.-Z.~Xianyu, \emph{{Large-Field Inflation and the
  Cosmological Collider}},  \href{https://arxiv.org/abs/2204.11869}{{\ttfamily
  2204.11869}}.

\bibitem{Chen:2022vzh}
X.~Chen, R.~Ebadi and S.~Kumar, \emph{{Classical Cosmological Collider Physics
  and Primordial Features}},
  \href{https://arxiv.org/abs/2205.01107}{{\ttfamily 2205.01107}}.

\bibitem{Qin:2022lva}
Z.~Qin and Z.-Z.~Xianyu, \emph{{Phase Information in Cosmological Collider
  Signals}},  \href{https://arxiv.org/abs/2205.01692}{{\ttfamily 2205.01692}}.

\bibitem{Craig:2024qgy}
N.~Craig, S.~Kumar and A.~McCune, \emph{{An effective cosmological collider}},
  \href{https://doi.org/10.1007/JHEP07(2024)108}{\emph{JHEP} {\bfseries 07}
  (2024) 108} [\href{https://arxiv.org/abs/2401.10976}{{\ttfamily
  2401.10976}}].

\bibitem{Quintin:2024boj}
J.~Quintin, X.~Chen and R.~Ebadi, \emph{{Fingerprints of a non-inflationary
  universe from massive fields}},
  \href{https://doi.org/10.1088/1475-7516/2024/09/026}{\emph{JCAP} {\bfseries
  09} (2024) 026} [\href{https://arxiv.org/abs/2405.11016}{{\ttfamily
  2405.11016}}].

\bibitem{Bodas:2024hih}
A.~Bodas, E.~Broadberry and R.~Sundrum, \emph{{Grand Unification at the
  Cosmological Collider with Chemical Potential}},
  \href{https://arxiv.org/abs/2409.07524}{{\ttfamily 2409.07524}}.

\bibitem{Gasparotto:2024bku}
F.~Gasparotto, P.~Mazloumi and X.~Xu, \emph{{Differential equations for
  tree--level cosmological correlators with massive states}},
  \href{https://arxiv.org/abs/2411.05632}{{\ttfamily 2411.05632}}.

\bibitem{Pajer:2024ckd}
E.~Pajer, D.-G.~Wang and B.~Zhang, \emph{{The UV Sensitivity of Axion Monodromy
  Inflation}},  \href{https://arxiv.org/abs/2412.05762}{{\ttfamily
  2412.05762}}.

\bibitem{Jazayeri:2023kji}
S.~Jazayeri, S.~Renaux-Petel, X.~Tong, D.~Werth and Y.~Zhu, \emph{{Parity
  Violation from Emergent Non-Locality During Inflation}},
  \href{https://arxiv.org/abs/2308.11315}{{\ttfamily 2308.11315}}.

\bibitem{Chen:2017ryl}
X.~Chen, Y.~Wang and Z.-Z.~Xianyu, \emph{{Schwinger-Keldysh Diagrammatics for
  Primordial Perturbations}},
  \href{https://doi.org/10.1088/1475-7516/2017/12/006}{\emph{JCAP} {\bfseries
  12} (2017) 006} [\href{https://arxiv.org/abs/1703.10166}{{\ttfamily
  1703.10166}}].

\bibitem{Salcedo:2024smn}
S.A.~Salcedo, T.~Colas and E.~Pajer, \emph{{The open effective field theory of
  inflation}}, \href{https://doi.org/10.1007/JHEP10(2024)248}{\emph{JHEP}
  {\bfseries 10} (2024) 248}
  [\href{https://arxiv.org/abs/2404.15416}{{\ttfamily 2404.15416}}].

\bibitem{Burgess:2024eng}
C.P.~Burgess, T.~Colas, R.~Holman, G.~Kaplanek and V.~Vennin, \emph{{Cosmic
  purity lost: perturbative and resummed late-time inflationary decoherence}},
  \href{https://doi.org/10.1088/1475-7516/2024/08/042}{\emph{JCAP} {\bfseries
  08} (2024) 042} [\href{https://arxiv.org/abs/2403.12240}{{\ttfamily
  2403.12240}}].

\bibitem{Burgess:2024heo}
C.P.~Burgess, T.~Colas, R.~Holman and G.~Kaplanek, \emph{{Does decoherence
  violate decoupling?}},  \href{https://arxiv.org/abs/2411.09000}{{\ttfamily
  2411.09000}}.

\bibitem{Green:2024cmx}
D.~Green and G.~Sun, \emph{{Effective Field Theory and In-In Correlators}},
  \href{https://arxiv.org/abs/2412.02739}{{\ttfamily 2412.02739}}.

\bibitem{upcomingYuhang}
S.~Renaux-Petel, X.~Tong, D.~Werth and Y.~Zhu, \emph{{in preparation}}, .

\bibitem{Bhowmick:2024kld}
S.~Bhowmick, D.~Ghosh and F.~Ullah, \emph{{Bispectrum at 1-loop in the
  Effective Field Theory of Inflation}},
  \href{https://doi.org/10.1007/JHEP10(2024)057}{\emph{JHEP} {\bfseries 10}
  (2024) 057} [\href{https://arxiv.org/abs/2405.10374}{{\ttfamily
  2405.10374}}].

\bibitem{Parker:2009uva}
L.E.~Parker and D.~Toms, \emph{{Quantum Field Theory in Curved Spacetime}:
  {Quantized Field and Gravity}}, Cambridge Monographs on Mathematical Physics,
  Cambridge University Press (8, 2009),
  \href{https://doi.org/10.1017/CBO9780511813924}{10.1017/CBO9780511813924}.

\bibitem{Baumann:2011nk}
D.~Baumann and D.~Green, \emph{{Signatures of Supersymmetry from the Early
  Universe}}, \href{https://doi.org/10.1103/PhysRevD.85.103520}{\emph{Phys.
  Rev. D} {\bfseries 85} (2012) 103520}
  [\href{https://arxiv.org/abs/1109.0292}{{\ttfamily 1109.0292}}].

\bibitem{Achucarro:2012sm}
A.~Achucarro, J.-O.~Gong, S.~Hardeman, G.A.~Palma and S.P.~Patil,
  \emph{{Effective theories of single field inflation when heavy fields
  matter}}, \href{https://doi.org/10.1007/JHEP05(2012)066}{\emph{JHEP}
  {\bfseries 05} (2012) 066} [\href{https://arxiv.org/abs/1201.6342}{{\ttfamily
  1201.6342}}].

\bibitem{Achucarro:2012yr}
A.~Achucarro, V.~Atal, S.~Cespedes, J.-O.~Gong, G.A.~Palma and S.P.~Patil,
  \emph{{Heavy fields, reduced speeds of sound and decoupling during
  inflation}}, \href{https://doi.org/10.1103/PhysRevD.86.121301}{\emph{Phys.
  Rev.} {\bfseries D86} (2012) 121301}
  [\href{https://arxiv.org/abs/1205.0710}{{\ttfamily 1205.0710}}].

\bibitem{Bunch:1979uk}
T.S.~Bunch and L.~Parker, \emph{{Feynman Propagator in Curved Space-Time: A
  Momentum Space Representation}},
  \href{https://doi.org/10.1103/PhysRevD.20.2499}{\emph{Phys. Rev. D}
  {\bfseries 20} (1979) 2499}.

\bibitem{Vassilevich:2003xt}
D.V.~Vassilevich, \emph{{Heat kernel expansion: User's manual}},
  \href{https://doi.org/10.1016/j.physrep.2003.09.002}{\emph{Phys. Rept.}
  {\bfseries 388} (2003) 279}
  [\href{https://arxiv.org/abs/hep-th/0306138}{{\ttfamily hep-th/0306138}}].

\bibitem{Avramidi:1986mj}
I.G.~Avramidi, \emph{{Covariant methods for the calculation of the effective
  action in quantum field theory and investigation of higher derivative quantum
  gravity}},  other thesis, 1986,
  [\href{https://arxiv.org/abs/hep-th/9510140}{{\ttfamily hep-th/9510140}}].

\bibitem{Nicolis:2013sga}
A.~Nicolis, R.~Penco, F.~Piazza and R.A.~Rosen, \emph{{More on gapped
  Goldstones at finite density: More gapped Goldstones}},
  \href{https://doi.org/10.1007/JHEP11(2013)055}{\emph{JHEP} {\bfseries 11}
  (2013) 055} [\href{https://arxiv.org/abs/1306.1240}{{\ttfamily 1306.1240}}].

\bibitem{Nicolis:2013lma}
A.~Nicolis, R.~Penco and R.A.~Rosen, \emph{{Relativistic Fluids, Superfluids,
  Solids and Supersolids from a Coset Construction}},
  \href{https://doi.org/10.1103/PhysRevD.89.045002}{\emph{Phys. Rev.}
  {\bfseries D89} (2014) 045002}
  [\href{https://arxiv.org/abs/1307.0517}{{\ttfamily 1307.0517}}].

\bibitem{Nicolis:2015sra}
A.~Nicolis, R.~Penco, F.~Piazza and R.~Rattazzi, \emph{{Zoology of condensed
  matter: Framids, ordinary stuff, extra-ordinary stuff}},
  \href{https://doi.org/10.1007/JHEP06(2015)155}{\emph{JHEP} {\bfseries 06}
  (2015) 155} [\href{https://arxiv.org/abs/1501.03845}{{\ttfamily
  1501.03845}}].

\bibitem{Joyce:2022ydd}
A.~Joyce, A.~Nicolis, A.~Podo and L.~Santoni, \emph{{Integrating out beyond
  tree level and relativistic superfluids}},
  \href{https://doi.org/10.1007/JHEP09(2022)066}{\emph{JHEP} {\bfseries 09}
  (2022) 066} [\href{https://arxiv.org/abs/2204.03678}{{\ttfamily
  2204.03678}}].

\bibitem{Aghanim:2018eyx}
{\scshape Planck} collaboration, \emph{{Planck 2018 results. VI. Cosmological
  parameters}},  \href{https://arxiv.org/abs/1807.06209}{{\ttfamily
  1807.06209}}.

\bibitem{Planck:2019kim}
{\scshape Planck} collaboration, \emph{{Planck 2018 results. IX. Constraints on
  primordial non-Gaussianity}},
  \href{https://doi.org/10.1051/0004-6361/201935891}{\emph{Astron. Astrophys.}
  {\bfseries 641} (2020) A9}
  [\href{https://arxiv.org/abs/1905.05697}{{\ttfamily 1905.05697}}].

\bibitem{Bordin:2018pca}
L.~Bordin, P.~Creminelli, A.~Khmelnitsky and L.~Senatore, \emph{{Light
  Particles with Spin in Inflation}},
  \href{https://doi.org/10.1088/1475-7516/2018/10/013}{\emph{JCAP} {\bfseries
  10} (2018) 013} [\href{https://arxiv.org/abs/1806.10587}{{\ttfamily
  1806.10587}}].

\bibitem{Pinol:2023oux}
L.~Pinol, S.~Renaux-Petel and D.~Werth, \emph{{The Cosmological Flow: A
  Systematic Approach to Primordial Correlators}},
  \href{https://arxiv.org/abs/2312.06559}{{\ttfamily 2312.06559}}.

\bibitem{Grall:2020tqc}
T.~Grall and S.~Melville, \emph{{Inflation in Motion: Unitarity Constraints in
  Effective Field Theories with Broken Lorentz Symmetry}},
  \href{https://arxiv.org/abs/2005.02366}{{\ttfamily 2005.02366}}.

\bibitem{Sohn:2024xzd}
W.~Sohn, D.-G.~Wang, J.R.~Fergusson and E.P.S.~Shellard, \emph{{Searching for
  cosmological collider in the Planck CMB data}},
  \href{https://doi.org/10.1088/1475-7516/2024/09/016}{\emph{JCAP} {\bfseries
  09} (2024) 016} [\href{https://arxiv.org/abs/2404.07203}{{\ttfamily
  2404.07203}}].

\bibitem{Cabass:2024wob}
G.~Cabass, O.H.E.~Philcox, M.M.~Ivanov, K.~Akitsu, S.-F.~Chen, M.~Simonovi\'c
  et~al., \emph{{BOSS Constraints on Massive Particles during Inflation: The
  Cosmological Collider in Action}},
  \href{https://arxiv.org/abs/2404.01894}{{\ttfamily 2404.01894}}.

\bibitem{Polchinski:1999ry}
J.~Polchinski, \emph{{S matrices from AdS space-time}},
  \href{https://arxiv.org/abs/hep-th/9901076}{{\ttfamily hep-th/9901076}}.

\bibitem{Giddings:1999jq}
S.B.~Giddings, \emph{{Flat space scattering and bulk locality in the AdS / CFT
  correspondence}},
  \href{https://doi.org/10.1103/PhysRevD.61.106008}{\emph{Phys. Rev. D}
  {\bfseries 61} (2000) 106008}
  [\href{https://arxiv.org/abs/hep-th/9907129}{{\ttfamily hep-th/9907129}}].

\bibitem{Gary:2009ae}
M.~Gary, S.B.~Giddings and J.~Penedones, \emph{{Local bulk S-matrix elements
  and CFT singularities}},
  \href{https://doi.org/10.1103/PhysRevD.80.085005}{\emph{Phys. Rev. D}
  {\bfseries 80} (2009) 085005}
  [\href{https://arxiv.org/abs/0903.4437}{{\ttfamily 0903.4437}}].

\bibitem{Heemskerk:2009pn}
I.~Heemskerk, J.~Penedones, J.~Polchinski and J.~Sully, \emph{{Holography from
  Conformal Field Theory}},
  \href{https://doi.org/10.1088/1126-6708/2009/10/079}{\emph{JHEP} {\bfseries
  10} (2009) 079} [\href{https://arxiv.org/abs/0907.0151}{{\ttfamily
  0907.0151}}].

\bibitem{Fitzpatrick:2010zm}
A.L.~Fitzpatrick, E.~Katz, D.~Poland and D.~Simmons-Duffin, \emph{{Effective
  Conformal Theory and the Flat-Space Limit of AdS}},
  \href{https://doi.org/10.1007/JHEP07(2011)023}{\emph{JHEP} {\bfseries 07}
  (2011) 023} [\href{https://arxiv.org/abs/1007.2412}{{\ttfamily 1007.2412}}].

\bibitem{Maldacena:2015iua}
J.~Maldacena, D.~Simmons-Duffin and A.~Zhiboedov, \emph{{Looking for a bulk
  point}}, \href{https://doi.org/10.1007/JHEP01(2017)013}{\emph{JHEP}
  {\bfseries 01} (2017) 013}
  [\href{https://arxiv.org/abs/1509.03612}{{\ttfamily 1509.03612}}].

\bibitem{Komatsu:2020sag}
S.~Komatsu, M.F.~Paulos, B.C.~Van~Rees and X.~Zhao, \emph{{Landau diagrams in
  AdS and S-matrices from conformal correlators}},
  \href{https://doi.org/10.1007/JHEP11(2020)046}{\emph{JHEP} {\bfseries 11}
  (2020) 046} [\href{https://arxiv.org/abs/2007.13745}{{\ttfamily
  2007.13745}}].

\bibitem{Li:2021snj}
Y.-Z.~Li, \emph{{Notes on flat-space limit of AdS/CFT}},
  \href{https://doi.org/10.1007/JHEP09(2021)027}{\emph{JHEP} {\bfseries 09}
  (2021) 027} [\href{https://arxiv.org/abs/2106.04606}{{\ttfamily
  2106.04606}}].

\bibitem{Gwyn:2012mw}
R.~Gwyn, G.A.~Palma, M.~Sakellariadou and S.~Sypsas, \emph{{Effective field
  theory of weakly coupled inflationary models}},
  \href{https://doi.org/10.1088/1475-7516/2013/04/004}{\emph{JCAP} {\bfseries
  1304} (2013) 004} [\href{https://arxiv.org/abs/1210.3020}{{\ttfamily
  1210.3020}}].

\bibitem{Gwyn:2014doa}
R.~Gwyn, G.A.~Palma, M.~Sakellariadou and S.~Sypsas, \emph{{On degenerate
  models of cosmic inflation}},
  \href{https://doi.org/10.1088/1475-7516/2014/10/005}{\emph{JCAP} {\bfseries
  1410} (2014) 005} [\href{https://arxiv.org/abs/1406.1947}{{\ttfamily
  1406.1947}}].

\bibitem{Marolf:2010zp}
D.~Marolf and I.A.~Morrison, \emph{{The IR stability of de Sitter: Loop
  corrections to scalar propagators}},
  \href{https://doi.org/10.1103/PhysRevD.82.105032}{\emph{Phys. Rev. D}
  {\bfseries 82} (2010) 105032}
  [\href{https://arxiv.org/abs/1006.0035}{{\ttfamily 1006.0035}}].

\end{thebibliography}\endgroup
\end{document}